\documentclass[aps,prb,twocolumn,superscriptaddress,10pt]{revtex4-2}

\usepackage{amsfonts}
\usepackage{amsmath}
\usepackage{amssymb}
\usepackage{amsthm}
\usepackage{mathrsfs}
\usepackage{latexsym}
\usepackage[table]{xcolor} 
\usepackage{multirow} 
\usepackage{dirtytalk} 
\usepackage{hyperref}
\hypersetup{colorlinks=true, linkcolor=blue, citecolor=red, filecolor=blue, menucolor = green, runcolor = cyan, urlcolor=blue}
\usepackage{subfig} 
\usepackage{enumerate} 
\usepackage{geometry} 
\usepackage{graphicx} 
\usepackage{tikz} 
\usetikzlibrary{quantikz}
\usetikzlibrary{decorations.pathreplacing}
\usetikzlibrary{3d,calc}
\usetikzlibrary{shapes.geometric}
\usepackage{tcolorbox} 
\usepackage{booktabs} 
\usepackage[font=footnotesize,labelfont=bf]{caption}
\usepackage{lipsum}
\usepackage{tikzpeople}
\usepackage{float} 
\usepackage[normalem]{ulem}
\usepackage{comment}
\usepackage{algorithm}
\usepackage{algpseudocode}
\usepackage{adjustbox}
\usepackage{framed}
\usepackage{ragged2e}
\usepackage{orcidlink}

\newcommand{\R}{\mathbb{R}}
\newcommand{\C}{\mathbb{C}}

\newcommand{\N}{\mathbb{N}}

\newcommand{\F}{\mathbb{F}}

\begin{document}

\title{Non-variational supervised quantum kernel methods: a review}

\author{John Tanner\orcidlink{0000-0002-7458-8295}}
\email{john.tanner@uwa.edu.au}
\affiliation{Centre for Quantum Information, Simulation and Algorithms, The University of Western Australia, 35 Stirling Hwy, Crawley WA, 6009, Australia}

\author{Chon-Fai Kam\orcidlink{0000-0002-0012-691X}}
\affiliation{Quantum Theory Group, Dipartimento di Fisica e Chimica Emilio Segrè,\\
Università degli Studi di Palermo, Via Archirafi 36, I-90123 Palermo, Italy}

\author{Jingbo Wang\orcidlink{0000-0001-7544-0084}}
\affiliation{Centre for Quantum Information, Simulation and Algorithms, The University of Western Australia, 35 Stirling Hwy, Crawley WA, 6009, Australia}

\date{\today}

\newgeometry{left=2cm,right=2cm}

\begin{abstract}
Quantum kernel methods (QKMs) have emerged as a prominent framework for supervised quantum machine learning. 
Unlike variational quantum algorithms, which rely on gradient-based optimisation and may suffer from issues such as barren plateaus, non-variational QKMs employ fixed quantum feature maps, with model selection performed classically via convex optimisation and cross-validation. 
This separation of quantum feature embedding from classical training ensures stable optimisation while leveraging quantum circuits to encode data in high-dimensional Hilbert spaces.
In this review, we provide a thorough analysis of non-variational supervised QKMs, covering their foundations in classical kernel theory, constructions of fidelity and projected quantum kernels, and methods for their estimation in practice. 
We examine frameworks for assessing quantum advantage, including generalisation bounds and necessary conditions for separation from classical models, and analyse key challenges such as exponential concentration, dequantisation via tensor-network methods, and the spectral properties of kernel integral operators. 
We further discuss structured problem classes that may enable advantage, and synthesise insights from comparative and hardware studies.
Overall, this review aims to clarify the regimes in which QKMs may offer genuine advantages, and to delineate the conceptual, methodological, and technical obstacles that must be overcome for practical quantum-enhanced learning.
\end{abstract}

\maketitle


\section{Introduction}
\label{sec:Introduction}

Quantum machine learning (QML)~\cite{Biamonte2017Learning,Arunachalam2017QMLSurvey} is considered one of the most promising techniques for exploiting quantum computing technologies.
As quantum processors advance in scale and fidelity, interest has grown in identifying machine learning (ML) models that exploit quantum resources to achieve practical advantages over classical algorithms.
To date, a variety of quantum algorithms have been shown to deliver exponential speed-ups for key ML tasks~\cite{Giovannetti2008QRAM,Harrow2009HHL,Wiebe2012DataFitting,Lloyd2013SupervisedUnsupervised,Lloyd2014QPCA,Rebentrost2014QSVM,Lloyd2016Topological,Cong2016Discriminant,Rebentrost2018QSVD,Zhao2019QGP}.
However, these algorithms often rest on strong assumptions about how data is supplied, leaving it unclear whether the speed-ups arise from data access or from the quantum resources they exploit. These concerns motivated the community to search for algorithms which can achieve speed-ups while only being given classical access to data.

A large majority of the initial attempts to formulate such algorithms were variational in nature~\cite{Benedetti2019Parameterised,Cerezo2021Variational}, employing parameterised quantum circuits trained to minimise a task-dependent loss function using a classical optimiser. 
Representative examples include the variational quantum eigensolver~\cite{Peruzzo2014VQE}, the quantum approximate optimisation algorithm~\cite{Farhi2014QAOA}, quantum neural networks~\cite{Beer2020QNN}, quantum convolutional neural networks (QCNNs)~\cite{Cong2019Convolutional}, and quantum autoencoders~\cite{Romero2017QAE}. 
These approaches are appealing because they enable flexible and expressive models, leverage well-established optimisation techniques such as gradient descent, and are often implementable on noisy intermediate-scale quantum (NISQ) devices~\cite{Preskill2018NISQ}. 
Nevertheless, despite their widespread adoption, variational QML models are increasingly recognised to encounter fundamental scalability barriers. 

A central challenge for variational QML models is the barren plateau phenomenon~\cite{McClean2018BPs,Larocca2025BPsReview}, wherein the gradients of the cost function vanish exponentially with system size. 
A variety of factors—including noise~\cite{Wang2021NoiseInducedBPs}, entanglement~\cite{Marrero2021EntanglementBPs}, expressibility~\cite{Holmes2022ConnectingExpressibility}, and the choice of cost function~\cite{Cerezo2021CostFunctionBPs}—have been demonstrated to induce barren plateaus, thereby rendering classical optimisation intractable for many circuit architectures. 
These developments have motivated the search for strategies and criteria to mitigate barren plateaus, typically by constraining models to operate within a polynomially sized subspace of the Hilbert space, thereby ensuring trainability. 

For QCNNs, it has been shown that the magnitude of gradients vanishes at most polynomially with system size~\cite{Pesah2021Absence}, indicating that QCNNs do not exhibit barren plateaus. 
However, subsequent work demonstrated that QCNNs can, in fact, be efficiently simulated on classical computers~\cite{Bermejo2024QCNN}.
This development raised the important question of whether \emph{any} variational QML model can be efficiently simulated classically, provided that one can establish the absence of barren plateaus~\cite{Cerezo2025ProvableAbsence}. 
Although this question remains open, these issues raise significant concerns regarding the prospect of achieving rigorous advantages with variational quantum algorithms.

These limitations underscore the importance of investigating non-variational QML models, which employ quantum circuits that challenge classical simulability while avoiding variational optimisation of circuit parameters.
Among these approaches, non-variational quantum kernel methods (QKMs)~\cite{Schuld2019Feature,Havlicek2019Supervised,Schuld2021Supervised} (which we refer to simply as QKMs in this review) have emerged as a particularly promising framework, especially for supervised learning tasks.
In this context, quantum circuits are regarded as maps that embed input data into a high-dimensional Hilbert space.
The main idea motivating this approach is that learning may be simplified by embedding data points into such a space.
Crucially, classically representing the resulting embeddings can be prohibitively expensive, while preparing the corresponding states on quantum devices may be comparatively efficient.

Similarities between embedded data points are then estimated using procedures such as the swap test, and the resulting kernel matrix is processed by a classical kernel machine, such as kernel ridge regression (KRR) or a support vector machine (SVM).
This enables stable classical training through convex optimisation, while eliminating the need for quantum-circuit parameter training.
This close integration with classical kernel theory renders non-variational QKMs practically appealing, and has facilitated the identification of necessary conditions for achieving quantum advantages with QKMs.

Despite their advantages, QKMs face notable challenges. One key concern is the exponential concentration (EC) of kernel values for certain families of data embeddings~\cite{Thanasilp2024Exponential}, a pathology that parallels the barren plateau phenomenon. 
EC can cause the kernel matrix to approach the identity, yielding a model that fails to capture features of the training data and therefore fails to generalise. 
Identifying when quantum kernels avoid this pathology and which structural features enable meaningful generalisation remains an active area of investigation. 

Beyond this, important open questions remain, including the sample complexity of estimating kernel values in practice~\cite{Gentinetta2024Complexity,Miroszewski2024QKMShots}, the impact of hardware noise~\cite{Wang2021Towards,Heyraud2022NoisyQKMs}, the role of expressivity~\cite{Jerbi2023Beyond}, and the identification of problem classes for which QKMs may provide computational advantages~\cite{Huang2021Power}. 
Analyses of the spectrum of the associated kernel integral operator~\cite{Kubler2021Inductive} reveal that flat spectra hinder the efficacy of QKMs. 
Although mitigation strategies such as quantum bandwidth tuning~\cite{Shaydulin2022Importance,Canatar2023Bandwidth} have been proposed, these techniques are now understood to compromise potential advantages in other respects~\cite{Slattery2023Numerical}. 
Additionally, dequantisation results further constrain such claims, demonstrating efficient classical simulations of QKMs under realistic data structures and reducing advantages to polynomial separations~\cite{shin2024dequantizing,sahebi2025dequantization}, thereby eliminating the possibility of exponential advantages.

Despite the challenges associated with QKMs, studies have demonstrated provable quantum advantages through QKMs and proposed problem instances that highlight promising prospects~\cite{Liu2021Rigorous,Muser2024Provable,Wu2023Phase,Huang2021Power,Havlicek2019Supervised,Glick2024Covariant}.
Similarly, extensive research has reported empirical studies of QKM performance on domain-specific tasks~\cite{Beaulieu2022ManufacturingDefects,Ragab2022Biomedical,Krunic2022Health,Zhuang2024NonHemolytic,Miroszewski2023Clouds,Wijata2024Soil,Miyabe2023Financial}, while other works have compared quantum and classical kernels on standard benchmarking datasets~\cite{Schnabel2025Scrutiny,AlvarezEstevez2025Benchmarking,Abdulsalam2025Comparative,Egginger2024Hyperparameter}.
In addition, experimental implementations of QKMs have been demonstrated on superconducting transmon~\cite{Peters2021Noisy,Wu2021Application,Agnihotri2026Practical}, trapped-ion~\cite{Suzuki2024TrappedIon}, and integrated photonic processors~\cite{Anai2024Continuous,Yin2025Experimental}, underscoring progress toward practical realizations.

This review synthesises insights from the recent QKM literature, covering theoretical analyses, empirical evaluations, and hardware demonstrations. 
We focus on the foundations and applications of non-variational QKMs, which employ fixed feature maps with hyperparameters tuned via cross-validation for supervised learning.
In particular, we consider their integration with SVMs and KRR for classification and regression, but exclude applications to Gaussian process regression, as this has not been a major focus of the QKM literature.
We discuss a framework for assessing the potential of achieving quantum advantages with QKMs, and provide details about key obstacles, including the EC phenomena, sensitivity to hardware noise, tensor-network based and other dequantisation methods, and considerations of the spectra of the kernel integral operator. 
We finish by discussing structured problems which provide promising routes to advantage, and highlight findings derived from empirical findings and hardware demonstrations. 

While a 2019 review of QKMs~\cite{Mengoni2019QKMReview} covered both variational and non-variational approaches, substantial progress since then motivates the need for an updated survey. 
Similarly, recent large-scale empirical studies have been performed~\cite{Egginger2024Hyperparameter,Bowles2024Benchmarking,Schnabel2025Scrutiny}, but a focused theoretical and practical synthesis of non-variational QKMs remains fragmented. 
This review bridges this gap, discussing in detail the connections between classical kernel foundations, quantum hardware realities, concentration challenges, and provable-advantage problem classes to clarify when and how such methods may yet yield meaningful quantum enhancements.
To this end, in this review we provide a thorough and pedagogical exposition of past and recent progress in non-variational QKMs for supervised learning, highlighting limitations, advances, open questions, and promising directions that may enable future breakthroughs in QML with QKMs.

The structure of this review is as follows. 
Section~\ref{sec:Classical Kernel Methods} introduces the foundations of classical kernel methods, including the classical algorithms that underpin QKMs. 
Section~\ref{sec:Quantum kernel methods} presents two standard classes of quantum kernels and outlines protocols for their estimation on quantum devices. 
Section~\ref{sec:Assessing quantum advantage in quantum kernel methods} examines the framework of Huang et al.~\cite{Huang2021Power}, which enables the assessment of problem instances that may exhibit quantum advantages with QKMs. 
Section~\ref{sec:Challenges for quantum kernel methods} addresses central challenges for QKMs, including EC and its causes, dequantisation methods, the spectrum of quantum kernel integral operators, and bandwidth tuning. 
Section~\ref{sec:Structured problems enabling potential advantage in quantum kernel methods} reviews prior work on problem instances that demonstrate provable quantum advantages or compatibility with problem-inspired quantum kernels. 
Section~\ref{sec:Benchmarking, comparative, and hardware implementation studies} discusses insights provided by benchmarking studies and hardware implementations.
Finally, Section~\ref{sec:Conclusions and perspectives} concludes the review and offers perspectives on promising directions for future research in QKMs.

\section{Classical Kernel Methods}
\label{sec:Classical Kernel Methods}

This section reviews the foundations of classical kernel theory, including basic definitions and results, followed by standard formulations of three kernel-based algorithms commonly used for regression and binary classification. 
Readers familiar with classical kernel methods may skip this section, though Subsection~\ref{sec:Basic kernel theory} defines notations adopted throughout the review.

\subsection{Basic kernel theory}
\label{sec:Basic kernel theory}

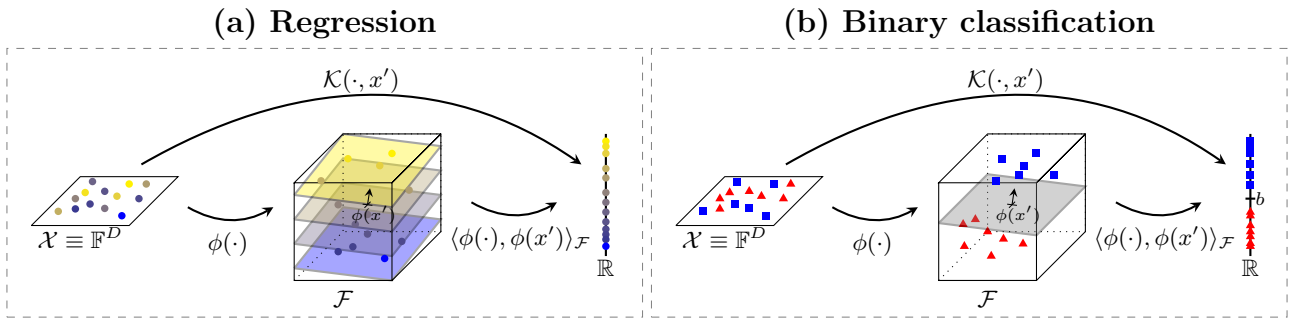
\begin{figure*}[ht]
\centering
\begin{tikzpicture}[scale=0.65]
\node at ({-0.75+0.5*13},{-1.25+5.5+0.5}) {\large\textcolor{black}{\textbf{(a) Regression}}};
\draw[black!50,dashed,line width=0.1mm] (-0.75,-1.25) --++ (13,0) --++ (0,5.5) --++ (-13,0) -- cycle;
\draw (0.5-0.75,0.5-0.125+0.25) -- ++(2,0) -- ++(1,1) -- ++(-2,0) -- ++(-1,-1) ;
\node[circle, fill=yellow!100!blue, inner sep = 0.35mm] at (1.75,1.475) {};
\node[circle, fill=yellow!95!blue, inner sep = 0.35mm] at (0.85,1.3) {};
\node[circle, fill=yellow!88.4!blue, inner sep = 0.35mm] at (1.75+0.5-0.75,1.225) {};
\node[circle, fill=yellow!73.9!blue, inner sep = 0.35mm] at (0.55+0.5-0.75,0.915) {};
\node[circle, fill=yellow!65!blue, inner sep = 0.35mm] at (2.35+0.5-0.75,1.475) {};
\node[circle, fill=yellow!51.2!blue, inner sep = 0.35mm] at (0.75+0.65-0.75,1.175) {};
\node[circle, fill=yellow!41.9!blue, inner sep = 0.35mm] at (1.5+0.5-0.75,0.975) {};
\node[circle, fill=yellow!32.6!blue, inner sep = 0.35mm] at (1.25+0.5-0.75,1.525) {};
\node[circle, fill=yellow!23.3!blue, inner sep = 0.35mm] at (1.5+0.5-0.75,1.325) {};
\node[circle, fill=yellow!16.3!blue, inner sep = 0.35mm] at (2.2+0.5-0.75,1.175) {};
\node[circle, fill=yellow!11.2!blue, inner sep = 0.35mm] at (1.2+0.5-0.75,1.065) {};
\node[circle, fill=yellow!6.5!blue, inner sep = 0.35mm] at (0.9+0.5-0.75,0.965) {};
\node[circle, fill=yellow!0!blue, inner sep = 0.35mm] at (1.85+0.5-0.75,0.825) {};
\node at (1.75+0.5-0.75-0.75,2-0.125-1.75+0.25) {$\mathcal{X}\equiv\F^D$};
\draw[thick,opacity=0.35,fill=yellow!0!blue] (5+0.125,-0.25) -- ++(2,-0.25) -- ++(1,1.25) -- ++(-2,0.25) -- ++(-1,-1.25) ;
\draw[thick,opacity=0.35,fill=yellow!33!blue] (5+0.125,0.25) -- ++(2,-0.25) -- ++(1,1.25) -- ++(-2,0.25) -- ++(-1,-1.25) ;
\draw[thick,opacity=0.35,fill=yellow!66!blue] (5+0.125,0.75) -- ++(2,-0.25) -- ++(1,1.25) -- ++(-2,0.25) -- ++(-1,-1.25) ;
\draw[thick,opacity=0.35,fill=yellow!100!blue] (5+0.125,1.25) -- ++(2,-0.25) -- ++(1,1.25) -- ++(-2,0.25) -- ++(-1,-1.25) ;
\draw[dotted] (5+0.125,-0.5) -- ++(1,1) -- ++(0,2) -- ++(2,0) -- ++(0,-2) -- ++(-2,0) ;
\draw (5+0.125,-0.5) -- ++ (2,0) -- ++(1,1) -- ++(0,2) -- ++(-2,0) -- ++(-1,-1) -- ++ (2,0) -- ++(1,1) -- ++(-1,-1) -- ++(0,-2) -- ++(-2,0) -- ++(0,2) ;
\node[circle, fill=yellow!100!blue, inner sep = 0.35mm] at (2+5+0.125,2.1) {};
\node[circle, fill=yellow!95!blue, inner sep = 0.35mm] at (2+5+0.125-0.9,2.1-0.1075) {};
\node[circle, fill=yellow!88.4!blue, inner sep = 0.35mm] at (1.75+5+0.125,1.85) {};
\node[circle, fill=yellow!73.9!blue, inner sep = 0.35mm] at (0.55+5+0.125,1.54) {};
\node[circle, fill=yellow!65!blue, inner sep = 0.35mm] at (2.35+5+0.125,1.35) {};
\node[circle, fill=yellow!51.2!blue, inner sep = 0.35mm] at (0.75+5.15+0.125,1.05) {};
\node[circle, fill=yellow!41.9!blue, inner sep = 0.35mm] at (1.5+5+0.125,0.85) {};
\node[circle, fill=yellow!32.6!blue, inner sep = 0.35mm] at (1.25+5+0.125,0.65) {};
\node[circle, fill=yellow!23.3!blue, inner sep = 0.35mm] at (1.5+5+0.125,0.45) {};
\node[circle, fill=yellow!16.3!blue, inner sep = 0.35mm] at (2.2+5+0.125,0.3) {};
\node[circle, fill=yellow!11.2!blue, inner sep = 0.35mm] at (1.2+5+0.125,0.19) {};
\node[circle, fill=yellow!6.5!blue, inner sep = 0.35mm] at (0.9+5+0.125,0.09) {};
\node[circle, fill=yellow!0!blue, inner sep = 0.35mm] at (1.85+5+0.125,-0.05) {};
\node at (6.75-0.75+0.125,2.875-3.75) {$\mathcal{F}$} ;
\draw[-stealth] (6.625,1.05) --++ (0.075*0.675,0.5*0.675);
\node at (6.75,0.85) {\fontsize{6}{7}\selectfont$\phi(x^\prime)$};
\draw[line width=0.1mm] (6.625-2*0.05,1.05+0.25*0.05) -- (6.625+2*0.05,1.05-0.25*0.05);
\draw[line width=0.1mm] (6.625-1*0.0625,1.05-1*0.0625) -- (6.625+1*0.0625,1.05+1*0.0625);
\draw[thick] (11-0.25+0.5+0.25,0) -- ++(0,2.5) ;
\node[circle, fill=yellow!100!blue, inner sep = 0.35mm] at (11-0.25+0.5+0.25,2.35) {};
\node[circle, fill=yellow!95!blue, inner sep = 0.35mm] at (11-0.25+0.5+0.25,2.35-0.1075) {};
\node[circle, fill=yellow!88.4!blue, inner sep = 0.35mm] at (11-0.25+0.5+0.25,2.1) {};
\node[circle, fill=yellow!73.9!blue, inner sep = 0.35mm] at (11-0.25+0.5+0.25,1.79) {};
\node[circle, fill=yellow!65!blue, inner sep = 0.35mm] at (11-0.25+0.5+0.25,1.6) {};
\node[circle, fill=yellow!51.2!blue, inner sep = 0.35mm] at (11-0.25+0.5+0.25,1.3) {};
\node[circle, fill=yellow!41.9!blue, inner sep = 0.35mm] at (11-0.25+0.5+0.25,1.1) {};
\node[circle, fill=yellow!32.6!blue, inner sep = 0.35mm] at (11-0.25+0.5+0.25,0.9) {};
\node[circle, fill=yellow!23.3!blue, inner sep = 0.35mm] at (11-0.25+0.5+0.25,0.7) {};
\node[circle, fill=yellow!16.3!blue, inner sep = 0.35mm] at (11-0.25+0.5+0.25,0.55) {};
\node[circle, fill=yellow!11.2!blue, inner sep = 0.35mm] at (11-0.25+0.5+0.25,0.44) {};
\node[circle, fill=yellow!6.5!blue, inner sep = 0.35mm] at (11-0.25+0.5+0.25,0.34) {};
\node[circle, fill=yellow!0!blue, inner sep = 0.35mm] at (11-0.25+0.5+0.25,0.2) {};
\node at (11-0.25+0.5+0.25,2.75-3.25+0.25) {$\R$};
\draw[thick,-stealth ] (3.25-0.3,1.75-0.5-0.25) to[out=-30,in=-150] ++(1.75,0);
\node at (4.0725-0.3,1.5-0.5-0.5-0.25) {$\phi(\cdot)$};
\draw[thick,-stealth] (8.5+0.25,1.75-0.5-0.25+0.1) to[out=-30,in=-150] ++(1.75,0);
\node at (9.5+0.25,1.5-0.5-0.5-0.25+0.1) {$\langle\phi(\cdot),\phi(x^\prime)\rangle_{\mathcal{F}}$};
\draw[thick,-stealth ] (3-1,2.125-0.25) to[out=30,in=145] ++(9,0);
\node at (6.5,4.25-0.4-0.25) {$\mathcal{K}(\cdot,x^\prime)$};
\end{tikzpicture}
\begin{tikzpicture}[scale=0.65]
\node at ({-0.75+0.5*13},{-1.25+5.5+0.5}) {\large\textcolor{black}{\textbf{(b) Binary classification}}};
\draw[black!50,dashed,line width=0.1mm] (-0.75,-1.25) --++ (13,0) --++ (0,5.5) --++ (-13,0) -- cycle;
\draw (0.5-0.75,0.5-0.125+0.25) -- ++(2,0) -- ++(1,1) -- ++(-2,0) -- ++(-1,-1) ;
\node[rectangle, fill=blue, inner sep = 0.5mm] at (1.75,1.475) {};
\node[regular polygon, regular polygon sides=3, fill=red, inner sep = 0.25mm] at (0.85,1.3) {};
\node[regular polygon, regular polygon sides=3, fill=red, inner sep = 0.25mm] at (1.75+0.5-0.75,1.225) {};
\node[rectangle, fill=blue, inner sep = 0.5mm] at (0.55+0.5-0.75,0.915) {};
\node[regular polygon, regular polygon sides=3, fill=red, inner sep = 0.25mm] at (2.35+0.5-0.75,1.475) {};
\node[regular polygon, regular polygon sides=3, fill=red, inner sep = 0.25mm] at (0.75+0.65-0.75,1.175) {};
\node[rectangle, fill=blue, inner sep = 0.5mm] at (1.5+0.5-0.75,0.975) {};
\node[rectangle, fill=blue, inner sep = 0.5mm] at (1.25+0.5-0.75,1.525) {};
\node[regular polygon, regular polygon sides=3, fill=red, inner sep = 0.25mm] at (1.5+0.5-0.75,1.325) {};
\node[regular polygon, regular polygon sides=3, fill=red, inner sep = 0.25mm] at (2.2+0.5-0.75,1.175) {};
\node[rectangle, fill=blue, inner sep = 0.5mm] at (1.2+0.5-0.75,1.065) {};
\node[regular polygon, regular polygon sides=3, fill=red, inner sep = 0.25mm] at (0.9+0.5-0.75,0.965) {};
\node[rectangle, fill=blue, inner sep = 0.5mm] at (1.85+0.5-0.75,0.825) {};
\node at (1.75+0.5-0.75-0.75,2-0.125-1.75+0.25) {$\mathcal{X}\equiv\F^D$};
\draw[dotted] (5+0.125,-0.5) -- ++(1,1) -- ++(0,2) -- ++(2,0) -- ++(0,-2) -- ++(-2,0) ;
\draw (5+0.125,-0.5) -- ++ (2,0) -- ++(1,1) -- ++(0,2) -- ++(-2,0) -- ++(-1,-1) -- ++ (2,0) -- ++(1,1) -- ++(-1,-1) -- ++(0,-2) -- ++(-2,0) -- ++(0,2) ;
\node[regular polygon, regular polygon sides=3, fill=red, inner sep = 0.25mm] at (1.5+5+0.125-0.75,0.5+0.25) {};
\node[regular polygon, regular polygon sides=3, fill=red, inner sep = 0.25mm] at (1.25+5+0.125-0.75,0.4+0.25) {};
\node[regular polygon, regular polygon sides=3, fill=red, inner sep = 0.25mm] at (1.5+5+0.125-0.5,0.25+0.25) {};
\node[regular polygon, regular polygon sides=3, fill=red, inner sep = 0.25mm] at (2.2+4.25,0.35) {};
\node[regular polygon, regular polygon sides=3, fill=red, inner sep = 0.25mm] at (1.2+5,-0.25+0.25) {};
\node[regular polygon, regular polygon sides=3, fill=red, inner sep = 0.25mm] at (0.9+5-0.25,-0.05+0.25) {};
\node[regular polygon, regular polygon sides=3, fill=red, inner sep = 0.25mm] at (1.85+5,0.25) {};
\draw[thick,opacity=0.35,fill=gray] (5+0.125,0.675) -- ++(2,-0.25) -- ++(1,1) -- ++(-2,0.25) -- ++(-1,-1) ;
\draw[-stealth] (6.625,1.05) --++ (0.075*0.675,0.5*0.675);
\node at (6.75,0.85) {\fontsize{6}{7}\selectfont$\phi(x^\prime)$};
\draw[line width=0.1mm] (6.625-2*0.05,1.05+0.25*0.05) -- (6.625+2*0.05,1.05-0.25*0.05);
\draw[line width=0.1mm] (6.625-1*0.0625,1.05-1*0.0625) -- (6.625+1*0.0625,1.05+1*0.0625);
\node[rectangle, fill=blue, inner sep = 0.5mm] at (2+5+0.125,2.1) {};
\node[rectangle, fill=blue, inner sep = 0.5mm] at (2+5+0.125-0.7,2.1-0.1075) {};
\node[rectangle, fill=blue, inner sep = 0.5mm] at (1.75+5+0.125,1.85) {};
\node[rectangle, fill=blue, inner sep = 0.5mm] at (0.55+5+0.75,1.54) {};
\node[rectangle, fill=blue, inner sep = 0.5mm] at (2.35+5+0.125,1.65) {};
\node[rectangle, fill=blue, inner sep = 0.5mm] at (0.75+5.15+0.875,1.65) {};
\node at (6.75-0.75+0.125,2.875-3.75) {$\mathcal{F}$} ;
\draw[thick] (11-0.25+0.5+0.25,0) -- ++(0,2.5) ;
\node[rectangle, fill=blue, inner sep = 0.5mm] at (11-0.25+0.5+0.25,2.35) {};
\node[rectangle, fill=blue, inner sep = 0.5mm] at (11-0.25+0.5+0.25,2.225) {};
\node[rectangle, fill=blue, inner sep = 0.5mm] at (11-0.25+0.5+0.25,2.1) {};
\node[rectangle, fill=blue, inner sep = 0.5mm] at (11-0.25+0.5+0.25,1.89) {};
\node[rectangle, fill=blue, inner sep = 0.5mm] at (11-0.25+0.5+0.25,1.65) {};
\node[rectangle, fill=blue, inner sep = 0.5mm] at (11-0.25+0.5+0.25,1.45) {};
\draw[thick] (11.4,1.175) --++ (0.2,0);
\node at (11.7,1.175) {\scriptsize $b$};
\node[regular polygon, regular polygon sides=3, fill=red, inner sep = 0.25mm] at (11-0.25+0.5+0.25,0.9) {};
\node[regular polygon, regular polygon sides=3, fill=red, inner sep = 0.25mm] at (11-0.25+0.5+0.25,0.825) {};
\node[regular polygon, regular polygon sides=3, fill=red, inner sep = 0.25mm] at (11-0.25+0.5+0.25,0.625) {};
\node[regular polygon, regular polygon sides=3, fill=red, inner sep = 0.25mm] at (11-0.25+0.5+0.25,0.5) {};
\node[regular polygon, regular polygon sides=3, fill=red, inner sep = 0.25mm] at (11-0.25+0.5+0.25,0.4) {};
\node[regular polygon, regular polygon sides=3, fill=red, inner sep = 0.25mm] at (11-0.25+0.5+0.25,0.26) {};
\node[regular polygon, regular polygon sides=3, fill=red, inner sep = 0.25mm] at (11-0.25+0.5+0.25,0.2) {};
\node at (11-0.25+0.5+0.25,2.75-3.25+0.25) {$\R$};
\draw[thick,-stealth ] (3.25-0.3,1.75-0.5-0.25) to[out=-30,in=-150] ++(1.75,0);
\node at (4.0725-0.3,1.5-0.5-0.5-0.25) {$\phi(\cdot)$};
\draw[thick,-stealth] (8.5+0.25,1.75-0.5-0.25+0.1) to[out=-30,in=-150] ++(1.75,0);
\node at (9.5+0.25,1.5-0.5-0.5-0.25+0.1) {$\langle\phi(\cdot),\phi(x^\prime)\rangle_{\mathcal{F}}$};
\draw[thick,-stealth ] (3-1,2.125-0.25) to[out=30,in=145] ++(9,0);
\node at (6.5,4.25-0.4-0.25) {$\mathcal{K}(\cdot,x^\prime)$};
\end{tikzpicture}
\caption{\footnotesize \justifying If a feature map $\phi$ associated with a kernel $\mathcal{K}$ embeds input samples into a high-dimensional feature space $\mathcal{F}$ in a suitable manner, then kernel methods based on $\mathcal{K}$ can convert a non-linear relationship between inputs and labels into a linear relationship in $\mathcal{F}$.
\textbf{(a)} In regression settings, suppose that $\phi$ maps inputs into $\mathcal{F}$ such that samples with similar continuous labels (indicated by their colour) lie on approximately parallel hyperplanes. 
In this case, the label of an unseen input $x\in\mathcal{X}$ can be predicted by projecting its embedded representation $\phi(x)$ linearly along an appropriate direction in $\mathcal{F}$.
\textbf{(b)} In binary classification, suppose instead that $\phi$ arranges inputs in $\mathcal{F}$ so that samples from different classes (represented by red triangles and blue squares) lie on opposite sides of a separating hyperplane. 
The label of an unseen input $x\in\mathcal{X}$ can then be predicted by projecting its embedded representation $\phi(x)$ onto the direction normal to this hyperplane. 
The sign of the resulting value, possibly after applying a fixed offset $b\in\R$, determines the predicted class.
In both \textbf{(a)} and \textbf{(b)}, the axis along which we project is shown as $\phi(x^\prime)$. 
However in general, this axis is usually given by a linear combination of the form $\sum_{i=1}^{M}\alpha_i\phi(\mathbf{x}_i)$, where the $\mathbf{x}_i$'s are the input training data samples (see Eq.~\ref{eq:RTModelInnerProduct}).}
\label{fig:Kernel_Methods}
\end{figure*}

Kernel methods~\cite{Scholkopf2001Kernels} enable modelling of complex data structures by implicitly computing inner products between embeddings of input data samples in a high- (possibly infinite-) dimensional feature space via a kernel function. 
This embedding can transform non-linear relationships in the original input space into linear ones in the feature space (see Fig.~\ref{fig:Kernel_Methods}). 
Moreover, for fixed hyperparameters, many kernel-based problems admit closed-form or deterministic solutions.

Formally, we consider a training dataset $\mathcal{D}=\{(\mathbf{x}_i,y_i)\}_{i=1}^{M}\subset\mathcal{X}\times\mathcal{Y}$.
Here $\mathcal{X}\equiv\F^D$ (with $\F=\C$ or $\F=\R$) is the \emph{input data domain} of \emph{dimension} $D\in\mathbb{N}$ (which is usually small, otherwise storing the dataset $\mathcal{D}$ could be expensive to begin with).
We denote by $\mathbf{x}_i\in\mathcal{X}$ the $i^{\textrm{th}}$ \emph{input training data sample} drawn from a distribution $\mathscr{D}$ over $\mathcal{X}$, $y_i\in\mathcal{Y}$ the \emph{label} for the $i^{\textrm{th}}$ training data sample, and $M\in\N$ the total \emph{number of training data samples}.
In the case of regression and binary classification, we have $\mathcal{Y}=\R$ and $\mathcal{Y}=\{\pm1\}$ respectively.

A \emph{kernel} is a symmetric function $\mathcal{K}:\mathcal{X}\times\mathcal{X}\to\R$ such that the \emph{Gram matrix} $K_{ij}\equiv\mathcal{K}(x_i,x_j)$ of $\mathcal{K}$ is positive semi-definite for all choices of $\{x_1,\ldots,x_m\}\subset\mathcal{X}$ and $m\in\N$. 
Associated with any kernel $\mathcal{K}$ is a map $\phi:\mathcal{X}\to\mathcal{F}$ called the \emph{feature map} for $\mathcal{K}$ whose codomain $\mathcal{F}$ is a Hilbert space over $\R$ called a \emph{feature space}. 
Using $\phi$, we can express $\mathcal{K}$ in the form
\begin{equation}
\label{eq:FeatureMapKernel}
\mathcal{K}(x,x^\prime)=\left\langle\phi(x),\phi(x^\prime)\right\rangle_{\mathcal{F}}.
\end{equation}
Kernel functions hence implicitly calculate an inner product between embeddings of inputs $x,x^\prime\in\mathcal{X}$ in $\mathcal{F}$.

Each kernel $\mathcal{K}$ similarly induces a real Hilbert space, referred to as the \emph{reproducing kernel Hilbert space} (RKHS). 
We denote this space by $\mathcal{R}_{\mathcal{K}}$. 
The RKHS consists of real-valued functions defined on the input data domain $\mathcal{X}$, and may be constructed as the closure of the real linear span of kernel sections $\mathcal{K}(\cdot, x^\prime)$ for $x^\prime \in \mathcal{X}$. 
That is,
\begin{equation*}
    \mathcal{R}_{\mathcal{K}} \equiv \overline{\textrm{span}}_{\mathbb{R}}\{\mathcal{K}(\cdot,x^\prime)|x^\prime\in\mathcal{X}\}.
\end{equation*}
Within supervised learning frameworks, applying kernel methods to a learning problem effectively amounts to selecting a function from $\mathcal{R}_{\mathcal{K}}$ that minimises a regularised empirical risk over the available training data. 
A natural question, then, is how such a function can be determined in practice.

Although generic elements of $\mathcal{R}_{\mathcal{K}}$ need not admit finite expansions in terms of kernel sections, the situation simplifies considerably under mild assumptions on the kernel $\mathcal{K}$ and the input space $\mathcal{X}$~\cite[Lemma 4.33]{Steinwart2008Support}. 
In this setting, the \emph{representer theorem}~\cite{Scholkopf2001Kernels,Mohri2018Foundations} guarantees that, for a broad class of regularised learning objectives, any minimiser can be expressed as a finite kernel expansion of the form
\begin{equation}
\label{eq:RTmodel}
\mathscr{F}(\cdot) = \sum_{i=1}^{M} \alpha_i \, \mathcal{K}(\cdot, \mathbf{x}_i),
\end{equation}
where $\{\mathbf{x}_i\}_{i=1}^{M}$ denote the training inputs and $\{\alpha_i\}_{i=1}^{M} \subset \mathbb{R}$ are scalar coefficients.

This result implies that optimisation over the full (potentially infinite-dimensional) RKHS can be reduced to an optimisation problem over the finite-dimensional coefficient vector $\boldsymbol{\alpha}=(\alpha_1,\ldots,\alpha_M)\in \mathbb{R}^M$. 
Such a reduction enables efficient and deterministic solution strategies, as employed by the kernel machines we discuss later in this section.

Finally, by combining the kernel expansion in Eq.~\eqref{eq:RTmodel} with the feature-map representation of the kernel from Eq.~\eqref{eq:FeatureMapKernel}, the learned function may equivalently be written as
\begin{equation}
\label{eq:RTModelInnerProduct}
\mathscr{F}(\cdot) = \Big\langle \phi(\cdot), \sum_{i=1}^{M} \alpha_i \, \phi(\mathbf{x}_i) \Big\rangle_{\mathcal{F}}.
\end{equation}
This expression highlights that kernel-based models (up to an optional bias term and, in classification settings, a final sign operation) act by first mapping inputs into the feature space $\mathcal{F}$ and then evaluating their inner product with a single learned vector determined by the training data (see Fig.~\ref{fig:Kernel_Methods}).

\subsection{Classical kernel machines}
\label{sec:Classical kernel machines}

We now discuss classical kernel-based machine learning algorithms, commonly known as \emph{classical kernel machines}. Motivated by statistical learning theory and proven empirically successful, these methods are widely used for regression and binary classification tasks. For further details, see Chapters 5.3 and 11.3 of~\cite{Mohri2018Foundations}.

\subsubsection{Support vector regression}

Support vector regression (SVR) is an SVM designed for regression problems that allows for a margin of tolerance $\epsilon$ within which prediction errors are not penalised. 
For this reason, SVR is also sometimes called $\epsilon$-insensitive SVR, where the choice of $\epsilon$ controls the bias-variance tradeoff.

Formally, SVR is the dual convex quadratic program given by
\begin{align}
    \nonumber
    &\min_{\boldsymbol{\alpha},\boldsymbol{\alpha}^\prime\in[0,C]^M}\Bigg(\frac{1}{2}\sum_{i,j=1}^{M}(\alpha^\prime_i-\alpha_i)K_{ij}(\alpha^\prime_j-\alpha_j)\\
    \label{eq:SVR}
    &\qquad\quad\quad\quad+\epsilon\sum_{k=1}^{M}(\alpha_k^\prime+\alpha_k)-\sum_{l=1}^{M}y_l(\alpha^\prime_l-\alpha_l)\Bigg)\\
    \nonumber
    &\textrm{subject to}\quad\sum_{m=1}^{M}(\alpha^\prime_m-\alpha_m)=0.
\end{align}
Here $K_{ij}=\mathcal{K}(\mathbf{x}_i,\mathbf{x}_j)$ denotes the real entries of the symmetric $M\times M$ \emph{kernel matrix} $K$, $C>0$ is a regularisation parameter which specifies the cost of a misclassification, and $\epsilon>0$ specifies the margin of tolerance within which continuous deviations from ground truth labels are ignored.
SVR hence uses the $\epsilon$-insensitive loss function given by $L_{\epsilon}(y,y^\prime)=\max(0,|y-y^\prime|-\epsilon)$ for all $y,y^\prime\in\mathcal{Y}$.

Once the solutions $\boldsymbol{\alpha},\boldsymbol{\alpha}^\prime\in[0,C]^M\subset\R^M$ have been determined, we can make predictions about the label for an unseen input $x\in\mathcal{X}$ with the model
\begin{equation}
    \label{eq:SVRModel}
    f_{\textrm{SVR}}(x)=\sum_{i=1}^{M}(\alpha_i^\prime-\alpha_i)\mathcal{K}(x,\mathbf{x}_i)+b,
\end{equation}
where the constant offset $b\in\R$ can be determined using the Karush-Kuhn-Tucker (KKT) conditions~\cite{Kuhn2014Nonlinear,Karush2014Minima}.
Specifically, given some training sample $\mathbf{x}_j$ such that $\alpha_i\in(0,C)$ or $\alpha^\prime_i\in(0,C)$ (note that for each $j$, it may be the case that $\alpha_j\neq0$ or $\alpha^\prime_j\neq0$, but not both) we can determine $b$ as
\begin{equation*}
    b=\begin{cases}
        -\sum_{i=1}^{M}(\alpha^\prime_i-\alpha_i)\mathcal{K}(\mathbf{x}_j,\mathbf{x}_i)+y_j+\epsilon,\quad\alpha_i\in(0,C),\\
        -\sum_{i=1}^{M}(\alpha^\prime_i-\alpha_i)\mathcal{K}(\mathbf{x}_j,\mathbf{x}_i)+y_j-\epsilon,\quad\alpha^\prime_i\in(0,C).
    \end{cases}
\end{equation*}

\subsubsection{Support vector classification}

Support vector classification (SVC) is another SVM designed for binary classification.
The main idea is to separate classes by a maximal-margin hyperplane in feature space, while allowing for misclassifications with a penalty proportional to the parameter $C$, which controls the trade-off between margin width and training error.
This formulation guarantees a unique solution even when the classes are not perfectly separable in the feature space $\mathcal{F}$. 

SVC is given by the convex quadratic program, called the soft-margin dual optimisation problem,
\begin{equation}
\begin{aligned}
\label{eq:SVC}
&\min_{{\boldsymbol{\alpha}}\in[0,C]^M}\frac{1}{2}\sum_{i,j=1}^{M}{\alpha}_i{\alpha}_jy_iy_jK_{ij}-\sum_{i=1}^{M}{\alpha}_i\\
&\textrm{subject to}\quad\sum_{i=1}^{M}{\alpha}_iy_i=0.
\end{aligned}
\end{equation}

Once the solution $\boldsymbol{\alpha}\in[0,C]^M\subset\R^M$ has been determined, we can predict the label for an unseen input $x\in\mathcal{X}$ using the model
\begin{equation}
    \label{eq:SVCModel}
    f_{\textrm{SVC}}(x)=\textrm{sign}\left(\sum_{i=1}^{M}y_i\alpha_i\mathcal{K}(x,\mathbf{x}_i)+b\right).
\end{equation}
As with SVR, we can determine $b\in\R$ using the KKT conditions. 
In particular, for a training example $\mathbf{x}_j$ with $\alpha_j\in(0,C)$, $b$ is given by
\begin{equation*}
    b=y_j-\sum_{i=1}^{M}y_i\alpha_i\mathcal{K}(\mathbf{x}_j,\mathbf{x}_i).
\end{equation*}

\subsubsection{Kernel ridge regression}

KRR is a regression method in which model parameters are learned by minimising a regularised squared loss. 
In this case, training reduces to simply solving a system of linear equations involving the kernel matrix and a regularisation parameter, which controls the trade-off between prediction accuracy and model complexity.

KRR is the convex optimisation problem
\begin{equation}
\label{eq:KRR}
\min_{\boldsymbol{\alpha}\in\mathbb{R}^M}
\sum_{i=1}^{M}\left(y_i-\sum_{j=1}^{M}\alpha_j K_{ij}\right)^2
+\lambda \sum_{k,l=1}^{M}\alpha_k K_{kl}\alpha_l .
\end{equation}
Here $\lambda>0$ is a regularisation parameter that controls the penalty on the RKHS norm of the learned function.

Standard results from convex optimisation show that the unique minimiser of Eq.~\eqref{eq:KRR} is given in closed form by
\begin{equation}
\label{eq:KRRSolution}
\boldsymbol{\alpha} = (K+\lambda \mathbb{I})^{-1}\mathbf{y},
\end{equation}
where $\mathbb{I}$ is the $M\times M$ identity matrix, and $\mathbf{y}=(y_1,\ldots,y_M)\in\mathbb{R}^M$ is the vector of training data labels.

Since the kernel matrix $K$ is positive semi-definite, adding $\lambda\mathbb{I}$ with $\lambda>0$ ensures that $K+\lambda\mathbb{I}$ is strictly positive definite and therefore invertible. Moreover, increasing the value of $\lambda$ improves the conditioning of this matrix, which in turn enhances the numerical stability of the computation.

After determining the solution $\boldsymbol{\alpha}\in\R^M$ via Eq.~\eqref{eq:KRRSolution}, predictions for a new input $x\in\mathcal{X}$ are obtained using
\begin{equation}
\label{eq:KRRModel}
f_{\textrm{KRR}}(x)=\sum_{i=1}^{M}\alpha_i\,\mathcal{K}(x,\mathbf{x}_i).
\end{equation}

\subsubsection{Runtime and scalability} 

Training for both SVR and SVC (Eqs.~\eqref{eq:SVR} and \eqref{eq:SVC}) is formulated as a convex quadratic program, ensuring the existence of a unique global optimum for fixed hyperparameters. 
Such programs are typically solved via the sequential minimal optimisation (SMO) algorithm~\cite{Platt1998Sequential}, with a runtime scaling of roughly $\mathcal{O}(M)$ to $\mathcal{O}(M^2)$ in the dataset size $M$. 
However, practical factors can increase this complexity.
For example, kernel matrix construction costs $\mathcal{O}(\kappa M^2)$ (where $\kappa$ is the kernel evaluation cost), large regularisation $C$ can cause ill-conditioning and slower convergence due to numerical issues, and kernel/solver choices can further vary performance. 
Precisely quantifying runtimes for SVM training in general is thus challenging. 
In contrast, predictions from the trained model (Eqs.~\eqref{eq:SVRModel} and~\eqref{eq:SVCModel}) require only $\mathcal{O}(\kappa M)$ time per prediction.

Contrarily, KRR (Eq.~\eqref{eq:KRR}) runtime is straightforward to quantify: $\boldsymbol{\alpha}$ is computed by inverting $K+\lambda \mathbb{I}$ (Eq.~\eqref{eq:KRRSolution}) in $\mathcal{O}(M^3)$ time, or by solving the linear system $(K+\lambda\mathbb{I})\boldsymbol{\alpha}=\boldsymbol{y}$ in $\mathcal{O}(M^{2.373})$ time~\cite{LeGall2014Powers}. 
Similarly, predictions from the trained model (Eq.~\eqref{eq:KRRModel}) require $\mathcal{O}(\kappa M)$ time per predcition. 
Thus, training complexity is $\mathcal{O}(\kappa M^2+M^3)$, while each prediction costs $\mathcal{O}(\kappa M)$. 
However, small $\lambda>0$ may cause numerical instability if $K$ is nearly singular, potentially increasing runtimes in practice.

Note that in both cases, since runtimes scale with the size of the dataset $M$, this limits the feasible number of data points which one can utilise with kernel methods. 
If $M$ needs to scale exponentially with the size of the problem in order to achieve good learning performance, then kernel methods will not be feasible.
Even in the case where $M$ scales polynomially in the size of the problem, kernel methods may become prohibitively expensive in practice.
Similarly, the cost $\kappa$ required to evaluate the kernel must be tractably small, otherwise obtaining the kernel matrix $K$ may impose unmanageable costs from the outset.

\section{Quantum kernel methods}
\label{sec:Quantum kernel methods}

QKMs~\cite{Schuld2019Feature,Havlicek2019Supervised,Schuld2021Supervised}, in the sense that we consider in this review, are hybrid quantum-classical algorithms in which a quantum computer is used to estimate a fixed (up to hyperparameters) kernel function via a procedure known as \emph{quantum kernel estimation} (QKE)~\cite{Liu2021Rigorous,Zhou2024QKE-QSVR}.
The central idea is to encode data samples into quantum states, thereby inducing a feature map from $\mathcal{X}$ into a high-dimensional Hilbert space given by the space of quantum states viewed as density operators (which are expensive to represent classically in general).
By preparing these states on quantum devices and measuring them using one of several protocols, the corresponding kernel function can be estimated.
Using the quantum computer to then estimate the full kernel matrix $K$, a classical kernel machine is trained with $K$ and, together with further applications of QKE, used to make predictions on new inputs.
The idea of combining multiple quantum kernels into a single model has also been explored in prior work~\cite{Vedaie2020Multiple}, but we do not discuss such approaches further in this review.

Note that in much of the QKM literature, kernel values are described as being \emph{evaluated} on a quantum computer, typically under idealised assumptions of exact computation. 
In practice, however, such values are obtained via finite sampling and are therefore statistical \emph{estimates} (see paragraph (c) in Section~\ref{subsec:Preliminaries}). 
While the terms \emph{evaluation} and \emph{estimation} are often used interchangeably, we adopt the convention of using \emph{evaluation} in theoretical contexts, and \emph{estimation} when referring to implementations on quantum devices in practice, to emphasise the role of shot noise.

In this section we introduce and discuss two standard classes of quantum kernels, including their definitions and standard methods for their estimation in practice.
For further details about quantum computing we refer the reader to Chapters 1, 2 and 4 of~\cite{Nielsen2000Quantum}.

\subsection{Fidelity and projected quantum kernels}
\label{sec:Fidelity and projected quantum kernels}

Quantum kernels are most naturally formulated in the language of density operators, which provide a unified description of both pure and mixed quantum states. 
In this review, however, we restrict attention to quantum kernels derived from pure quantum states, reflecting the focus of the vast majority of the QKM literature. 
We therefore begin with the density operator formalism before specialising to the pure state setting of primary interest.

For an $n$-qubit quantum system, a quantum state is described by a density operator, that is, a $2^n\times2^n$ positive semi-definite matrix with unit trace. 
These operators belong to a real Hilbert space of $2^n\times2^n$ Hermitian matrices, denoted $\mathcal{H}_n$, equipped with the Frobenius inner product $\langle A,B\rangle_{\mathcal{H}_n}=\mathrm{tr}(AB)$. 
Viewing $\mathcal{H}_n$ as a feature space, an $n$-qubit \emph{quantum feature map} is a map $\rho:\mathcal{X}\to\mathcal{H}_n$ such that $\rho(x)$ is a density operator for all $x\in\mathcal{X}$.

When the feature map assigns a pure state to each input, the density operator may be written as $\rho(x)=|\psi(x)\rangle\langle\psi(x)|$ for some unit vector $\ket{\psi(x)}\in\mathbb{C}^{2^n}$. 
For this reason, quantum feature maps are sometimes viewed as mappings $\mathcal{X}\to\mathbb{C}^{2^n}$ of the form $x\mapsto\ket{\psi(x)}$. 
The dependence of $\ket{\psi(x)}$ on the input $x\in\mathcal{X}$ can be made explicit by assuming that the state is prepared by applying a \emph{data-encoding unitary} $U:\mathcal{X}\to\mathbb{U}(2^n)$, where $\mathbb{U}(2^n)$ is the group of $2^n\times2^n$ unitary matrices, to the initial state $\ket{0}^{\otimes n}$.
That is, $\ket{\psi(x)}=U(x)\ket{0}^{\otimes n}$. 

In many works on QKMs, the feature map additionally depends on variational parameters $\theta\in\R^{\widetilde{D}}$, yielding states of the form $\rho_{\theta}(x)=|\psi(x,\theta)\rangle\langle\psi(x,\theta)|$, where $\theta$ is tuned to minimise a chosen loss function.
In this review however, we will only consider non-variational QKMs, which do not involve such parameters.
Accordingly, throughout this review we consider quantum feature maps of the form $\rho(x)=|\psi(x)\rangle\langle\psi(x)|$ with $\ket{\psi(x)}=U(x)\ket{0}^{\otimes n}$.

Given a quantum feature map $\rho$, the \emph{fidelity quantum kernel} (FQK), denoted $\mathcal{K}_F:\mathcal{X}\times\mathcal{X}\to\mathbb{R}$, is defined by
\begin{equation}
    \label{eq:FQK}
    \mathcal{K}_F(x,x^\prime)=\langle\rho(x),\rho(x^\prime)\rangle_{\mathcal{H}_n}.
\end{equation}
In the pure state setting described above, the FQK defined in Eq.~\eqref{eq:FQK} admits the equivalent expression
\begin{equation}
    \label{eq:FQKPureState}
    \mathcal{K}_F(x,x^\prime)=\left|\langle0|^{\otimes n}U^\dagger(x)U(x^\prime)|0\rangle^{\otimes n}\right|^2.
\end{equation}

In addition to the FQK, a class of quantum kernels known as \emph{projected quantum kernels} (PQKs)~\cite{Huang2021Power} have been applied extensively in the QKM literature.
To define PQKs in general, it is natural to introduce the idea of a $k$-qubit reduced density matrix ($k$-RDM).
A $k$-RDM, denoted $\rho_{S_k}(x)$, is defined such that 
\begin{equation}
    \label{eq:k-RDM}
    \rho_{S_k}(x)=\textrm{tr}_{\overline{S_k}}\rho(x),
\end{equation}
where $S_k$ is a subset of containing $k\leq n$ qubits, and $\textrm{tr}_{\overline{S_k}}$ denotes the partial trace over all qubits not in $S_k$ (see Chapter 8.3.1 of~\cite{Nielsen2000Quantum} for more details about the partial trace).
PQKs induced by a quantum feature map $\rho$ are then derived by measuring observables from the $k$-RDMs with different choices of $S_k$. 
The expectation values of these observables are used as classical features which are fed to a classical kernel.
From this idea, it is clear that there are many kinds of PQKs (see Supplementary Section 10 of~\cite{Huang2021Power} for more details). 

Despite the rich variety of PQKs that one may construct, the definition that has received the most attention~\cite{Huang2021Power,Kubler2021Inductive,Thanasilp2024Exponential}, and the definition that we adopt in this review, is given by
\begin{equation}
    \label{eq:PQK}
    \mathcal{K}_P(x,x^\prime)=\exp\left(-\gamma\sum_{i=1}^{n}\|\rho_i(x)-\rho_i(x^\prime)\|_{F}^2\right).
\end{equation}
Here $\rho_i(x)$ denotes the 1-RDM of $\rho(x)$ with $S_1=\{i\}$ (see Eq.~\eqref{eq:k-RDM}), $\|\cdot\|_F$ denotes the Frobenius norm, and $\gamma>0$ is a hyperparameter.
Throughout this review, when we consider a PQK, we are specifically referring to the PQK given in Eq.~\eqref{eq:PQK}.

As a final remark, we note that despite their structural differences, FQKs and PQKs can be viewed as special cases of a more general construction. 
In particular, Gan et al.~\cite{Gan2023Unified} show that these kernels can be embedded within a unified framework of trace-induced quantum kernels. 
In this review, however, we restrict our attention to FQKs and PQKs.

\subsection{Estimating quantum kernels in practice}
\label{sec:Estimating quantum kernels in practice}

We now discuss standard protocols for estimating both the FQK and PQK in practice with a quantum computer.

\subsubsection{Fidelity quantum kernels}

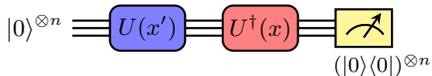
\begin{figure}[H]
    \centering
    \begin{tikzpicture}
        \node at (0.5,0) {$\ket{0}^{\otimes n}$};
        \draw[thick] (1,-0.1) --++ (3.5,0);
        \draw[thick] (1,0) --++ (3.5,0);
        \draw[thick] (1,0.1) --++ (3.5,0);
        \draw[thick,fill=blue!50,rounded corners] (1.5,0.3) --++ (1,0) --++ (0,-0.6) --++ (-1,0) -- cycle;
        \node at (2,0) {$U(x^\prime)$};
        \draw[thick,fill=red!50,rounded corners] (3,0.3) --++ (1,0) --++ (0,-0.6) --++ (-1,0) -- cycle;
        \node at (3.5,0) {$U^\dagger(x)$};
        \draw[thick,fill=yellow!50] (4.5,-0.25) -- ++(0.75,0) -- ++(0,0.5) -- ++(-0.75,0) -- cycle;
        \draw[thick] (4.645,-0.085) arc (145:35:0.3) ;
        \draw[thick,-stealth] (4.875,-0.125) -- ++({0.4*cos(55)},{0.4*sin(55)});
        \node at (5.125,-0.5) {\footnotesize$(|0\rangle\langle0|)^{\otimes n}$};
    \end{tikzpicture}
    \caption{The circuit diagram used to estimate $\mathcal{K}_F$ with the Loschmidt echo test.
    In this case $\mathcal{K}_F(x,x^\prime)$ is given by the expected value of the global observable $(|0\rangle\langle0|)^{\otimes n}$, which can equivalently be interpreted as the probability of measuring every qubit to be in the state $\ket{0}$ at the end of the computation.}
    \label{fig:FQK-LoschmidtEcho}
\end{figure}

There are three main protocols for estimating the FQK in practice.
The first protocol is known as the Loschmidt echo test, deriving from its close relation to the Loschmidt echo~\cite{Wisniacki2012Loschmidt} used in the study of quantum chaos. 
This protocol requires just $n$ qubits to implement and can be described by noting from Eq.~\eqref{eq:FQKPureState} that the FQK can be written as
\begin{equation}
    \label{eq:FQKPureState2}
    \mathcal{K}_F(x,x^\prime)
    =\langle0|^{\otimes n}U^\dagger(x)U(x^\prime)|0\rangle^{\otimes n}\langle0|^{\otimes n}U^\dagger(x^\prime)U(x)|0\rangle^{\otimes n}.
\end{equation}
From Eq.~\eqref{eq:FQKPureState2}, we see that the FQK is given by the probability of measuring the all-0 state $\ket{0}^{\otimes n}$ from the pure state $U^\dagger(x)U(x^\prime)\ket{0}^{\otimes n}$.
A circuit diagram of this protocol can be see in Fig.~\ref{fig:FQK-LoschmidtEcho}.

\begin{figure}[H]
    \centering
    \begin{tikzpicture}
        \node at (0.2+0.5,0) {$\ket{0}$};
        \draw[thick] (1,0) --++ (3.75,0); 
        \draw[thick,fill=gray!25,rounded corners] (1.5+0.1875,0.3) --++ (0.625,0) --++ (0,-0.6) --++ (-0.625,0) -- cycle;
        \node at (2,0) {$H$};
        \draw[thick] (3,0) --++ (0,-2);
        \draw[fill=black] (3,0) circle (0.05cm);
        \draw[fill=black] (3,-1) circle (0.075cm);
        \draw[thick] (3,-1) --++ (0.1875,0.1875) --++ (-0.375,-0.375) --++ (0.1875,0.1875) --++ (-0.1875,0.1875) --++ (0.375,-0.375) --++ (-0.1875,0.1875);
        \draw[fill=black] (3,-2) circle (0.075cm);
        \draw[thick] (3,-2) --++ (0.1875,0.1875) --++ (-0.375,-0.375) --++ (0.1875,0.1875) --++ (-0.1875,0.1875) --++ (0.375,-0.375) --++ (-0.1875,0.1875);
        \draw[thick,fill=gray!25,rounded corners] (3.5+0.1875,0.3) --++ (0.625,0) --++ (0,-0.6) --++ (-0.625,0) -- cycle;
        \node at (4,0) {$H$};
        \draw[thick,fill=yellow!50] (4.75,-0.25) -- ++(0.75,0) -- ++(0,0.5) -- ++(-0.75,0) -- cycle;
        \draw[thick] (4.645+0.25,-0.085) arc (145:35:0.3) ;
        \draw[thick,-stealth] (4.875+0.25,-0.125) -- ++({0.4*cos(55)},{0.4*sin(55)});
        \node at (6-0.5-0.375,-0.5) {\footnotesize$Z$};
        \node at (0+0.5,0-1) {$\ket{0}^{\otimes n}$};
        \draw[thick] (1,0-1-0.1) --++ (3.75,0);
        \draw[thick] (1,0-1) --++ (3.75,0);
        \draw[thick] (1,0-1+0.1) --++ (3.75,0);
        \draw[thick,fill=red!50,rounded corners] (1.5,0.3-1) --++ (1,0) --++ (0,-0.6) --++ (-1,0) -- cycle;
        \node at (2,0-1) {$U(x)$};
        \node at (0+0.5,0-1-1) {$\ket{0}^{\otimes n}$};
        \draw[thick] (1,0-1-1-0.1) --++ (3.75,0);
        \draw[thick] (1,0-1-1) --++ (3.75,0);
        \draw[thick] (1,0-1-1+0.1) --++ (3.75,0);
        \draw[thick,fill=blue!50,rounded corners] (1.5,0.3-1-1) --++ (1,0) --++ (0,-0.6) --++ (-1,0) -- cycle;
        \node at (2,0-1-1) {$U(x^\prime)$};
    \end{tikzpicture}
    \caption{The circuit diagram used to estimate $\mathcal{K}_F$ with the swap test.
    In this case $\mathcal{K}_F(x,x^\prime)$ is given by the expected value of the 1-qubit Pauli-$Z$ operator measured from the top ancilla qubit at the end of the computation.}
    \label{fig:FQK-SwapTest}
\end{figure}
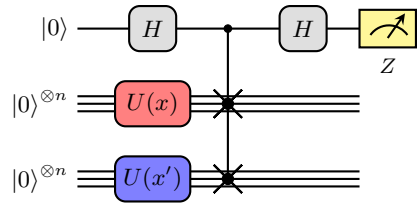

The second protocol for estimating FQKs in practice is called the swap test.
The swap test requires $2n+1$ qubits to implement and involves preparing both $\ket{\psi(x)}$ and $\ket{\psi(x^\prime)}$ on separate registers. 
At the same time, we prepare a 1-qubit ancilla register in the state $\ket{0}$ to which a Hadamard gate is applied.
Next, we perform a controlled swap operation, controlling on the ancilla register, and swapping the states of the two $n$-qubit registers.
Finally we apply another Hadamard operation to the ancilla register and measure the expected value of the Pauli-$Z$ operator.
This expectation value then yields the value of the FQK.
A circuit diagram illustrating this protocol can be found in Fig.~\ref{fig:FQK-SwapTest}.

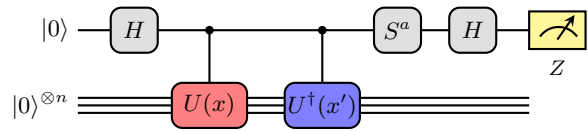
\begin{figure}[H]
    \centering
    \begin{tikzpicture}
        \node at (0.2+0.5,0) {$\ket{0}$};
        \draw[thick] (1,0) --++ (6,0); 
        \draw[thick,fill=gray!25,rounded corners] (1.25+0.1875,0.3) --++ (0.625,0) --++ (0,-0.6) --++ (-0.625,0) -- cycle;
        \node at (1.75,0) {$H$};
        \draw[thick] (3-0.25,0) --++ (0,-1);
        \draw[fill=black] (3-0.25,0) circle (0.05cm);
        \draw[thick] (4.5-0.25,0) --++ (0,-1);
        \draw[fill=black] (4.5-0.25,0) circle (0.05cm);
        \draw[thick,fill=gray!25,rounded corners] (5+0.1875-0.25,0.3) --++ (0.625,0) --++ (0,-0.6) --++ (-0.625,0) -- cycle;
        \node at (5.5-0.25,0) {$S^a$};
        \draw[thick,fill=gray!25,rounded corners] (6+0.1875-0.25,0.3) --++ (0.625,0) --++ (0,-0.6) --++ (-0.625,0) -- cycle;
        \node at (6.5-0.25,0) {$H$};
        \draw[thick,fill=yellow!50] (6.75+0.25,-0.25) -- ++(0.75,0) -- ++(0,0.5) -- ++(-0.75,0) -- cycle;
        \draw[thick] (6.645+0.25+0.25,-0.085) arc (145:35:0.3) ;
        \draw[thick,-stealth] (6.875+0.25+0.25,-0.125) -- ++({0.4*cos(55)},{0.4*sin(55)});
        \node at (7.375,-0.5) {\footnotesize$Z$};
        \node at (0+0.5,0-1) {$\ket{0}^{\otimes n}$};
        \draw[thick] (1,0-1-0.1) --++ (6,0);
        \draw[thick] (1,0-1) --++ (6,0);
        \draw[thick] (1,0-1+0.1) --++ (6,0);
        \draw[thick,fill=red!50,rounded corners] (2.5-0.25,0.3-1) --++ (1,0) --++ (0,-0.6) --++ (-1,0) -- cycle;
        \node at (3-0.25,0-1) {$U(x)$};
        \draw[thick,fill=blue!50,rounded corners] (4-0.25,0.3-1) --++ (1,0) --++ (0,-0.6) --++ (-1,0) -- cycle;
        \node at (4.5-0.25,0-1) {$U^\dagger(x^\prime)$};
    \end{tikzpicture}
    \caption{The circuit diagram used to estimate $\mathcal{K}_F$ with the Hadamard test.
    In this case $\mathcal{K}_F(x,x^\prime)$ is given by the sum of the squares of the expected value of the 1-qubit Pauli-$Z$ operator measured from the top ancilla qubit for the $a=0$ and $a=1$ cases.
    In other words $\mathcal{K}_F(x,x^\prime)=\langle Z_1\rangle_{a=0}^2+\langle Z_1\rangle_{a=1}^2$.}
    \label{fig:FQK-HadamardTest}
\end{figure}

The third and final protocol that we discuss for estimating FQKs is called the Hadamard test.
This protocol requires $n+1$ qubits and involves the estimation of two expectation values corresponding to the real and imaginary parts of $\langle\psi(x)|\psi(x^\prime)\rangle$, from which we can obtain $\mathcal{K}_F(x,x^\prime)$ by summing the squares of these parts.
The protocol involves preparing a 1-qubit ancilla register in the state $\ket{0}$, and at the same time preparing an $n$-qubit register in the state $\ket{0}^{\otimes n}$.
After applying a Hadamard gate to the ancilla qubit, we then apply two controlled operations with the ancilla qubit as the control. 
Firstly a controlled $U(x)$ then a controlled $U^\dagger(x^\prime)$.
Next we apply $S^a$ to the ancilla, where $S$ is a standard phase gate and $a\in\{0,1\}$ depends on whether we are seeking to estimate the real ($a=0$) or imaginary part ($a=1$) of $\langle\psi(x)|\psi(x^\prime)\rangle$.
Finally we apply another Hadamard gate to the ancilla register and measure the expected value of the Pauli-$Z$ operator from the ancilla.
A circuit diagram illustrating this protocol can be seen in Figure~\ref{fig:FQK-HadamardTest}.

\subsubsection{Projected quantum kernels}

We now discuss two methods for estimating the PQK on quantum devices.
In the first protocol, one endeavours to construct the $1$-RDM $\rho_i(x)$ via full state-tomography.
In the case of $k$-RDMs with $k>1$, the RDMs are usually estimated from localised random measurements using the classical shadow formalism~\cite{Huang2020Predicting} rather than state tomography, since the latter requires a number of measurements scaling exponentially in $k$
(see Supplementary Section 10 of~\cite{Huang2021Power} for more details on the case of $k>1$).

\begin{figure}
    \centering
    \begin{tikzpicture}
        \draw[black!50,dashed,line width=0.1mm] (-0.35,0.5) --++ (4,0) --++ (0,-8.25) --++ (-4,0) -- cycle;
        \node at (1.65,0.75) {\footnotesize Single qubit state tomography};
        \node at (0,0) {$\ket{0}$};
        \node at (0,-0.5) {$\ket{0}$};
        \node at (0,-0.9) {$\vdots$};
        \node at (0,-1.5) {$\ket{0}$};
        \draw[thick] (0.5,0) --++ (2,0);
        \draw[thick] (0.5,-0.5) --++ (2,0);
        \draw[thick] (0.5,-1.5) --++ (2,0);
        \draw[thick,fill=white,rounded corners] (1,0.25) --++ (1,0) --++ (0,-2) --++ (-1,0) -- cycle;
        \node at (1.5,-0.25) {\textcolor{blue!50}{$U(x)$}};
        \node at (1.5,-0.75) {or};
        \node at (1.5,-1.25) {\textcolor{red!50}{$U(x^\prime)$}};
        \draw[thick,fill=yellow!50] (4.75-2.25,-0.25) -- ++(0.75,0) -- ++(0,0.5) -- ++(-0.75,0) -- cycle;
        \draw[thick] (4.645+0.25-2.25,-0.085) arc (145:35:0.3) ;
        \draw[thick,-stealth] (4.875+0.25-2.25,-0.125) -- ++({0.4*cos(55)},{0.4*sin(55)});
        \node at (6-0.5-2.375,-0.5) {\footnotesize$X,Y,Z$};
        \node at (0,0-2.5) {$\ket{0}$};
        \node at (0,-0.5-2.5) {$\ket{0}$};
        \node at (0,-0.9-2.5) {$\vdots$};
        \node at (0,-1.5-2.5) {$\ket{0}$};
        \draw[thick] (0.5,0-2.5) --++ (2,0);
        \draw[thick] (0.5,-0.5-2.5) --++ (2,0);
        \draw[thick] (0.5,-1.5-2.5) --++ (2,0);
        \draw[thick,fill=white,rounded corners] (1,0.25-2.5) --++ (1,0) --++ (0,-2) --++ (-1,0) -- cycle;
        \node at (1.5,-0.25-2.5) {\textcolor{blue!50}{$U(x)$}};
        \node at (1.5,-0.75-2.5) {or};
        \node at (1.5,-1.25-2.5) {\textcolor{red!50}{$U(x^\prime)$}};
        \draw[thick,fill=yellow!50] (4.75-2.25,-0.25-2.5-0.5) -- ++(0.75,0) -- ++(0,0.5) -- ++(-0.75,0) -- cycle;
        \draw[thick] (4.645+0.25-2.25,-0.085-2.5-0.5) arc (145:35:0.3) ;
        \draw[thick,-stealth] (4.875+0.25-2.25,-0.125-2.5-0.5) -- ++({0.4*cos(55)},{0.4*sin(55)});
        \node at (6-0.5-2.375,-0.5-2.5-0.5) {\footnotesize$X,Y,Z$};
        \node at (1.5,-4.65) {$\vdots$};
        \node at (4.5,-4.65) {$\vdots$};
        \node at (0,0-2.5-2.5-0.5) {$\ket{0}$};
        \node at (0,-0.5-2.5-2.5-0.5) {$\ket{0}$};
        \node at (0,-0.9-2.5-2.5-0.5) {$\vdots$};
        \node at (0,-1.5-2.5-2.5-0.5) {$\ket{0}$};
        \draw[thick] (0.5,0-2.5-2.5-0.5) --++ (2,0);
        \draw[thick] (0.5,-0.5-2.5-2.5-0.5) --++ (2,0);
        \draw[thick] (0.5,-1.5-2.5-2.5-0.5) --++ (2,0);
        \draw[thick,fill=white,rounded corners] (1,0.25-2.5-2.5-0.5) --++ (1,0) --++ (0,-2) --++ (-1,0) -- cycle;
        \node at (1.5,-0.25-2.5-2.5-0.5) {\textcolor{blue!50}{$U(x)$}};
        \node at (1.5,-0.75-2.5-2.5-0.5) {or};
        \node at (1.5,-1.25-2.5-2.5-0.5) {\textcolor{red!50}{$U(x^\prime)$}};
        \draw[thick,fill=yellow!50] (4.75-2.25,-0.25-2.5-2.5-0.5-1.5) -- ++(0.75,0) -- ++(0,0.5) -- ++(-0.75,0) -- cycle;
        \draw[thick] (4.645+0.25-2.25,-0.085-2.5-2.5-0.5-1.5) arc (145:35:0.3) ;
        \draw[thick,-stealth] (4.875+0.25-2.25,-0.125-2.5-2.5-0.5-1.5) -- ++({0.4*cos(55)},{0.4*sin(55)});
        \node at (6-0.5-2.375,-0.5-2.5-2.5-0.5-1.5) {\footnotesize$X,Y,Z$};
        \draw[black!50,dashed,line width=0.1mm] (3.85,0.5) --++ (3.875-0.25,0) --++ (0,-8.25) --++ (-3.875+0.25,0) -- cycle;
        \node at (5.6625,0.75) {\footnotesize Classical post-processing};
        \node at (3.775,0-0.05) {$\Longrightarrow$};
        \node at (4.5,0.25) {$\textcolor{blue!50}{\rho_1(x)}$};
        \node at (4.5,-0.25) {$\textcolor{red!50}{\rho_1(x^\prime)}$};
        \node at (3.775,-0.5-2.5-0.05) {$\Longrightarrow$};
        \node at (4.5,-0.5-2.5+0.25) {$\textcolor{blue!50}{\rho_2(x)}$};
        \node at (4.5,-0.5-2.5-0.25) {$\textcolor{red!50}{\rho_2(x^\prime)}$};
        \node at (3.775,-2-2.5-2.5-0.05) {$\Longrightarrow$};
        \node at (4.5,-2-2.5-2.5+0.25) {$\textcolor{blue!50}{\rho_n(x)}$};
        \node at (4.5,-2-2.5-2.5-0.25) {$\textcolor{red!50}{\rho_n(x^\prime)}$};
        \draw[thick,rounded corners,-stealth] (5,0) --++ (1-0.25,0) --++ (0,-3.5+0.125) --++(0.25,0);
        \draw[thick,rounded corners, -stealth] (5,-3) --++ (0.75-0.25,0) --++ (0,-0.5) --++ (0.5,0);
        \draw[thick,rounded corners, -stealth] (5,-7) --++ (1-0.25,0) --++ (0,3.5-0.125) --++ (0.25,0);
        \node at (7-0.25,-3.5) {$\mathcal{K}_P(x,x^\prime)$};
    \end{tikzpicture}
    \caption{Applying 1-qubit state tomography and classical post-processing to estimate the PQK $\mathcal{K}_P(x,x^\prime)$.}
    \label{fig:PQK-StateTomography}
\end{figure}
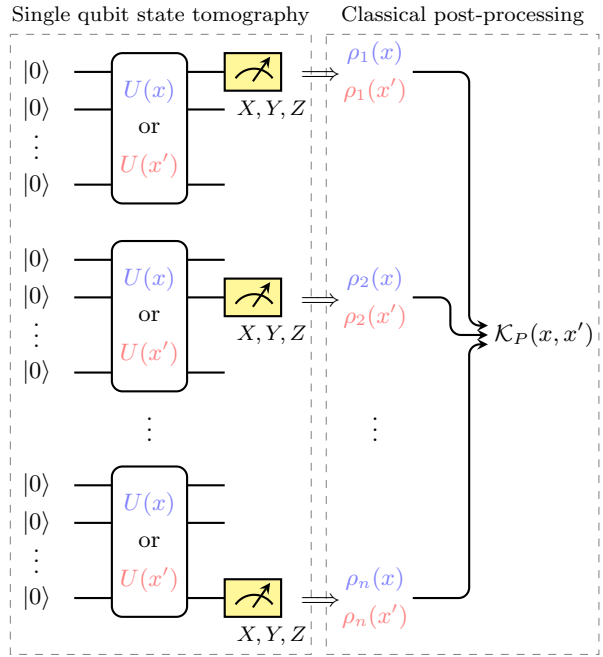

Since the number of measurements required to construct 1-RDMs via state-tomography is constant with respect to the number of qubits in the underlying system, this protocol is efficient as long as the time required to prepare the state $\rho(x)$ scales efficiently for all $x\in\mathcal{X}$.
In particular, each 1-RDM can be written as
\begin{equation}
    \label{eq:1-RDMStateTomography}
    \rho_i(x)=\frac{1}{2}\left(\mathbb{I}+\langle X_i\rangle_{\rho(x)} X+\langle Y_i\rangle_{\rho(x)} Y+\langle Z_i\rangle_{\rho(x)} Z\right),
\end{equation}
where $\langle \sigma_i\rangle_{\rho(x)}$ denotes the expected value of the Pauli operator $\sigma\in\{X,Y,Z\}$ measured from qubit $i$ when the total $n$-qubit system occupies the state $\rho(x)$.
Accordingly we see that in order to determine $\mathcal{K}_P(x,x^\prime)$, we need to determine $6n$ coefficients given by the expected value of $X$, $Y$, and $Z$ measured from each of the $n$ qubits for both $\rho(x)$ and $\rho(x^\prime)$.
These coefficients are then classically post-processed to determine the 1-RDMs $\rho_i(x)$ and $\rho_i(x^\prime)$, which in turn are used to calculate $\mathcal{K}_P(x,x^\prime)$.
See Fig.~\ref{fig:PQK-StateTomography} for an illustration of this procedure.

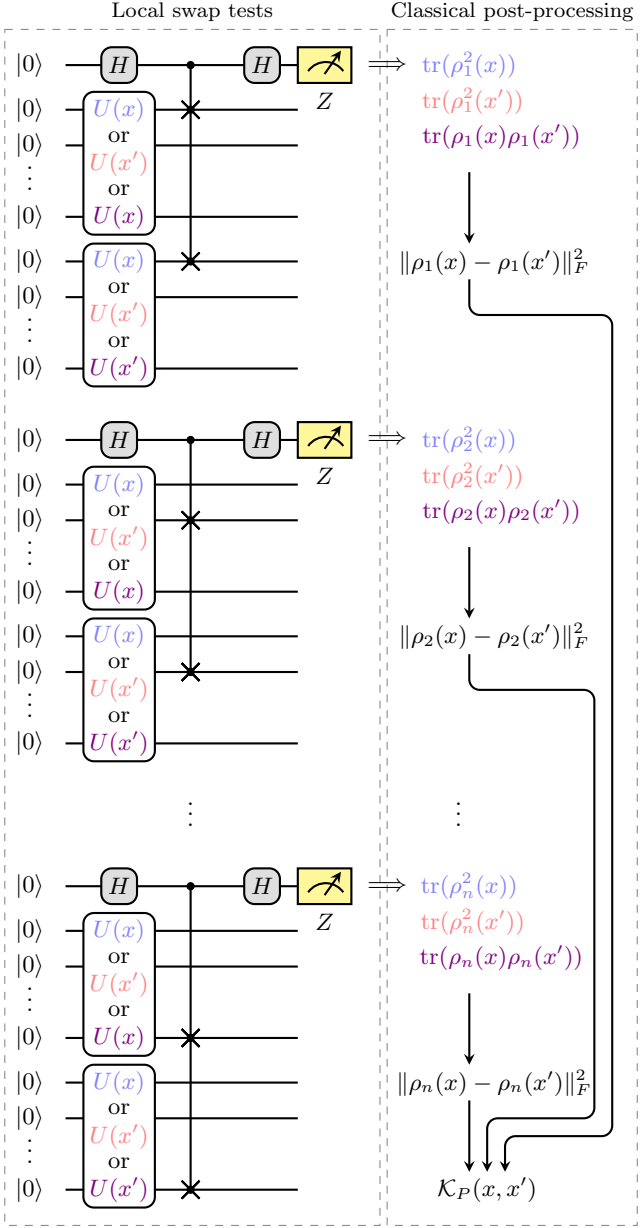
\begin{figure}
    \centering
    \begin{tikzpicture}[scale=0.95]
        \draw[black!50,dashed,line width=0.1mm] (-0.35,0.5) --++ (5.25,0) --++ (0,-16.75) --++ (-5.25,0) -- cycle;
        \node at (2.275,0.75) {\footnotesize Local swap tests};
        \node at (0,0) {$\ket{0}$};
        \draw[thick] (0.5,0) --++ (3.25,0);
        \draw[thick,fill=gray!25,rounded corners] (1,0.25) --++ (0.5,0) --++ (0,-0.5) --++ (-0.5,0) -- cycle;
        \node at (1.25,0) {$H$};
        \draw[fill=black] (2.25,0) circle (0.05cm);
        \draw[thick] (2.25,0) --++ (0,-3.5+0.375+0.375);
        \draw[fill=black] (2.25,-1+0.375) circle (0.05cm);
        \draw[thick] (2.25,-1+0.375) --++ (0.125,0.125) --++ (-0.25,-0.25) --++ (0.125,0.125) --++ (-0.125,0.125) --++ (0.25,-0.25) --++ (-0.125,0.125);
        \draw[fill=black] (2.25,-3.5+0.375+0.375) circle (0.05cm);
        \draw[thick] (2.25,-3.5+0.375+0.375) --++ (0.125,0.125) --++ (-0.25,-0.25) --++ (0.125,0.125) --++ (-0.125,0.125) --++ (0.25,-0.25) --++ (-0.125,0.125);
        \draw[thick,fill=gray!25,rounded corners] (3,0.25) --++ (0.5,0) --++ (0,-0.5) --++ (-0.5,0) -- cycle;
        \node at (3.25,0) {$H$};
        \draw[thick,fill=yellow!50] (4.75-1,-0.25) -- ++(0.75,0) -- ++(0,0.5) -- ++(-0.75,0) -- cycle;
        \draw[thick] (4.645+0.25-1,-0.085) arc (145:35:0.3) ;
        \draw[thick,-stealth] (4.875+0.25-1,-0.125) -- ++({0.4*cos(55)},{0.4*sin(55)});
        \node at (6-0.5-0.375-1,-0.5) {$Z$};
        \node at (0,-1+0.375) {$\ket{0}$};
        \node at (0,-1.5+0.375) {$\ket{0}$};
        \node at (0,-1.85+0.375) {$\vdots$};
        \node at (0,-2.5+0.375) {$\ket{0}$};
        \draw[thick] (0.5,-1+0.375) --++ (3.25,0);
        \draw[thick] (0.5,-1.5+0.375) --++ (3.25,0);
        \draw[thick] (0.5,-2.5+0.375) --++ (3.25,0);
        \draw[thick,fill=white,rounded corners] (0.75,-0.75+0.375) --++ (1,0) --++ (0,-2) --++ (-1,0) -- cycle;
        \node at (1.25,-1+0.375) {$\textcolor{blue!50}{U(x)}$};
        \node at (1.25,-1.375+0.375) {or};
        \node at (1.25,-1.75+0.375) {$\textcolor{red!50}{U(x^\prime)}$};
        \node at (1.25,-2.125+0.375) {or};
        \node at (1.25,-2.5+0.375) {$\textcolor{blue!50!red}{U(x)}$};
        \node at (0,-3.5+0.375+0.375) {$\ket{0}$};
        \node at (0,-4+0.375+0.375) {$\ket{0}$};
        \node at (0,-4.35+0.375+0.375) {$\vdots$};
        \node at (0,-5+0.375+0.375) {$\ket{0}$};
        \draw[thick] (0.5,-3.5+0.375+0.375) --++ (3.25,0);
        \draw[thick] (0.5,-4+0.375+0.375) --++ (3.25,0);
        \draw[thick] (0.5,-5+0.375+0.375) --++ (3.25,0);
        \draw[thick,fill=white,rounded corners] (0.75,-3.25+0.375+0.375) --++ (1,0) --++ (0,-2) --++ (-1,0) -- cycle;
        \node at (1.25,-3.5+0.375+0.375) {$\textcolor{blue!50}{U(x)}$};
        \node at (1.25,-3.875+0.375+0.375) {or};
        \node at (1.25,-4.25+0.375+0.375) {$\textcolor{red!50}{U(x^\prime)}$};
        \node at (1.25,-4.625+0.375+0.375) {or};
        \node at (1.25,-5+0.375+0.375) {$\textcolor{blue!50!red}{U(x^\prime)}$};
        \node at (0,0-5.25) {$\ket{0}$};
        \draw[thick] (0.5,0-5.25) --++ (3.25,0);
        \draw[thick,fill=gray!25,rounded corners] (1,0.25-5.25) --++ (0.5,0) --++ (0,-0.5) --++ (-0.5,0) -- cycle;
        \node at (1.25,0-5.25) {$H$};
        \draw[fill=black] (2.25,0-5.25) circle (0.05cm);
        \draw[thick] (2.25,0-5.25) --++ (0,-3.5+0.375+0.375-0.5);
        \draw[fill=black] (2.25,-1+0.375-5.25-0.5) circle (0.05cm);
        \draw[thick] (2.25,-1+0.375-5.25-0.5) --++ (0.125,0.125) --++ (-0.25,-0.25) --++ (0.125,0.125) --++ (-0.125,0.125) --++ (0.25,-0.25) --++ (-0.125,0.125);
        \draw[fill=black] (2.25,-3.5+0.375+0.375-5.25-0.5) circle (0.05cm);
        \draw[thick] (2.25,-3.5+0.375+0.375-5.25-0.5) --++ (0.125,0.125) --++ (-0.25,-0.25) --++ (0.125,0.125) --++ (-0.125,0.125) --++ (0.25,-0.25) --++ (-0.125,0.125);
        \draw[thick,fill=gray!25,rounded corners] (3,0.25-5.25) --++ (0.5,0) --++ (0,-0.5) --++ (-0.5,0) -- cycle;
        \node at (3.25,0-5.25) {$H$};
        \draw[thick,fill=yellow!50] (4.75-1,-0.25-5.25) -- ++(0.75,0) -- ++(0,0.5) -- ++(-0.75,0) -- cycle;
        \draw[thick] (4.645+0.25-1,-0.085-5.25) arc (145:35:0.3) ;
        \draw[thick,-stealth] (4.875+0.25-1,-0.125-5.25) -- ++({0.4*cos(55)},{0.4*sin(55)});
        \node at (6-0.5-0.375-1,-0.5-5.25) {$Z$};
        \node at (0,-1+0.375-5.25) {$\ket{0}$};
        \node at (0,-1.5+0.375-5.25) {$\ket{0}$};
        \node at (0,-1.85+0.375-5.25) {$\vdots$};
        \node at (0,-2.5+0.375-5.25) {$\ket{0}$};
        \draw[thick] (0.5,-1+0.375-5.25) --++ (3.25,0);
        \draw[thick] (0.5,-1.5+0.375-5.25) --++ (3.25,0);
        \draw[thick] (0.5,-2.5+0.375-5.25) --++ (3.25,0);
        \draw[thick,fill=white,rounded corners] (0.75,-0.75+0.375-5.25) --++ (1,0) --++ (0,-2) --++ (-1,0) -- cycle;
        \node at (1.25,-1+0.375-5.25) {$\textcolor{blue!50}{U(x)}$};
        \node at (1.25,-1.375+0.375-5.25) {or};
        \node at (1.25,-1.75+0.375-5.25) {$\textcolor{red!50}{U(x^\prime)}$};
        \node at (1.25,-2.125+0.375-5.25) {or};
        \node at (1.25,-2.5+0.375-5.25) {$\textcolor{blue!50!red}{U(x)}$};
        \node at (0,-3.5+0.375+0.375-5.25) {$\ket{0}$};
        \node at (0,-4+0.375+0.375-5.25) {$\ket{0}$};
        \node at (0,-4.35+0.375+0.375-5.25) {$\vdots$};
        \node at (0,-5+0.375+0.375-5.25) {$\ket{0}$};
        \draw[thick] (0.5,-3.5+0.375+0.375-5.25) --++ (3.25,0);
        \draw[thick] (0.5,-4+0.375+0.375-5.25) --++ (3.25,0);
        \draw[thick] (0.5,-5+0.375+0.375-5.25) --++ (3.25,0);
        \draw[thick,fill=white,rounded corners] (0.75,-3.25+0.375+0.375-5.25) --++ (1,0) --++ (0,-2) --++ (-1,0) -- cycle;
        \node at (1.25,-3.5+0.375+0.375-5.25) {$\textcolor{blue!50}{U(x)}$};
        \node at (1.25,-3.875+0.375+0.375-5.25) {or};
        \node at (1.25,-4.25+0.375+0.375-5.25) {$\textcolor{red!50}{U(x^\prime)}$};
        \node at (1.25,-4.625+0.375+0.375-5.25) {or};
        \node at (1.25,-5+0.375+0.375-5.25) {$\textcolor{blue!50!red}{U(x^\prime)}$};
        \node at (2.25,-10.375) {$\vdots$};
        \node at (6,-10.375) {$\vdots$};
        \node at (0,0-5.25-6.25) {$\ket{0}$};
        \draw[thick] (0.5,0-5.25-6.25) --++ (3.25,0);
        \draw[thick,fill=gray!25,rounded corners] (1,0.25-5.25-6.25) --++ (0.5,0) --++ (0,-0.5) --++ (-0.5,0) -- cycle;
        \node at (1.25,0-5.25-6.25) {$H$};
        \draw[fill=black] (2.25,0-5.25-6.25) circle (0.05cm);
        \draw[thick] (2.25,0-5.25-6.25) --++ (0,-3.5+0.375+0.375-1.5);
        \draw[fill=black] (2.25,-1+0.375-5.25-6.25-1.5) circle (0.05cm);
        \draw[thick] (2.25,-1+0.375-5.25-6.25-1.5) --++ (0.125,0.125) --++ (-0.25,-0.25) --++ (0.125,0.125) --++ (-0.125,0.125) --++ (0.25,-0.25) --++ (-0.125,0.125);
        \draw[fill=black] (2.25,-3.5+0.375+0.375-5.25-6.25-1.5) circle (0.05cm);
        \draw[thick] (2.25,-3.5+0.375+0.375-5.25-6.25-1.5) --++ (0.125,0.125) --++ (-0.25,-0.25) --++ (0.125,0.125) --++ (-0.125,0.125) --++ (0.25,-0.25) --++ (-0.125,0.125);
        \draw[thick,fill=gray!25,rounded corners] (3,0.25-5.25-6.25) --++ (0.5,0) --++ (0,-0.5) --++ (-0.5,0) -- cycle;
        \node at (3.25,0-5.25-6.25) {$H$};
        \draw[thick,fill=yellow!50] (4.75-1,-0.25-5.25-6.25) -- ++(0.75,0) -- ++(0,0.5) -- ++(-0.75,0) -- cycle;
        \draw[thick] (4.645+0.25-1,-0.085-5.25-6.25) arc (145:35:0.3) ;
        \draw[thick,-stealth] (4.875+0.25-1,-0.125-5.25-6.25) -- ++({0.4*cos(55)},{0.4*sin(55)});
        \node at (6-0.5-0.375-1,-0.5-5.25-6.25) {$Z$};
        \node at (0,-1+0.375-5.25-6.25) {$\ket{0}$};
        \node at (0,-1.5+0.375-5.25-6.25) {$\ket{0}$};
        \node at (0,-1.85+0.375-5.25-6.25) {$\vdots$};
        \node at (0,-2.5+0.375-5.25-6.25) {$\ket{0}$};
        \draw[thick] (0.5,-1+0.375-5.25-6.25) --++ (3.25,0);
        \draw[thick] (0.5,-1.5+0.375-5.25-6.25) --++ (3.25,0);
        \draw[thick] (0.5,-2.5+0.375-5.25-6.25) --++ (3.25,0);
        \draw[thick,fill=white,rounded corners] (0.75,-0.75+0.375-5.25-6.25) --++ (1,0) --++ (0,-2) --++ (-1,0) -- cycle;
        \node at (1.25,-1+0.375-5.25-6.25) {$\textcolor{blue!50}{U(x)}$};
        \node at (1.25,-1.375+0.375-5.25-6.25) {or};
        \node at (1.25,-1.75+0.375-5.25-6.25) {$\textcolor{red!50}{U(x^\prime)}$};
        \node at (1.25,-2.125+0.375-5.25-6.25) {or};
        \node at (1.25,-2.5+0.375-5.25-6.25) {$\textcolor{blue!50!red}{U(x)}$};
        \node at (0,-3.5+0.375+0.375-5.25-6.25) {$\ket{0}$};
        \node at (0,-4+0.375+0.375-5.25-6.25) {$\ket{0}$};
        \node at (0,-4.35+0.375+0.375-5.25-6.25) {$\vdots$};
        \node at (0,-5+0.375+0.375-5.25-6.25) {$\ket{0}$};
        \draw[thick] (0.5,-3.5+0.375+0.375-5.25-6.25) --++ (3.25,0);
        \draw[thick] (0.5,-4+0.375+0.375-5.25-6.25) --++ (3.25,0);
        \draw[thick] (0.5,-5+0.375+0.375-5.25-6.25) --++ (3.25,0);
        \draw[thick,fill=white,rounded corners] (0.75,-3.25+0.375+0.375-5.25-6.25) --++ (1,0) --++ (0,-2) --++ (-1,0) -- cycle;
        \node at (1.25,-3.5+0.375+0.375-5.25-6.25) {$\textcolor{blue!50}{U(x)}$};
        \node at (1.25,-3.875+0.375+0.375-5.25-6.25) {or};
        \node at (1.25,-4.25+0.375+0.375-5.25-6.25) {$\textcolor{red!50}{U(x^\prime)}$};
        \node at (1.25,-4.625+0.375+0.375-5.25-6.25) {or};
        \node at (1.25,-5+0.375+0.375-5.25-6.25) {$\textcolor{blue!50!red}{U(x^\prime)}$};
        \draw[black!50,dashed,line width=0.1mm] (5,0.5) --++ (3.5,0) --++ (0,-16.75) --++ (-3.5,0) -- cycle;
        \node at (6.75,0.75) {\footnotesize Classical post-processing};
        \node at (5,0) {$\Longrightarrow$};
        \node at (6.25-0.1,0.5-0.5) {$\textcolor{blue!50}{\textrm{tr}(\rho_1^2(x))}$};
        \node at (6.3-0.1,0-0.5) {$\textcolor{red!50}{\textrm{tr}(\rho_1^2(x^\prime))}$};
        \node at (6.7-0.1,-0.5-0.5) {$\textcolor{blue!50!red}{\textrm{tr}(\rho_1(x)\rho_1(x^\prime))}$};
        \draw[thick,-stealth] (6.15,-1.5) --++ (0,-1);
        \node at (6.5,-2.75) {$\|\rho_1(x)-\rho_1(x^\prime)\|_F^2$};
        \node at (5,0-5.25) {$\Longrightarrow$};
        \node at (6.25-0.1,0.5-5.25-0.5) {$\textcolor{blue!50}{\textrm{tr}(\rho_2^2(x))}$};
        \node at (6.3-0.1,0-5.25-0.5) {$\textcolor{red!50}{\textrm{tr}(\rho_2^2(x^\prime))}$};
        \node at (6.7-0.1,-0.5-5.25-0.5) {$\textcolor{blue!50!red}{\textrm{tr}(\rho_2(x)\rho_2(x^\prime))}$};
        \draw[thick,-stealth] (6.15,-1.5-5.25) --++ (0,-1);
        \node at (6.5,-2.75-5.25) {$\|\rho_2(x)-\rho_2(x^\prime)\|_F^2$};
        \node at (5,0-5.25-1-5.25) {$\Longrightarrow$};
        \node at (6.25-0.1,0.5-5.25-1-5.25-0.5) {$\textcolor{blue!50}{\textrm{tr}(\rho_n^2(x))}$};
        \node at (6.3-0.1,0-5.25-1-5.25-0.5) {$\textcolor{red!50}{\textrm{tr}(\rho_n^2(x^\prime))}$};
        \node at (6.7-0.1,-0.5-5.25-1-5.25-0.5) {$\textcolor{blue!50!red}{\textrm{tr}(\rho_n(x)\rho_n(x^\prime))}$};
        \draw[thick,-stealth] (6.15,-1.5-5.25-6.25) --++ (0,-1);
        \node at (6.5,-2.75-5.25-6.25) {$\|\rho_n(x)-\rho_n(x^\prime)\|_F^2$};
        \node at (6.15+0.25,-15.75) {$\mathcal{K}_P(x,x^\prime)$};
        \draw[thick,-stealth, rounded corners] (6.15,-3) --++ (0,-0.5) --++ (2,0) --++ (0,-11.5) --++ (-2+0.5,0) --++ (0,-0.75+0.25);
        \draw[thick,-stealth, rounded corners] (6.15,-3-5.25) --++ (0,-0.5) --++ (2-0.25,0) --++ (0,-11.5+5.25+0.25) --++ (-2+0.5,0) --++ (0,-0.75);
        \draw[thick,-stealth, rounded corners] (6.15,-3-5.25-6.25) --++ (0,-1);
    \end{tikzpicture}
    
    \caption{Applying local swap tests and classical post-processing to estimate the PQK $\mathcal{K}_P(x,x^\prime)$.}
    \label{fig:PQK-LocalSwapTest}
\end{figure}

To describe the second protocol for estimating the PQK, we first expand the norms appearing in the sum in Eq.~\eqref{eq:PQK} as
\begin{align}
    \nonumber
    &\|\rho_i(x)-\rho_i(x^\prime)\|_{F}^2\\
    \nonumber
    &=\textrm{tr}\left((\rho_i(x)-\rho_i(x^\prime))(\rho_i(x)-\rho_i(x^\prime))\right)\\
    \label{eq:PQK-ExponentialConcentratedOverlaps}
    &=\textrm{tr}(\rho_i^2(x))+\textrm{tr}(\rho_i^2(x^\prime))-2\textrm{tr}(\rho_i(x)\rho_i(x^\prime)).
\end{align}
Thus, computing $\|\rho_i(x)-\rho_i(x^\prime)\|_{F}^2$ reduces to estimating the purities $\textrm{tr}(\rho_i^2(x))$ and $\textrm{tr}(\rho_i^2(x^\prime))$, and the overlap $\textrm{tr}(\rho_i(x)\rho_i(x^\prime))$.
These quantities can be obtained using a local version of the swap test in which the controlled swap, still conditioned on an ancilla qubit, acts only to swap the local subsystems of a particular pair of qubits.
After estimating these quantities for each of the $n$ qubit pairs, the results are classically post-processed to obtain $\|\rho_i(x)-\rho_i(x^\prime)\|_F^2$ via Eq.~\eqref{eq:PQK-ExponentialConcentratedOverlaps} and hence calculate $\mathcal{K}_P(x,x^\prime)$.
An illustration of this procedure is shown in Fig.~\ref{fig:PQK-LocalSwapTest}.

\section{Assessing quantum advantage in quantum kernel methods}
\label{sec:Assessing quantum advantage in quantum kernel methods}

We now discuss a key framework developed by Huang et al.~\cite{Huang2021Power} for assessing whether quantum kernels can offer prediction advantages over classical kernels, especially on datasets with inherent quantum structure.
While the comparison is performed only between kernels, in principle this allows for comparisons between a rich class of classical and quantum ML models.
For example, arbitrary-depth classical or quantum neural networks can be represented in terms of a special kind of kernel known as a neural tangent kernel~\cite{Arthur2018NeuralTangent,Liu2022QNeuralTangent}.
Note that throughout this section, unless stated otherwise, we will not notationally distinguish between classical and quantum kernels since the results largely hold for arbitrary kernels.

The framework was motivated by early claims of quantum advantage based on the existence of quantum circuits whose output distributions are hard to sample from classically~\cite{Bremner2016Average,Harrow2017QSupremacy}.
In~\cite{Huang2021Power} the authors show that when classical machine learning models are granted access to data generated by such circuits, they can become competitive with, and in some cases outperform, the corresponding quantum models.
This observation suggests that a more meaningful comparison to assess prediction advantages is between classical and quantum models trained on the same classically-hard data.

Accordingly, the framework considers datasets $\mathcal{D}=\{(\mathbf{x}_i,y_i)\}_{i=1}^{M}$, where the inputs $\mathbf{x}_i$ are drawn from an unknown distribution $\mathscr{D}$ over $\mathcal{X}$ and the labels $y_i\in\mathbb{R}$ are generated by a quantum process.
Specifically, we consider pure states $\tilde{\rho}(\mathbf{x}_i)$ evolved under a fixed unitary operator $U$, followed by a measurement of an observable $O$.
For each input $\mathbf{x}_i\in\mathcal{X}$, the associated label is hence given by $y_i=\mathrm{tr}(U^{\dagger}OU\tilde{\rho}(\mathbf{x}_i))$.
Note that while this explicit quantum data structure is not required to apply all techniques used in the framework, it enables the derivation of informative relationships that yield useful insight in the quantum kernel setting.

Throughout this section, we assume that all kernel matrices $K$ satisfy the normalisation condition $\textrm{tr}(K)=M$.
For quantum kernels arising from feature maps that output pure quantum states, this condition is automatically satisfied.
In other cases, this normalisation is implicitly enforced via the rescaling $K\mapsto MK/\mathrm{tr}(K)$.

Finally, for pedagogical clarity we focus on the unregularised setting $\lambda=0$, which already captures the central insights of the framework.
In~\cite{Huang2021Power}, the authors extend their analysis to finite regularization $\lambda>0$ in the Supplementary Information.
And although regularisation modifies the precise generalisation bounds and inequalities, the qualitative conclusions of the framework remain largely unchanged, with the main differences relating to small non-zero training errors.

\subsection{Three important quantities}

The framework involves three quantities which capture various properties of kernel-based models.
The first quantity of interest is the \emph{dimension}, denoted $d$, of the subspace given by the span of the pure quantum states used to derive the labels for the training data. 
That is,
\begin{equation}
    \label{eq:Dimension-DataEncodedStates}
    d\equiv\dim\left(\textrm{span}\left(\{\tilde{\rho}(\mathbf{x}_i)\}_{i=1}^{M}\right)\right).
\end{equation}
Note that $d$ can equivalently be defined as the $d=\textrm{rank}(\tilde{K})$ where $\tilde{K}_{ij}=\textrm{tr}(\tilde{\rho}(\mathbf{x}_i),\tilde{\rho}(\mathbf{x}_j))$.

The second quantity of interest is the \emph{model complexity}, which measures the compatibility of the chosen kernel $\mathcal{K}$ with the dataset $\mathcal{D}$.
Specifically, given a kernel $\mathcal{K}$ we denote the associated feature map by $\phi:\mathcal{X}\to\mathcal{F}$ and the kernel matrix by $K_{ij}=\mathcal{K}(\mathbf{x}_i,\mathbf{x}_j)$. 
The model complexity, denoted $s_K(M)$, is then defined such that
\begin{equation}
    \label{eq:ModelComplexityRegularised-s}
    s_K(M)\equiv\sum_{i,j=1}^{M}\left(K^{-1}\right)_{ij}\textrm{tr}(O^U\tilde{\rho}(\mathbf{x}_i))\textrm{tr}(O^U\tilde{\rho}(\mathbf{x}_j)),
\end{equation}
where $O^U\equiv U^\dagger OU$ is the observable $O$ evolved in the Heisenberg picture of quantum mechanics.
This, of course, assumes that $K$ is invertible, otherwise we replace $K^{-1}$ with its Moore-Penrose pseudoinverse $K^+$.

The model complexity in Eq.~\eqref{eq:ModelComplexityRegularised-s} can be equivalently defined as $s_K(M)=\|\mathbf{w}\|_\mathcal{F}^2$, where $\mathbf{w}\in\mathcal{F}$ is given by
\begin{equation*}
    \mathbf{w}=\sum_{i,j=1}^{M}\textrm{tr}(O^U\tilde{\rho}(\mathbf{x}_j))(K)^{-1}_{ij}\phi(\mathbf{x}_i).
\end{equation*} 
After applying KRR with $\lambda=0$ to the dataset $\mathcal{D}$ with the kernel $\mathcal{K}$, $\mathbf{w}\in\mathcal{F}$ defines the trained model $f\in\mathcal{R}_{\mathcal{K}}$ via $f(x)=\langle\mathbf{w},\phi(x)\rangle_{\mathcal{F}}$.
The expression $\|\mathbf{w}\|_\mathcal{F}^2$ (which is the model complexity $s_K(M)$) hence relates to both the Rademacher complexity (see Theorem 6.12 of~\cite{Mohri2018Foundations}) and the Vapnik-Chervonenkis dimension (see Theorem 5.5 of~\cite{Scholkopf2001Kernels}), making it a natural measure of complexity.

Finally, the third and most important quantity is the \emph{geometric difference}, which is an asymmetric measure of the difference between the geometries induced in $\mathcal{X}$ by two distinct kernels.
Given two kernels $\mathcal{K}_1$ and $\mathcal{K}_2$ with kernel matrices $(K_1)_{ij}=\mathcal{K}_1(\mathbf{x}_i,\mathbf{x}_j)$ and $(K_2)_{ij}=\mathcal{K}_2(\mathbf{x}_i,\mathbf{x}_j)$, the geometric difference between the kernels, denoted $g_{12}$, is defined such that
\begin{equation}
    \label{eq:GeometricDifference-g}
    g_{12}\equiv\sqrt{\|\sqrt{K_2}(K_1)^{-1}\sqrt{K_2}\|_{\infty}},
\end{equation}
where $\|\cdot\|_{\infty}$ denotes the spectral norm, and $(K_1)^{-1}$ denotes the inverse of $K_1$ if it exists, or the Moore-Penrose pseudoinverse of $K_1$.
Note that since $g_{12}$ is asymmetric, $g_{12}\neq g_{21}$ in general.

All three quantities, $d$, $s_K(M)$, and $g_{12}$ can be calculated via singular value decompositions applied to $\tilde{K}$, $K$, and both $K_1$ and $K_2$, respectively.
Using standard numerical packages this can be achieved in $\mathcal{O}(M^3)$ time after the necessary $M\times M$ kernel matrices have been obtained, possibly with the use of quantum devices if we use a quantum kernel.

\subsection{Inequalities and generalisation bounds}

The framework relies primarily on the following prediction error bound.
Consider an observable $O$ satisfying $\|O\|_{\infty}\leq1$, then with probability $0.99=1-\delta$ (i.e. $\delta$ fixed), there exists a ML algorithm which provides a trained model $f\in\mathcal{R}_{\mathcal{K}}$ such that
\begin{equation}
    \label{eq:PredictionErrorBound}
    \underset{x\sim\mathscr{D}}{\mathbb{E}}\left|f(x)-\textrm{tr}\left(O^U\tilde{\rho}(x)\right)\right|\leq\mathcal{O}\left(\sqrt{\frac{s_K(M)}{M}}\right).
\end{equation}
This bound is a simplified version of the general bound supplied in Theorem 1 of the Supplementary Information in~\cite{Huang2021Power}, since it assumes that $\lambda=0$ and that $\delta$ is fixed.

From the inequality in Eq.~\eqref{eq:PredictionErrorBound}, we see that using a kernel with a \emph{smaller} model complexity will likely \emph{improve} prediction accuracy, which is especially important for generalisation.
Accordingly, if the model complexity is small for some classical kernel, then it is likely that the labels can be predicted accurately using this kernel, even if the labels are hard to compute classically.

Generalisation in QML has also been studied from perspectives that are not restricted to kernel methods. 
For example, Peters et al.~\cite{Peters2023Generalization} relate generalisation in overparameterised quantum models to spectral properties and Fourier-type analyses, characterising regimes of benign overfitting arising from circuit structure and data encoding. 
From a complementary information-theoretic viewpoint, Banchi et al.~\cite{Banchi2021Generalization} connect generalisation performance to the mutual information between classical data or labels and the quantum state space, analysing how properties such as noise, Hilbert space dimension, and information compression affect learnability. 
These works highlight alternative mechanisms governing generalisation in QML, but they will not be a focus in this review.

The bound in Eq.~\eqref{eq:PredictionErrorBound} shows us that we can assess whether one kernel $\mathcal{K}_1$ is likely to provide a better prediction accuracy than another $\mathcal{K}_2$ for a given dataset by looking at the separation between $s_{K_1}(M)$ and $s_{K_2}(M)$.
Such a separation can be analysed using the following inequality which relates the model complexities of $\mathcal{K}_1$ and $\mathcal{K}_2$, and the geometric difference $g_{12}$ defined in Eq.~\eqref{eq:GeometricDifference-g}.
Specifically,
\begin{equation}
    \label{eq:GeometricDifference-ModelComplexitySeparation}
    s_{K_1}(M)\leq (g_{12})^2s_{K_2}(M),
\end{equation}
which naturally implies that $\sqrt{s_{K_1}\slash M}\leq g_{12}\sqrt{s_{K_2}\slash M}$.

From Eq.~\eqref{eq:GeometricDifference-ModelComplexitySeparation}, we see that when $g_{12}$ is small ($\approx1$), $\mathcal{K}_1$ will have a similar or smaller model complexity than $\mathcal{K}_2$, and hence likely predict equally as well or better than $\mathcal{K}_2$.
Contrarily, when $g_{12}$ is large it is \emph{possible} that the model complexity of $\mathcal{K}_1$ will be much larger, allowing for possible prediction advantages using $\mathcal{K}_2$ instead of $\mathcal{K}_1$.

To finish off this subsection, we consider the FQK $\mathcal{K}_{\tilde{\rho}}(x,x^\prime)=\langle\tilde{\rho}(x),\tilde{\rho}(x^\prime)\rangle_{\mathcal{H}_n}$ with kernel matrix $(K_{\tilde{\rho}})_{ij}=\mathcal{K}_{\tilde{\rho}}(\mathbf{x}_i,\mathbf{x}_j)$.
In this case, we can gain further insight into the model complexity.
Specifically, it can be shown that
\begin{equation}
    \label{eq:Huang-QuantumModelComplexityBound}
    s_{K_{\tilde{\rho}}}(M)\leq\min(d,\textrm{tr}(O^2)).
\end{equation}
Together with Eq.~\eqref{eq:PredictionErrorBound}, Eq.~\eqref{eq:Huang-QuantumModelComplexityBound} shows that the prediction error, in the case where we use the problem-inspired kernel $\mathcal{K}_{\tilde{\rho}}$ derived from the same quantum feature map $\tilde{\rho}$ used to determine the dataset labels $y_i$, is dictated by the minimum of the dimension $d$ defined in Eq.~\eqref{eq:Dimension-DataEncodedStates} and the squared Frobenius norm of the observable $O$.

This provides an alternative lens for assessing how well a problem-inspired FQK is likely to perform on a given problem, and may be used to guide FQK designs in some scenarios.
Note that an estimate of $\textrm{tr}(O^2)$ can be obtained by sampling random states from a quantum 2-design, measuring $O$ from the random states, and then post-processing the measurement outcomes~\cite{Huang2020Predicting}.

\subsection{Necessary conditions for quantum advantage}
\label{subsec:Necessary conditions for quantum advantage}

We now put everything together and discuss necessary conditions for finding a prediction advantage using a quantum kernel in place of a classical one.
Accordingly, in this discussion we denote an arbitrary classical and quantum kernel by $\mathcal{K}_C$ and $\mathcal{K}_Q$ respectively.
The main quantity to consider is the geometric difference 
\begin{equation*}
    g_{CQ}=\sqrt{\left\|\sqrt{K_Q}(K_C)^{-1}\sqrt{K_Q}\right\|_{\infty}},
\end{equation*}
where $(K_C)_{ij}=\mathcal{K}_C(\mathbf{x}_i,\mathbf{x}_j)$ and $(K_Q)_{ij}=\mathcal{K}_Q(\mathbf{x}_i,\mathbf{x}_j)$ denote the associated kernel matrices.
Note that $g_{CQ}$ depends only the input training data $\mathbf{x}_i\in\mathcal{X}$, but not the labels $y_i$ (allowing for $g_{CQ}$ to be calculated for classical datasets), and indicates whether there is a difference in the geometries induced in $\mathcal{X}$ by $\mathcal{K}_C$ and $\mathcal{K}_Q$.

With this in mind, the following conditions are necessary for the possibility of a prediction advantage.
\begin{framed}
\noindent\emph{Necessary conditions for a prediction advantage with QKMs:}
\begin{enumerate}[(i)]
    \item \emph{The geometric difference $g_{CQ}$ is large, growing proportional to $\sqrt{M}$,}
    \item \emph{The classical model complexity $s_{K_C}(M)$ is large, growing proportional to $M$, and}
    \item \emph{The quantum model complexity $s_{K_Q}(M)$ is small compared with $M$.}
\end{enumerate}
\end{framed}

Note that these conditions are necessary, but not sufficient, for a prediction advantage.
If any of the three conditions are not satisfied, then it will likely not be possible to achieve a prediction advantage using QKMs in place of a classical model.
This may be due to the fact that classical KMs will provide equally as good, or possibly better, prediction accuracy, or to the fact that both QKMs and classical KMs will likely perform poorly (see Fig. 1 of~\cite{Huang2021Power}).
And even if all three conditions are satisfied, it may be the case that a prediction advantage cannot be obtained. 

Worth noting is that in the case where condition (i) is satisfied, Huang et al.~\cite{Huang2021Power} show how to construct a dataset for which $s_{K_C}(M)=(g_{CQ})^2s_{K_Q}(M)$.
However this outcome is specific to their constructed dataset, and in general will not be true.
This means that the conditions need to be explicitly checked for a given dataset, and serve as a good starting point to filter out problem instances which will not provide prediction advantages.

To check the conditions, ideally one would evaluate the model complexity for all classical kernel-based models, and take the minimum value to assess whether the labels can be accurately predicted by a classical ML model.
In practice however, the model complexity is usually evaluated over an ensemble of classical kernel-based models whose hyperparameters have been tuned for the dataset.

As a final note, Huang et al.~\cite{Huang2021Power} observed in their numerical simulations that the PQK exhibited a larger $g_{CQ}$ than the FQK. 
This provides an empirical indication that PQKs might hold more promise for obtaining a prediction advantage than FQKs.

\section{Challenges for quantum kernel methods}
\label{sec:Challenges for quantum kernel methods}

We now discuss some of the central challenges surrounding QKMs.
We begin by considering the well-known exponential concentration (EC) phenomenon and its relation to shot noise.
We then examine how expressivity, entanglement, global measurements, and hardware noise can induce EC and create practical implementation barriers for QKMs.
We subsequently review dequantisation methods, with a particular focus on tensor-network based dequantuisation, which have pushed the limits of classical simulation capabilities for quantum systems and therefore challenge claims of quantum advantage based on certain hardware architectures.
Next, we consider issues related to the spectral properties of the associated kernel integral operator.
Finally, we address a method known as quantum bandwidth tuning, which was originally thought to be a promising approach for mitigating some of these challenges, but is now known to introduce its own difficulties.

This section is especially important because deriving practical advantages from QKMs requires not only that the necessary conditions discussed in the previous section are satisfied, but also that the challenges outlined here are taken into account when designing quantum feature maps.
For example, one must ensure that the employed quantum kernels do not suffer from EC as system size increases, and that the associated quantum models lie beyond the reach of efficient classical simulation.

\subsection{Exponential concentration and shot noise}
\label{sec:Exponential concentration and shot noise}

The EC phenomena was first observed in~\cite{Huang2021Power}, further analysed in~\cite{Kubler2021Inductive,Shaydulin2022Importance,Canatar2023Bandwidth}, and then explored in great detail by Thanasilp et al.~\cite{Thanasilp2024Exponential}.
In this section we will primarily discuss the results from~\cite{Thanasilp2024Exponential}.
The phenomena is similar to the barren plateau phenomena in quantum neural networks and occurs when the value of a quantum kernel concentrates exponentially with increasing qubit counts to some fixed value.
In fact, the authors of~\cite{Thanasilp2024Exponential} identified four sources of EC for quantum kernels, namely expressivity, entanglement, hardware noise, and the use of a global measurement, which are discussed in subsequent subsections (see Table~\ref{tab:ExponentialConcentrationSources}).
Perhaps not surprisingly, these four sources have also been shown to induce barren plateaus in quantum neural networks~\cite{Holmes2022ConnectingExpressibility,Cerezo2021CostFunctionBPs,Marrero2021EntanglementBPs,Wang2021NoiseInducedBPs}.

\begin{table*}
\begin{adjustbox}{center,max width=\textwidth}
\begin{tabular}{ | c | c | c | c |}
\cline{1-4}
\textbf{Source}&\textbf{Affected kernels}&\textbf{Relevant section}&\textbf{Corresponding result in~\cite{Thanasilp2024Exponential}}\\
\cline{1-4}
Expressivity&FQKs \& PQKs&\ref{sec:Expressivity}&Theorem 1\\
Entanglement&PQKs&\ref{sec:Entanglement}&Theorem 2 \& Corollary 2\\
Global measurements&FQKs&\ref{sec:Global measurements}&Proposition 3\\
Hardware noise&FQKs \& PQKs&\ref{sec:Hardware noise}&Theorem 3\\
\cline{1-4}
\end{tabular}
\end{adjustbox}
\caption{\footnotesize \justifying A summary of the sources identified in~\cite{Thanasilp2024Exponential} which cause EC, the type of quantum kernels these sources impact, the sections of this paper in which they are discussed, and the associated results from the original paper.}
\label{tab:ExponentialConcentrationSources}
\end{table*}

\subsubsection{Preliminaries}
\label{subsec:Preliminaries}

To aid our discussion of EC, here we will establish precisely what is meant by EC and statistical indistinguishability, and finish by discussing how kernel values are estimated in practice on quantum computers.

\paragraph{Exponential concentration definition.}

Consider a quantity $q(x)$ that depends on some variables $x$ and can be measured from a quantum computer with $n$ qubits.
The quantity $q(x)$ is said to \emph{deterministically exponentially concentrate} in the number of qubits $n$ to a fixed value $\mu\in\R$ if
\begin{equation*}
    \left|q(x)-\mu\right|\leq\omega,
\end{equation*}
for some $\omega\in\mathcal{O}(1\slash c^n)$ and $c>1$, and for all $x$.
The quantity $q(x)$ is said to \emph{probabilistically exponentially concentrate} in $n$ to a fixed value $\mu$ if
\begin{equation*}
    \textrm{P}_x\big[\left|q(x)-\mu\right|\geq\delta\big]\leq\frac{\omega}{\delta^2},
\end{equation*}
for some $\omega\in\mathcal{O}(1\slash c^n)$ and $c>1$, where $\textrm{P}_{x}$ denotes the probability taken over all choices of $x$ sampled from the distribution associated with the variables $x$.
In other words, the probability that $q(x)$ deviates from $\mu$ by a small amount $\delta$ is exponentially small in $n$.
Further, if $\mu$ decays exponentially with the number of qubits, i.e. $\mu\in\mathcal{O}(1\slash (c^\prime)^n)$ for some $c^\prime>1$, then we say that the quantity $q(x)$ \emph{exponentially concentrates to an exponentially small value}.
Note that throughout the discussion we will refer to either type of EC (deterministic or probabilistic) as just EC.
The specific kind of EC being considered should be obvious from notation.

\paragraph{Statistical indistinguishability.}

Consider two probability distributions $\mathcal{P}_1$ and $\mathcal{P}_2$ over a common set of outcomes $\mathscr{O}$.
Suppose that we are given $m\in\N$ samples from $\mathscr{O}$, either all drawn independently according to $\mathcal{P}_1$, or all drawn independently according to $\mathcal{P}_2$, with equal probability.
We consider the following hypotheses:
\begin{itemize}
    \item $\mathscr{H}_0$: the $m$ samples are drawn from $\mathcal{P}_1$,
    \item $\mathscr{H}_1$: the $m$ samples are drawn from $\mathcal{P}_2$.
\end{itemize}

The distributions $\mathcal{P}_1$ and $\mathcal{P}_2$ are said to be \emph{statistically indistinguishable from $m$ samples} if, for any algorithm, the probability $P[\textrm{success}]$ of correctly identifying the true hypothesis, $\mathscr{H}_0$ or $\mathscr{H}_1$, satisfies
\begin{equation*}
P[\textrm{success}] \leq \tfrac{1}{2} + \epsilon,
\end{equation*}
where $\epsilon>0$ is small.
The precise value of $\epsilon$ is unimportant. 
Rather, indistinguishability means that access to $m$ samples does not permit performance significantly better than random guessing.

If two distributions are statistically indistinguishable from $m$ samples, then the distribution of the output of any statistic computed from the samples is, up to small probabilistic deviations, the same under $\mathcal{P}_1$ and $\mathcal{P}_2$.
Formally, let $\mathcal{P}_1$ and $\mathcal{P}_2$ be statistically indistinguishable from $m$ samples, and consider a map $F:\mathscr{O}^m\to\R$.
Given $m$ samples drawn independently from each distribution $S_{\mathcal{P}_1}$ and $S_{\mathcal{P}_2}$, we say that the values $F(S_{\mathcal{P}_1})$ and $F(S_{\mathcal{P}_2})$ are \emph{statistically indistinguishable}.

\paragraph{Statistical estimates of quantum kernels.}

In practice, the value of a quantum kernel or a quantity used to determine its value—such as the overlap $\textrm{tr}(\rho_i(x)\rho_i(x^\prime))$ for the PQK—is estimated on a quantum computer by taking $m$ measurements of some observable $O$ from some quantum state $\rho^\prime$ and averaging the measurement outcomes.

Taking the spectral decomposition of $O$ we can write $O=\sum_io_i|o_i\rangle\langle o_i|$ where $o_i$ and $|o_i\rangle$ are the eigenvalues and eigenvectors of $O$ respectively.
The value of the quantity of interest $q(x,x^\prime)$ estimated from the $m$ measurement outcomes is then given by
\begin{equation}
    \label{eq:StatisticalEstimate-q}
    q(x,x^\prime)=\frac{1}{m}\sum_{j=1}^{m}\lambda_j,
\end{equation}
where $\lambda_j$ is the outcome of the $j^{\textrm{th}}$ measurement, taking on one of the values in $\{o_i\}$ with probability $\langle o_i|\rho^\prime|o_i\rangle$.

With the exception of estimating the FQK via the Loschmidt echo test, the observable $O$ will be a Pauli operator $\sigma\in\{X,Y,Z\}$ measured from a single qubit (see Section \ref{sec:Estimating quantum kernels in practice}).
In this case, the measurement outcomes $\lambda_j$ are $\pm1$ occurring with probabilities
\begin{equation*}
    P[\sigma=\pm 1] = \frac{1 \pm q(x,x^\prime)}{2}.
\end{equation*}
This induces a binary distribution
\begin{equation*}
\mathcal{P}_{q(x,x^\prime)}
\equiv \left\{ \tfrac{1+q(x,x^\prime)}{2}, \tfrac{1-q(x,x^\prime)}{2} \right\}.
\end{equation*}

With all components now in place, the central observation is as follows.
If we assume that $q(x,x^\prime)$ exponentially concentrates to a fixed value $\mu\in\R$, then $\mathcal{P}_{q(x,x^\prime)}$ and the constant distribution
\begin{equation*}
\mathcal{P}_{\mu}
\equiv \left\{ \tfrac{1+\mu}{2}, \tfrac{1-\mu}{2} \right\}
\end{equation*}
will be statistically indistinguishable from any polynomial number of shots $m\in\mathcal{O}(\textrm{poly}(n))$.
Consequently, with only polynomially many shots, the statistical estimate of $q(x,x^\prime)$ given in Eq.~\eqref{eq:StatisticalEstimate-q} will be statistically indistinguishable from the same statistic estimated according to $\mathcal{P}_\mu$. 
This means that our estimate of $q(x,x^\prime)$ will be ultimately independent of the inputs $x,x^\prime\in\mathcal{X}$.
Much of the following discussion about why EC is an issue for QKMs rests on this idea, since we can only afford polynomially many shots in practice, which is not sufficient to ensure that estimates of $q(x,x^\prime)$ depend on the inputs $x$ and $x^\prime$ in a meaningful way.

\subsubsection{Why is exponential concentration an issue?}

We now discuss why EC is a problem.
Specifically, we will assume that the quantum kernel in question exponentially concentrates, possibly to an exponentially small value, and show that this results in models which have not been meaningfully informed by the training data, and hence will not be capable of generalising.

\paragraph{Effects of exponential concentration for FQKs.}

Consider the FQK $\mathcal{K}_F(x,x^\prime)$ estimated via the Loschmidt echo test shown in Fig.~\ref{fig:FQK-LoschmidtEcho}.
For two distinct inputs $x \neq x^\prime$, the kernel value equals the probability of observing the all-zero state $\ket{0}^{\otimes n}$ upon measurement.
When the kernel exponentially concentrates to an exponentially small value, this probability is likewise exponentially small.
With only polynomially many shots, it is therefore exponentially unlikely to observe the all-zero outcome even once, and the empirical estimate of each off-diagonal entry of the kernel matrix will be zero.
Consequently, the estimated kernel matrix equals the $M \times M$ identity with probability exponentially close to one.
That is,
\begin{equation*}
    P\left[ K = \mathbb{I} \right] \geq 1 - \delta,
\end{equation*}
for some $\delta\in\mathcal{O}(1/c^n)$ and $c>1$.
Thus, the kernel matrix is exponentially likely to be independent of the training data.

Now consider the FQK $\mathcal{K}_F(x,x^\prime)$ estimated via the swap test shown in Fig.~\ref{fig:FQK-SwapTest}.
In this case, the FQK is given by the expectation value of the Pauli-$Z$ operator measured on the ancilla qubit.
If $\mathcal{K}_F(x,x^\prime)$ exponentially concentrates to an exponentially small value, then the induced binary distribution $\mathcal{P}_{\mathcal{K}_F(x,x^\prime)}$ will be statistically indistinguishable from the distribution $\mathcal{P}_0$.
Consequently, the kernel estimates obtained from the swap test with polynomially many shots are statistically indistinguishable from the estimates obtained with $\mathcal{P}_0$, which is independent of the inputs $x,x^\prime\in\mathcal{X}$. 
This means that the kernel matrix, once again, will not capture any structure from the training data.

In their work, Thanasilp et al.~\cite{Thanasilp2024Exponential} do not explicitly consider the Hadamard test for estimating $\mathcal{K}_F(x,x^\prime)$, however the same issues naturally arise since
\begin{equation*}
    \mathcal{K}_F(x,x^\prime)=\textrm{Re}[\langle\psi(x)|\psi(x^\prime)\rangle]^2+\textrm{Im}[\langle\psi(x)|\psi(x^\prime)\rangle]^2.
\end{equation*}
From the above equation, if $\mathcal{K}_F(x,x^\prime)$ exponentially concentrates to an exponentially small value then so will $\textrm{Re}[\langle\psi(x)|\psi(x^\prime)\rangle]$ and $\textrm{Im}[\langle\psi(x)|\psi(x^\prime)\rangle]$, resulting in the statistical indistinguishability of the distributions $\mathcal{P}_{\textrm{Re}[\langle\psi(x)|\psi(x^\prime)\rangle]}$ and $\mathcal{P}_{\textrm{Im}[\langle\psi(x)|\psi(x^\prime)\rangle]}$ from $\mathcal{P}_0$ with polynomially many shots.
Hence the associated kernel estimates obtained with polynomially many shots will again encode no information about the training data.

\paragraph{Effects of exponential concentration for PQKs.}

Analysing the consequences of EC for PQKs is somewhat less straight-forward, since in order to estimate $\mathcal{K}_P(x,x^\prime)$ we first need to obtain estimates of the Frobenius norms $\|\rho_i(x)-\rho_i(x^\prime)\|_F^2$ for all $i\in\{1,2\ldots,n\}$.
Accordingly, in order to understand the effects of EC for PQKs it is more natural to assume that
\begin{equation}
    \label{eq:PQK-ReducedStateExponentialConcentration}
    \underset{x\sim\mathscr{D}}{\mathbb{E}}\left\|\rho_i(x)-\tfrac{\mathbb{I}}{2}\right\|_F\leq\omega
\end{equation}
for all $i\in\{1,2,\ldots,n\}$ and some $\omega\in\mathcal{O}(1\slash c^n)$ with $c>1$, where $\mathbb{I}$ denotes the $2\times2$ identity matrix.
To reinforce why this is a natural starting point, the authors of~\cite{Thanasilp2024Exponential} show (in Supplemental Lemma 4 of their Supplementary Information) that this assumption implies that PQKs experience EC, i.e.
\begin{equation*}
    \underset{x,x^\prime\sim\mathscr{D}}{P}\left[\left|\mathcal{K}_P(x,x^\prime)-\mu\right|\geq\delta\right]\leq\frac{\omega}{\delta^2}
\end{equation*}
for some $\omega\in\mathcal{O}(1\slash c^n)$ with $c>1$.

Let us first consider estimating PQKs via the state tomography protocol illustrated in Fig.~\ref{fig:PQK-StateTomography}. 
Recall that the main idea underlying this method is to expand the 1-RDMs as in Eq.~\eqref{eq:1-RDMStateTomography}, estimate the expectation values of the local observables $\sigma_i$ from the state $\rho(x)$ to build the 1-RDMs, and post-process the results to calculate the PQK.
To achieve this we measure the Pauli operators $\sigma\in\{X,Y,Z\}$ from qubit $i$.
By substituting Eq.~\eqref{eq:1-RDMStateTomography} into Eq.~\eqref{eq:PQK-ReducedStateExponentialConcentration}, we have that
\begin{equation*}
    \langle X_i\rangle_{\rho(x)}^2+\langle Y_i\rangle_{\rho(x)}^2+\langle Z_i\rangle_{\rho(x)}^2\leq\omega
\end{equation*}
where $\omega\in\mathcal{O}(1\slash c^n)$ for some $c>1$, and hence all the expectation values $\langle\sigma_i\rangle_{\rho(x)}$ exponentially concentrate too.
This means that the induced binary distribution $\mathcal{P}_{\langle\sigma_i\rangle_{\rho(x)}}$
are statistically indistinguishable from $\mathcal{P}_0$ for all $x\in\mathcal{X}$.
As before, this implies that the statistical estimates of $\rho_i(x)$ will be independent of the training data, resulting in a model which is ultimately independent of the training data.

Now consider estimating the PQK $\mathcal{K}_P(x,x^\prime)$ via the local swap test protocol shown in Fig.~\ref{fig:PQK-LocalSwapTest}.
Recall that the main idea underlying this method is to expand the Frobenius norms $\|\rho_i(x)-\rho_i(x^\prime)\|_F^2$ as in Eq.~\eqref{eq:PQK-ExponentialConcentratedOverlaps} and estimate the purities $\textrm{tr}(\rho_i(x))^2$ and $\textrm{tr}(\rho_i(x^\prime))^2$, as well as the overlap $\textrm{tr}(\rho_i(x)\rho_i(x^\prime))$, using the local swap tests.
Here the quantity of interest is $q_i(x,x^\prime)=\textrm{tr}(\rho_i(x)\rho_i(x^\prime))$, which can capture both the purities (with $x=x^\prime$) and the overlap. 
In order to estimate $q_i(x,x^\prime)$ we measure the Pauli-$Z$ operator from the ancilla qubit in the local swap protocol.
If we assume that Eq.~\eqref{eq:PQK-ReducedStateExponentialConcentration} holds, then the authors of~\cite{Thanasilp2024Exponential} show (in Supplemental Lemma 5 of their Supplementary Information)that $q_i(x,x^\prime)$ exponentially concentrates to $\mu=\tfrac{1}{2}$.
Accordingly the binary distribution $\mathcal{P}_{q_i(x,x^\prime)}$
is statistically indistinguishable from the distribution $\mathcal{P}_{1\slash2}$.
Once again this implies that the resulting model will be insensitive to the training data.

\subsubsection{Shot noise, implications, and mitigation of exponential concentration}

The EC phenomenon poses a significant challenge for QKMs, since the trained models extract information about the input training data solely through the kernel matrix $K$.
However, when EC occurs, polynomially many measurements are insufficient to construct $K$ in a way that meaningfully reflects the training data.
Consequently, with only polynomially many shots, the trained models are highly unlikely to generalise.
Instead, achieving a model that genuinely learns from the training data will require exponentially many measurements to overcome shot noise, suggesting that no computational speed-ups are likely in this regime.

Previous works have also considered the effects of shot noise in QKMs to understand their practical limitations. 
Gentinetta et al.~\cite{Gentinetta2024Complexity} analyse the complexity of quantum SVMs, deriving bounds on the number of measurements required to train these models within a target accuracy. 
Specifically they show that under some basic assumptions, with a training dataset of size $M$ and a solution accuracy of $\tilde{\epsilon}\equiv\max_{x\in\mathcal{X}}|f(x)-\tilde{f}(x)|$, where $f$ denotes the trained model obtained with infinitely many measurement shots and $\tilde{f}$ the trained model obtained with finitely many shots, that the quantum SVM can be trained with a total of $\mathcal{O}\left(M^{4.67}\slash\tilde{\epsilon}^2\right)$ measurements.
However, their work does not account for EC, which presents worries about whether this result will hold in practice. 

In contrast, Miroszewski et al.~\cite{Miroszewski2024QKMShots} develop a framework for estimating the number of shots required to obtain reliable estimates of quantum kernel entries, in both noiseless and error-corrected scenarios, based on a prescribed estimation precision. 
Their analysis explicitly incorporates both EC and the spread effect, which pertains to the distribution of kernel values across the input data domain. 
They instantiate this framework for both FQKs estimated via the Loschmidt echo test (Fig.~\ref{fig:FQK-LoschmidtEcho}), and PQKs estimated using state tomography protocol (Fig.~\ref{fig:PQK-StateTomography}), deriving separate shot requirements in each case. 

By combining the constraints imposed by spread and concentration into a unified criterion, they provide practical rules for translating accuracy requirements into circuit run counts. 
To maintain the focus and flow of the presentation, we defer a detailed discussion of their framework to the original work and summarise only the key ideas here.
Nevertheless, because this framework directly addresses EC and explicitly accounts for resource costs, it is particularly relevant for assessing the feasibility of QKMs in realistic settings.

In an attempt to mitigate EC, Suzuki et al.~\cite{Suzuki2024QuantumFisher} propose a class of quantum kernels which incorporate the geometric structure of input data in the feature map, showing that these kernels can avoid the EC issue with log-depth alternating layered ansatzes while retaining expressivity comparable to standard fidelity‑based kernels. 
However, the same circuits once extended to linear-depth once again exhibit the EC phenomena.
Another proposal for mitigating EC is a technique known as quantum bandwidth tuning, which involves scaling input data features to reduce the expressivity of the data-encoding unitary.
We defer detailed discussions of this approach to Section~\ref{sec:Bandwidth tuning}.

\subsection{Expressivity}
\label{sec:Expressivity}

Expressivity in QML has been a topic of focus for a number of years.
For example, Schuld et al.~\cite{Schuld2021Effect} studies the expressivity of a class of variational QML models in their 2021 paper.
At this time, it seemed that highly expressive QML models were considered desirable, especially given the many results about benign overfitting from classical ML.
However, the idea that that overly expressive QML models could present issues quickly became apparent as the barren plateau phenomena began receiving more attention, and was shown to be rigorously connected to the expressivity of the data-encoding unitary.

In the context of QKMs, Jerbi et al.~\cite{Jerbi2023Beyond} show, via the explicit example of basis encoding, that using a highly expressive data-encoding unitary, despite achieving good training performance, could result in a model which generalises poorly.
This was ultimately a result of the fact that the kernel matrix in this case becomes the identity matrix due to the orthogonal nature of distinct computational basis vectors.
This idea clearly exhibits similarities with some of the issues that we have discussed in the context of EC.
In fact, we will now discuss how expressivity can \emph{cause} EC.

Given a data-encoding unitary $U:\mathcal{X}\to\mathbb{U}(2^n)$, the unitary ensemble $U(\mathcal{X})$ generated by $U$ is defined as
\begin{equation}
    \label{eq:DataEncodingUnitary-SetOfUnitaries}
    U(\mathcal{X})\equiv\{U(x):x\in\mathcal{X}\}.
\end{equation}
The expressivity of $U(\mathcal{X})$ is captured by the superoperator $\mathcal{A}_U$ defined such that
\begin{equation}
    \label{eq:ExpressivitySuperOperator}
    \mathcal{A}_U(\cdot)=\mathcal{V}_{\textrm{Haar}}(\cdot)-\int_{U(\mathcal{X})}d\mu(U)U^{\otimes2}(\cdot)^{\otimes2}(U^\dagger)^{\otimes2},
\end{equation}
where $\mathcal{V}_{\textrm{Haar}}(\cdot)=\int_{\mathbb{U}(2^n)}d\mu(V)V^{\otimes2}(\cdot)^{\otimes2}(V^\dagger)^{\otimes2}$ is an integral over the unitary group $\mathbb{U}(2^n)$ with the Haar measure, and the second term is an integral over the ensemble $U(\mathcal{X})$ with measure induced by the distribution $\mathscr{D}$ on $\mathcal{X}$~\cite{Holmes2022ConnectingExpressibility}.
Intuitively, the superoperator $\mathcal{A}_U$ captures how close the ensemble $U(\mathcal{X})$ is to uniformly covering the unitary group $\mathbb{U}(2^n)$, as the Haar measure does.

To quantify how close the induced measure from $U(\mathcal{X})$ is to the Haar measure, the authors of~\cite{Thanasilp2024Exponential} introduce the quantity $\epsilon_{U(\mathcal{X})}$ which, for a given fixed initial state $\rho_0$, is defined such that 
\begin{equation}
    \label{eq:epsilon_U(x)}
    \epsilon_{U(\mathcal{X})}=\left\|\mathcal{A}_{U(\mathcal{X})}(\rho_0)\right\|_1.
\end{equation}
For example, it will often be the case that $\rho_0=(|0\rangle\langle0|)^{\otimes n}$.
The value of $\epsilon_{U(\mathcal{X})}$ reflects both the expressivity of $U$ and the distribution $\mathcal{D}$ over $\mathcal{X}$, making it a data-dependent measure of expressivity which provides insight into specific problems.
When $\epsilon_{U(\mathcal{X})}$ is small (close to 0), this indicates that $U$ is highly expressive and covers the unitary group with a high degree of uniformity given the distribution $\mathcal{D}$.
Contrarily, when $\epsilon_{U(\mathcal{X})}$ is large, this indicates that $U$ is far from covering $\mathbb{U}(2^n)$ uniformly, and hence is not very expressive.

We will now discuss how the quantity $\epsilon_{U(\mathcal{X})}$ is related to EC.
Given a quantum kernel $\mathcal{K}$, we have that
\begin{equation}
    \label{eq:ExponentialConcentration-Expressivity}
    \underset{x,x^\prime\sim\mathscr{D}}{P}[|\mathcal{K}(x,x^\prime)-\mu|\geq\delta]\leq\frac{G_n(\epsilon_{U(\mathcal{X})})}{\delta^2}
\end{equation}
where $G_n(\epsilon_{U(\mathcal{X})})$ is defined:
\begin{enumerate}
    \item For the FQK $\mathcal{K}=\mathcal{K}_F$ as,
    \begin{equation*}
        G_n(\epsilon_{U(\mathcal{X})})=\omega_{\textrm{Haar}}+\epsilon_{U(\mathcal{X})}\left(\epsilon_{U(\mathcal{X})}+\sqrt{\omega_{\textrm{Haar}}}\right),
    \end{equation*}
    where $\omega_{\textrm{Haar}}=\tfrac{1}{2^{n-1}(2^n+1)}$.
    \item For the PQK $\mathcal{K}=\mathcal{K}_P$ as,
    \begin{equation*}
        G_n(\epsilon_{U(\mathcal{X})})=4\gamma n(\tilde{\omega}_{\textrm{Haar}}+\epsilon_{U(\mathcal{X})}),
    \end{equation*}
    where $\tilde{\omega}_{\textrm{Haar}}=\tfrac{3}{2^{n+1}+2}$.
\end{enumerate}

The above result shows that when a data-encoding unitary is more expressive (i.e. $\epsilon_{U(\mathcal{X})}$ is close to 0), then the associated kernel will experience a larger degree of concentration.
For example, in the case where $U(\mathcal{X})$ is exponentially close to covering $\mathbb{U}(2^n)$ uniformly (i.e. $\epsilon_{U(\mathcal{X})}\in\mathcal{O}(1\slash c^n)$ for some $c>1$), then the associated kernel will concentrate exponentially.

The take-away message from this result is that the expressivity of the chosen data encoding unitary should be intentionally restricted.
This is especially important since the result makes no assumptions about the specific form of $U$, which emphasises that problem-agnostic encodings—which are typically highly expressive—can be problematic.
In~\cite{Thanasilp2024Exponential} the authors provide supporting numerical evidence for this result. 
Specifically, they show that using a hardware efficient ansatz~\cite{Kandala2017HEA,Leone2024HEA} with a greater number of layers (which roughly results in a more expressive encoding) results in a greater level of EC as $n$ increases.

In the case of the FQK, we can reason further about why expressivity might be a problem. 
Specifically, with a very expressive encoding, distinct inputs will be mapped to ``far away'' points in the exponentially large Hilbert space $\mathcal{H}_n$, resulting in vectors that are likely to be (approximately) orthogonal, with inner products that are exponentially close to 0.
Hence we obtain kernel matrices close to the identity that capture no data-specific structures and only predict accurately on the training samples.

The idea that the FQKs induced by highly expressive embeddings could create problems was also considered by Huang et al.~\cite{Huang2021Power}, and was one of the main motivations which lead to the introduction of PQKS.
Specifically, with reference to Eqs.~\eqref{eq:PredictionErrorBound} and \eqref{eq:Huang-QuantumModelComplexityBound}, if the effective dimension $d$ is large then the associated problem-inspired kernel $\mathcal{K}_{\tilde{\rho}}$ will also be highly expressive and hence may not generalise well. 
However, by using a PQK instead, we can project the highly ``spread out'' embeddings of the training data into a lower dimensional classical representation and reduce the overall expressivity, which is expected to bolster generalisation capabilities.

\subsection{Entanglement}
\label{sec:Entanglement}

The role of entanglement in QML has been another topic of great interest, since entanglement, as one of the few unique properties of QML models, must necessarily be exploited in order to derive rigorous advantages over classical ML.
In this subsection, however, we will discuss how using a data-encoding unitary that prepares states with great levels of entanglement can cause EC for the PQK.
We do not consider the FQK in this section.

In the case of PQKs, if we consider varying the feature map $\rho$ so that the prepared states $\rho(x)$ are increasingly entangled, then the 1-RDMs $\rho_i(x)$ will generally become more mixed, tending to the maximally mixed state $\tfrac{\mathbb{I}}{2}$ in the limit.
Informally, this makes it somewhat clear why entanglement might create issues in this context, since the maximally mixed state is constant and clearly contains no data-specific structure.
Additionally, if both $\rho_i(x)$ and $\rho_i(x^\prime)$ tend to the maximally mixed state, then the Frobenius norms $\|\rho_i(x)-\rho_i(x^\prime)\|_F^2$ will be 0 for all $i\in\{1,2,\ldots,n\}$, resulting in a PQK equal to 1 for all inputs $x,x^\prime\in\mathcal{X}$.

The relationship between EC for the PQK and the entanglement in the states produced by the feature map can be formalised as follows~\cite{Thanasilp2024Exponential}.
Given inputs $x,x^\prime\in\mathcal{X}$, we have that
\begin{equation}
    \label{eq:ExponentialConcentration-Entanglement}
    |\mathcal{K}_P(x,x^\prime)-1|\leq2\ln(2)\Gamma(x,x^\prime)
\end{equation}
where $\Gamma(x,x^\prime)$ is defined such that
\begin{equation*}
    \Gamma(x,x^\prime)=\sum_{i=1}^{n}\left[\sqrt{S\left(\rho_i(x)\bigg\|\frac{\mathbb{I}}{2}\right)}+\sqrt{S\left(\rho_i(x^\prime)\bigg\|\frac{\mathbb{I}}{2}\right)}\right]^2,
\end{equation*}
and $S(\cdot\|\cdot)$ denotes the quantum relative entropy (see Section 11.3.1 of~\cite{Nielsen2000Quantum}).

This result shows that the PQK exponentially concentrates to $1$ if the relative entropies $S(\rho_i(x)\|\tfrac{\mathbb{I}}{2})$ and $S(\rho_i(x')\|\tfrac{\mathbb{I}}{2})$ themselves exponentially concentrate for all $i \in \{1,2,\ldots,n\}$. 
For states obeying a volume-law entanglement scaling, one has $S(\rho_i(x)\|\tfrac{\mathbb{I}}{2}),S(\rho_i(x^\prime)\|\tfrac{\mathbb{I}}{2}) \in \mathcal{O}(1/2^{n-1})$ for all $i$, implying that $\mathcal{K}_P(x,x')$ will exponentially concentrate. 
In contrast, for states satisfying an area-law scaling~\cite{Eisert2010AreaLaw}, $S(\rho_i(x)\|\tfrac{\mathbb{I}}{2}),S(\rho_i(x^\prime)\|\tfrac{\mathbb{I}}{2}) \in \mathcal{O}(1)$, and EC of $\mathcal{K}_P(x,x')$ is not implied. 
However, since the result establishes only an inequality rather than an equality, EC cannot be excluded even in this regime.

These observations indicate that encodings generating extensive (volume-law) entanglement promote EC in PQKs. 
In contrast, FQKs do not involve partial traces, and entanglement alone does not necessarily induce concentration effects. 
This further underscores the importance of carefully designing data-encoding unitaries, particularly in avoiding highly entangling, problem-agnostic constructions that may inadvertently suppress input dependent structure.

\subsection{Global measurements}
\label{sec:Global measurements}

In this section, we discuss how global measurements, which are measurement of an observable $O$ which acts non-trivially on all qubits, may cause quantum kernels to exponentially concentrate.
In this section we will only discuss FQKs, since FQKs inherently require global measurements, while PQKs do not.
Even in the case of the swap test, despite the fact that the measurement is taken from the single qubit in the ancilla register, the causal cone for the ancilla qubit covers the entire system as a result of the global controlled swap gate, and so the arguments here still hold.

To illustrate how such a measurement can lead to EC, in~\cite{Thanasilp2024Exponential} the authors discuss the following example.
Consider the data-encoding unitary $U$ defined such that $U(x)=\otimes_{j=1}^{n}e^{-ix_jY}$, where $x_j$ denotes the $j^{\textrm{th}}$ component of the input $x\in\mathcal{X}\subseteq\R^n$.
In other words, the data-encoding unitary is a tensor product of single qubit rotations about the $y$-axis of the Bloch sphere, with the angle of rotation for each qubit being given by the corresponding component of the input $x$.
Assuming that each of the components of $x$ are independent and sampled from the uniform distribution over $[-\pi,\pi]$, and that the initial state is $\rho_0=(|0\rangle\langle0|)^{\otimes n}$, we have that
\begin{equation}
    \label{eq:ExponentialConcentration-GlobalMeasurement}
    \underset{x,x^\prime\sim\textrm{Unif}[-\pi,\pi]}{P}\left[\left|\mathcal{K}_F(x,x^\prime)-\frac{1}{2^n}\right|\geq\delta\right]\leq\left(\frac{3}{8}\right)^n\frac{1}{\delta^2}.
\end{equation}
Clearly $(3\slash8)^n\in\mathcal{O}(1\slash c^n)$ with $c=8\slash3>1$, so the FQK exponentially concentrates in this case.

In the Supplementary Information of~\cite{Thanasilp2024Exponential}, Thanasilp et al. generalise the result to consider data-encoding unitaries given by tensor products of general single qubit unitaries.
Since such data-encoding unitaries involve no entanglement, are only weakly expressive, and may capture a wide variety of underlying data distributions (by absorbing non-uniform product distributions into the definition of the local unitaries), we see that global measurements can induce EC.

Thanasilp et al.~\cite{Thanasilp2024Exponential} provide numerical evidence supporting the above result for the tensor product encodings in the main text.
They also show numerical evidence supporting the idea that more rapid concentration should be expected when using more general problem-agnostic encodings (such as hardware efficient ansatzes).
This is because the non-product encodings increase both expressivity and entanglement, which we expect to increase the rate of concentration in the kernel values.

Worth noting though, is that global measurements won't always cause EC.
If the feature map embeds inputs into the Hilbert space in such a way that distinct points have fidelity scaling at least $\Omega(1\slash\textrm{poly}(n))$, then the FQK can be efficiently determined.
Thus the main message here is again to try and construct encodings which encode information about the problem structure, or to artificially reduce expressivity in order to maintain the average fidelity between embedded inputs.

\subsection{Hardware noise}
\label{sec:Hardware noise}

Hardware noise plays a central role in determining the practical performance of QML models, as it alters the implemented feature maps and measurement statistics relative to their ideal noiseless descriptions. 
Unlike limitations arising purely from model design or theoretical considerations, hardware noise reflects underlying engineering constraints and is therefore one of the most challenging barriers to mitigate, directly impacting expressivity, executable circuit depths, and other prospects for practical quantum advantage.
Here we discuss how hardware noise affects the efficacy and practicality of QKMs, both in terms of generalisation and EC.

The presence of noise generally causes quantum states to become more mixed, tending in the limit to the maximally mixed state.
With this in mind, Heyraud et al.~\cite{Heyraud2022NoisyQKMs} investigate the impact of noise on the expressivity and generalisation performance of quantum kernel machines. 
By modelling decoherence and dissipation using Lindblad master equations (see Chapter 8.4.1 of~\cite{Nielsen2000Quantum}), they derive generalisation error bounds that depend explicitly on the average purity of the encoded quantum states.
A detailed discussion of their methods is beyond the scope of this review, but their results ultimately demonstrate that greater levels of noise degrade learning performance. 

Similarly, Wang et al.~\cite{Wang2021Towards} consider the power of QKMs under realistic NISQ-era constraints, focusing on the combined effects of noise, finite sampling, and dataset size.
Specifically, they choose to model noise using a global depolarising channel $\mathcal{N}$ defined such that 
\begin{align*}
    \mathcal{N}(\rho)=(1-p)\rho+p\frac{\mathbb{I}}{2^n},
\end{align*} 
for all $n$-qubit density operators $\rho$, and a noise parameter $p\in[0,1]$ known as the \emph{depolarising rate}, where $\mathbb{I}$ denotes the $2^n\times2^n$ identity matrix.
In this case, they prove a theorem (Theorem 1 of the main text) which shows that the generalisation error can actually increase with the number of training data samples used to train the model, which contrasts with the general results of Huang et al.~\cite{Huang2021Power} which suggest providing more training data should decrease the generalisation error.
Similarly, the theorem suggests that the minimum number of measurements $m$ should scale at least as $M^3$, and that shallow circuit depths are required to achieve quantum advantage, since if the level of noise gets too large (i.e. $p\mapsto1^-$) then the model performance will plummet.

We will now discuss how hardware noise can, in addition to threatening generalisation performance directly, also induce EC in the value of a quantum kernel.
To analyse the effects of noise in this context, Thanasilp et al.~\cite{Thanasilp2024Exponential} consider decomposing the data-encoding unitary into a product of $L\in\N$ layers such that $U(x)=\Pi_{l=1}^{L}U_l(x_l)$, where $x_l$ denotes a vector that depends on the input $x$.
For example, $x_l$ might be the $l^{\textrm{th}}$ component of $x$, or another vector which depends on $x$.
While more general decompositions exist, this form covers many standard encoding schemes.

To model the noise, the authors consider a local Pauli noise channel which is applied before and after every layer $U_l(x_l)$.
Accordingly, the final quantum state $\tilde{\rho}(x)$ is given by
\begin{equation*}
    \tilde{\rho}(x)=\mathcal{N}\circ U_l(x_l)\circ\mathcal{N}\circ\ldots\circ\mathcal{N}\circ U_1(x_1)\circ\mathcal{N}(\rho_0),
\end{equation*}
where $\rho_0$ is the initial state of the system, usually $(|0\rangle\langle0|)^{\otimes n}$, and $\mathcal{N}=\mathcal{N}_1\otimes\ldots\otimes\mathcal{N}_n$ denotes the local Pauli noise channel. 
Note that we use the notation $\tilde{\rho}(x)$ simply to indicate that $\tilde{\rho}$ could be mixed, it should not be confused with the use of this notation in Section~\ref{sec:Assessing quantum advantage in quantum kernel methods}.
Here each $\mathcal{N}_j$ denotes a unital channel such that for the Pauli operator $\sigma\in\{X,Y,Z\}$, $\mathcal{N}_j$ acts as $\mathcal{N}_j(\sigma)=\tilde{q}_\sigma\sigma$ for some $\tilde{q}_\sigma\in(-1,1)$.
To characterise the overall strength of the noise, we then consider $\tilde{q}=\max\{|\tilde{q}_X|,|\tilde{q}_Y|,|\tilde{q}_Z|\}$.

In this case, the concentration of a quantum kernel $\mathcal{K}$ to a fixed value $\mu$ after $L$ layers of the channel with noise parameter $\tilde{q}$ can be bounded such that 
\begin{equation}
    \label{eq:ExponentialConcentration-Noise}
    |\mathcal{K}(x,x^\prime)-\mu|\leq F(\tilde{q},L)
\end{equation}
where:
\begin{enumerate}
    \item For the FQK $\mathcal{K}=\mathcal{K}_F$, $F(\tilde{q},L)$ is given by
    \begin{equation*}
        F(\tilde{q},L)=\tilde{q}^{2L+1}\left\|\rho_0-\frac{\mathbb{I}}{2^n}\right\|,
    \end{equation*}
    with $\mu=1\slash2^n$,
    \item For the PQK $\mathcal{K}=\mathcal{K}_P$, $F(\tilde{q},L)$ is given by
    \begin{equation*}
        F(\tilde{q},L)=8\ln(2)\gamma n\tilde{q}^{\frac{L+1}{2\ln2}}S_2\left(\rho_0\bigg\|\frac{\mathbb{I}}{2^n}\right),
    \end{equation*}
    where $S_2(\cdot\|\cdot)$ denotes the sandwhiched 2-R\'enyi entropy (see Eq. 4 of~\cite{Wilde2014SandwhichedRenyi}).
\end{enumerate}

This result shows that the concentration experienced by quantum kernels, both FQKs and PQKs, is exponential in the number of layers $L$.
This is especially worrying since in order for QKMs to provide quantum advantages, the associated kernels must hard to classically simulate, and hence generally require a number of layers $L$ scaling with the size of the system $n$.
However, the above result shows that in this case, unless $L\in\mathcal{O}(\log(n))$, the associated quantum kernels will experience EC and thus a polynomial number of shots will not be sufficient to estimate the kernels, rendering this approach prohibitively expensive.
This is consistent with the idea that the kernels introduced by Suzuki et al.~\cite{Suzuki2024QuantumFisher} can avoid EC at log-depth, but will experience it when extended to linear-depth, as discussed in the Supplementary Information of~\cite{Thanasilp2024Exponential}.

The authors of~\cite{Thanasilp2024Exponential} once again provide numerical evidence supporting the above result, where increasing the level of noise (i.e. using a smaller $q$) and the number of layers $L$ both result in more rapid exponential decay in the average values of the kernels.

\subsection{Dequantisation in non-variational quantum kernel methods}
\label{sec:Dequantisation in non-variational quantum kernel methods}

Dequantisation refers to the development of efficient classical algorithms that replicate or approximate the performance of quantum models, particularly those processing classical data, thereby questioning the necessity of quantum resources for claimed advantages. 
In the context of non-variational QKMs, classical simulability provides a critical lens for assessing potential quantum utility.

In this subsection, we focus primarily on tensor-network (TN) based dequantisation results, which offer a structurally transparent and physically motivated framework for analysing the classical simulability of quantum feature maps~\cite{shin2024dequantizing,berezutskii2025tensor}. 
Tensor networks make explicit how entanglement geometry, locality, and circuit structure constrain expressive power and computational complexity, thereby providing rigorous conditions under which kernel evaluations admit efficient classical approaches.
Other dequantisation approaches, such as random Fourier feature approximations and kernel-based low-rank reconstructions, will be briefly discussed at the end of the subsection.

\subsubsection{Tensor network-based dequantisation}
\label{sec:Tensor network-based dequantisation}

TN methods~\cite{Orus2014Practical,Bridgeman2017Hand,Cirac2021MPSandPEPS} provide a powerful classical framework for representing and manipulating high-dimensional quantum states by exploiting their underlying entanglement structure.
From this perspective, it becomes clear that the classical simulability of QKMs is governed less by the sheer magnitude of entanglement, and more by the underlying geometric structure of the entanglement, and tensor networks provide a natural language for making this distinction precise. 
Tensor network representations hence provide a principled means of assessing whether such embeddings genuinely evade efficient classical simulation or instead admit polynomial-time classical contraction due to structural constraints~\cite{berezutskii2025tensor}.
In the latter case, the opportunities for advantages are reduced to mere polynomial gaps at best, which are significantly less desirable than the highly sought exponential advantages that quantum computing promises to provide.

\begin{figure}
    \centering
    \begin{tikzpicture}[scale=0.9]
        \node at (-2.5,0.65) {Explicit array};
        \node at (-2.5,0) {\large$\psi_{i_1i_2i_3i_4}$};
        \node at (1.5,0.65) {Tensor network};
        \draw[fill=blue!25,rounded corners] (-0.25,-0.25) rectangle (3.25,0.25);
        \node at (1.5,0) {$\psi$};
        \draw[thick] (0,-0.25) --++ (0,-0.375);
        \node at (0,-0.85) {$i_1$};
        \draw[thick] (1,-0.25) --++ (0,-0.375);
        \node at (1,-0.85) {$i_2$};
        \draw[thick] (2,-0.25) --++ (0,-0.375);
        \node at (2,-0.85) {$i_3$};
        \draw[thick] (3,-0.25) --++ (0,-0.375);
        \node at (3,-0.85) {$i_4$};
    \end{tikzpicture}
    \caption{An illustration of a 4-qubit quantum state represented in terms of an explicit array (left) and a tensor network (right).}
    \label{fig:TNNotation}
\end{figure}
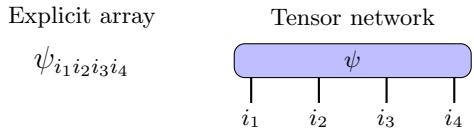

The usual starting point for any discussion of TN methods in the context of quantum computing is to note that a general $n$-qubit pure quantum state $\ket{\psi}\in\C^{2^n}$ can be written as
\begin{equation*}
    \ket{\psi}=\sum_{i_1,\ldots,i_n=0}^{1}\psi_{i_1\ldots i_n}\ket{i_1}\otimes\ldots\otimes\ket{i_n}.
\end{equation*}
In TN notation, the multi-dimensional array $\psi_{i_1\ldots i_n}$ is represented by a geometric shape (usually a rectangle or a circle) with $n$ legs, each representing one of the $n$ indices $i_1,\ldots,i_n$ (see Fig.~\ref{fig:TNNotation}).
The main idea underlying TN methods is to then decompose the TN representing $\psi_{i_1\ldots i_n}$, which completely describes the state of the physical $n$-qubit system, into a network of low-rank tensors connected by contracted indices (denoted by connecting the legs representing the indices). 
The resulting computational cost associated with representing the state in this manner is then determined by both the geometry and amount of entanglement present. 

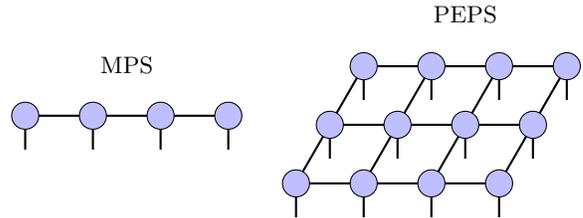
\begin{figure}
    \centering
    \begin{tikzpicture}[scale=0.9]
        \node at (2.5-5,1.75) {MPS};
        \draw[thick] (0-4,0+1) --++ (3,0);
        \draw[thick] (0-4,0+1) --++ (0,-0.5);
        \draw[thick] (1-4,0+1) --++ (0,-0.5);
        \draw[thick] (2-4,0+1) --++ (0,-0.5);
        \draw[thick] (3-4,0+1) --++ (0,-0.5);
        \draw[fill=blue!25] (0-4,0+1) circle (0.2cm);
        \draw[fill=blue!25] (1-4,0+1) circle (0.2cm);
        \draw[fill=blue!25] (2-4,0+1) circle (0.2cm);
        \draw[fill=blue!25] (3-4,0+1) circle (0.2cm);
        \node at (2.5,2.5) {PEPS};
        \draw[thick] (0,0) --++ (3,0);
        \draw[thick] ({cos(60)},{sin(60)}) --++ (3,0);
        \draw[thick] ({2*cos(60)},{2*sin(60)}) --++ (3,0);
        \draw[thick] (0,0) --++ ({2*cos(60)},{2*sin(60)});
        \draw[thick] (1,0) --++ ({2*cos(60)},{2*sin(60)});
        \draw[thick] (2,0) --++ ({2*cos(60)},{2*sin(60)});
        \draw[thick] (3,0) --++ ({2*cos(60)},{2*sin(60)});
        \draw[thick] (0,0) --++ (0,-0.5);
        \draw[thick] (1,0) --++ (0,-0.5);
        \draw[thick] (2,0) --++ (0,-0.5);
        \draw[thick] (3,0) --++ (0,-0.5);
        \draw[thick] ({cos(60)},{sin(60)}) --++ (0,-0.5);
        \draw[thick] ({cos(60)+1},{sin(60)}) --++ (0,-0.5);
        \draw[thick] ({cos(60)+2},{sin(60)}) --++ (0,-0.5);
        \draw[thick] ({cos(60)+3},{sin(60)}) --++ (0,-0.5);
        \draw[thick] ({2*cos(60)},{2*sin(60)}) --++ (0,-0.5);
        \draw[thick] ({2*cos(60)+1},{2*sin(60)}) --++ (0,-0.5);
        \draw[thick] ({2*cos(60)+2},{2*sin(60)}) --++ (0,-0.5);
        \draw[thick] ({2*cos(60)+3},{2*sin(60)}) --++ (0,-0.5);
        \draw[fill=blue!25] (0,0) circle (0.2cm);
        \draw[fill=blue!25] (1,0) circle (0.2cm);
        \draw[fill=blue!25] (2,0) circle (0.2cm);
        \draw[fill=blue!25] (3,0) circle (0.2cm);
        \draw[fill=blue!25] ({0+cos(60)},{0+sin(60)}) circle (0.2cm);
        \draw[fill=blue!25] ({1+cos(60)},{0+sin(60)}) circle (0.2cm);
        \draw[fill=blue!25] ({2+cos(60)},{0+sin(60)}) circle (0.2cm);
        \draw[fill=blue!25] ({3+cos(60)},{0+sin(60)}) circle (0.2cm);
        \draw[fill=blue!25] ({0+2*cos(60)},{0+2*sin(60)}) circle (0.2cm);
        \draw[fill=blue!25] ({1+2*cos(60)},{0+2*sin(60)}) circle (0.2cm);
        \draw[fill=blue!25] ({2+2*cos(60)},{0+2*sin(60)}) circle (0.2cm);
        \draw[fill=blue!25] ({3+2*cos(60)},{0+2*sin(60)}) circle (0.2cm);
    \end{tikzpicture}
    \caption{An illustration of a 4-qubit MPS with open boundary conditions (left) and a 12-qubit PEPS on a $3\times4$ rectangular lattice with open boundary conditions (right).}
    \label{fig:MPSandPEPS}
\end{figure}

The configuration and arrangement of the low-rank tensors dictates what kind of entanglement structures can be naturally captured by the TN.
In the case of a one-dimensional (1D) entanglement structure, where qubits lie on a path graph and interact only with adjacent neighbours, the most natural TN to use is called a matrix product state (MPS)~\cite{PerezGarcia2007MPS}.
In the case of a 2D lattice entanglement structure, where qubits lie on a square or rectangular lattice and interact only with their adjacent neighbours, the most natural TN to use is called a projected entangled-pair state (PEPS)~\cite{Verstraete2004PEPS}.
An illustration of an MPS and a PEPS can be found in Fig.~\ref{fig:MPSandPEPS}.
For a more pedagogical introduction to MPS and PEPS, we refer the reader first to~\cite{Orus2014Practical} and then to~\cite{Bridgeman2017Hand}.

In addition to MPS and PEPS, other entanglement structures can be naturally represented with the multi-entanglement renormalisation anstaz~\cite{Vidal2008MERA} and tree tensor networks~\cite{Shi2006TTN}.
However, here we will focus predominantly on MPS and PEPS, as has been the focus of much of the QML dequantisation literature.
Further, we will only discuss MPS and PEPS with open boundary conditions.
When periodic boundary conditions are introduced, some of the qualitative conclusions change, usually resulting in greater runtime complexities and theoretical difficulties.
For more details about the cases with periodic boundary conditions, we refer the reader to~\cite{Orus2014Practical}.

While the choice of TN ansatz dictates the kind of entanglement structure that can be naturally represented, a critical parameter for all TN methods is the bond dimension, which we denote by $\chi$. 
The bond dimension governs both the computational cost of simulating quantum systems and the amount of entanglement that can be represented with a given TN, and directly affects the accuracy of the computations. 
A larger value of $\chi$ allows more entanglement to be captured and improves computational accuracy, but generally incurs higher runtime costs as we will discuss shortly. 

In principle, exact simulations of generic quantum systems require $\chi$ to scale exponentially with the number of qubits $n$. 
TN methods, however, usually involve deliberately restricting $\chi$ to maintain computational efficiency at the expense of some approximation. 
A key empirical insight from TN studies is that for many physically relevant systems, $\chi$ need not scale exponentially with $n$ to achieve highly accurate or even exact results.
And these are precisely the scenarios where TN methods are most effective.

\paragraph{Simulating QKMs with TNs.}

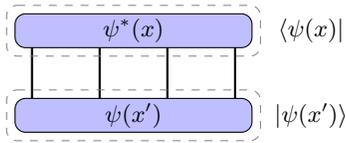
\begin{figure}
    \centering
    \begin{tikzpicture}[scale=0.9]
        \draw[fill=blue!25,rounded corners] (-0.25,-0.25) rectangle (3.25,0.25);
        \node at (1.5,0) {$\psi^*(x)$};
        \draw[thick] (0,-0.25) --++ (0,-0.75);
        \draw[thick] (1,-0.25) --++ (0,-0.75);
        \draw[thick] (2,-0.25) --++ (0,-0.75);
        \draw[thick] (3,-0.25) --++ (0,-0.75);
        \draw[black!50,dashed,line width=0.1mm,rounded corners] (-0.375,0.375) --++ (3.75,0) --++ (0,-0.75) --++ (-3.75,0) -- cycle;
        \node at (4.125,0) {$\bra{\psi(x)}$};
        \draw[fill=blue!25,rounded corners] (-0.25,-0.25-1.25) rectangle (3.25,0.25-1.25);
        \node at (1.5,-1.25) {$\psi(x^\prime)$};
        \draw[black!50,dashed,line width=0.1mm,rounded corners] (-0.375,-0.875) --++ (3.75,0) --++ (0,-0.75) --++ (-3.75,0) -- cycle;
        \node at (4.125,-1.25) {$\ket{\psi(x^\prime)}$};
        \node at (-1.25,0) {\textcolor{white}{X}};
    \end{tikzpicture}
    \caption{An illustration of the inner-product  $\langle\psi(x)|\psi(x^\prime)\rangle$ between 4-qubit quantum states $\ket{\psi(x)}$ and $\ket{\psi(x^\prime)}$ in TN notation used to calculate the FQK.
    Note that $\psi^*(x)$ represents the complex conjugate of $\psi(x)$.}
    \label{fig:TNsForFQK}
\end{figure}

In the context of QKMs, in order to simulate the FQK $\mathcal{K}_F(x,x^\prime)=|\langle\psi(x)|\psi(x^\prime)\rangle|^2$, we need to both calculate the quantum states $\ket{\psi(x)}$ and $\ket{\psi(x^\prime)}$ and then take the inner product $\langle\psi(x)|\psi(x^\prime)\rangle$, the result of which can be used to determine the FQK by taking the modulus squared.
See Fig.~\ref{fig:TNsForFQK} for a graphical representation of the inner product $\langle\psi(x)|\psi(x^\prime)\rangle$ in TN notation.

\begin{figure}
    \centering
    \begin{tikzpicture}[scale=0.9]
        \draw[fill=blue!25,rounded corners] (-0.25,0.75) rectangle (3.25,1.25);
        \node at (1.5,1) {$\psi(x)$};
        \draw[fill=blue!25,rounded corners] (-0.25,-0.25) rectangle (3.25,0.25);
        \node at (1.5,0) {$\psi^*(x)$};
        \draw[thick] (0,-0.25) --++ (0,-1);
        \draw[thick] (0,1.25) --++ (0,1);
        \draw[thick] (1,-0.25) --++ (0,-0.75) --++ (3,0) --++ (0,3) --++ (-3,0) --++ (0,-0.75);
        \draw[thick] (2,-0.25) --++ (0,-0.5) --++ (1.75,0) --++ (0,2.5) --++ (-1.75,0) --++ (0,-0.5);
        \draw[thick] (3,-0.25) --++ (0,-0.25) --++ (0.5,0) --++ (0,2) --++ (-0.5,0) --++ (0,-0.25);
        \draw[black!50,dashed,line width=0.1mm,rounded corners] (-0.375,2.125) --++ (4.5,0) --++ (0,-3.25) --++ (-4.5,0) -- cycle;
        \node at (4.75,0.5) {$\rho_1(x)$};
        \draw[fill=blue!25,rounded corners] (-0.25,0.75-3.5) rectangle (3.25,1.25-3.5);
        \node at (1.5,1-3.5) {$\psi(x^\prime)$};
        \draw[fill=blue!25,rounded corners] (-0.25,-0.25-3.5) rectangle (3.25,0.25-3.5);
        \node at (1.5,0-3.5) {$\psi^*(x^\prime)$};
        \draw[thick] (0,-0.25-3.5) --++ (0,-1);
        \draw[thick] (0,1.25-3.5) --++ (0,1);
        \draw[thick] (1,-0.25-3.5) --++ (0,-0.75) --++ (3,0) --++ (0,3) --++ (-3,0) --++ (0,-0.75);
        \draw[thick] (2,-0.25-3.5) --++ (0,-0.5) --++ (1.75,0) --++ (0,2.5) --++ (-1.75,0) --++ (0,-0.5);
        \draw[thick] (3,-0.25-3.5) --++ (0,-0.25) --++ (0.5,0) --++ (0,2) --++ (-0.5,0) --++ (0,-0.25);
        \draw[black!50,dashed,line width=0.1mm,rounded corners] (-0.375,-1.375) --++ (4.5,0) --++ (0,-3.25) --++ (-4.5,0) -- cycle;
        \node at (4.8,-3) {$\rho_1(x^\prime)$};
        \draw[thick] (0,2.25) --++ (-0.5,0) --++ (0,-7) --++ (0.5,0);
        \node at (-1.5,0) {\textcolor{white}{X}};
    \end{tikzpicture}
    \caption{An illustration of the overlap $\textrm{tr}(\rho_i(x)\rho_i(x^\prime))$ for $i=1$ in TN notation used to calculate the PQK with 4-qubit quantum systems.
    Note that with $n$-qubit quantum systems, the same computation needs to be performed for all $i\in\{1,\ldots,n\}$ to compute the PQK.}
    \label{fig:TNsForPQK}
\end{figure}
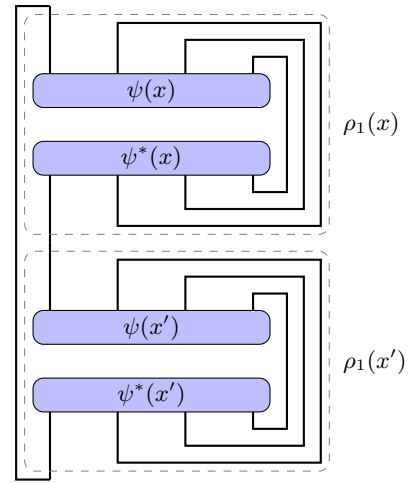

Similarly, for the PQK $\mathcal{K}_P(x,x^\prime)=\exp\left(-\gamma\sum_{i=1}^{n}\|\rho_i(x)-\rho_i(x^\prime)\|^2_F\right)$, we need to first calculate the 1-RDMs $\rho_i(x)=\textrm{tr}_{j\neq i}\left(|\psi(x)\rangle\langle\psi(x)|\right)$ and $\rho_i(x^\prime)=\textrm{tr}_{j\neq i}\left(|\psi(x^\prime)\rangle\langle\psi(x^\prime)|\right)$, and then evaluate the purities $\textrm{tr}(\rho_i(x)^2)$ and $\textrm{tr}(\rho_i(x^\prime)^2)$, and the overlap $\textrm{tr}(\rho_i(x)\rho_i(x^\prime))$ for each $i\in\{1,\ldots,n\}$.
Using these quantities we can then calculate the PQK as discussed in Section~\ref{sec:Estimating quantum kernels in practice}.
An illustration of this procedure for calculating the overlap $\textrm{tr}(\rho_i(x)\rho_i(x^\prime))$ with $i=1$ can be found in Fig.~\ref{fig:TNsForPQK}.
The calculation of the purities and overlaps for other values of $i\in\{1,\ldots,n\}$ are similar.

\paragraph{Matrix product states.}

Let us now consider using a MPS TN ansatz (see Fig.~\ref{fig:MPSandPEPS}).
A variety of numerical algorithms have been developed for MPS, most relevantly the time-evolving block decimation (TEBD) algorithm~\cite{Vidal2003Efficient,Vidal2004Efficient} for evolving quantum states under local 1D Hamiltonians in polynomial time.
Specifically, given a 2-local 1D Hamiltonian $H$ (i.e. a Hamiltonian for which every term acts non-trivially on at most two qubits which are adjacent in a path graph), suppose we want to apply the time-evolution operator $e^{-iHt}$ to an MPS.
Using a Trotter step size of $\Delta t$, the time complexity of TEBD in this case is
\begin{equation}
    \label{eq:TEBDComplexity}
    \mathcal{O}\left(n\chi^3t\slash\Delta t\right)
\end{equation}
(see Lemma 2 of~\cite{Vidal2003Efficient} and further discussion in~\cite{Vidal2004Efficient}).

Eq.~\eqref{eq:TEBDComplexity} shows us that the quantum states $\ket{\psi(x)}=e^{-i\hat{H}t}|\tilde{\psi}\rangle$ given by evolving $|\tilde{\psi}\rangle$ under 1D 2-local Hamiltonians can be simulated using MPS in polynomial time, contingent on the initial state $|\tilde{\psi}\rangle$ being efficiently representable by an MPS.
This idea naturally extends to provide a polynomial time classical algorithm for simulating quantum circuits containing 2-qubit gates that act only on adjacent qubits of a path graph.

The above result shows that a variety of quantum circuit and Hamiltonian evolutions can be simulated with an MPS for 1D systems.
So in order to simulate FQKs and PQKs generated by such evolutions and circuits, we simply need to be able to evaluate the relevant quantities for calculating the associated kernels as in Figs.~\ref{fig:TNsForFQK} and \ref{fig:TNsForPQK}.
As discussed in many standard references, such as~\cite{Orus2014Practical}, the time complexity for taking the inner product between two MPS (which allows us to calculate the FQK), each with bond dimension $\chi$, scales as
\begin{equation}
    \label{eq:MPSInnerProductComplexity}
    \mathcal{O}\left(n\chi^3\right).
\end{equation}
Evaluating the relevant purities and overlaps for calculating the PQK shares the same runtime scaling.

Together, Eqs.~\eqref{eq:TEBDComplexity} and \eqref{eq:MPSInnerProductComplexity} imply that FQKs and PQKs generated by polynomial-time local 1D Hamiltonian evolutions, or equivalently by polynomial-depth local 1D circuits, are classically simulable in polynomial time provided the MPS bond dimension $\chi$ grows at most polynomially in $n$. 
This is consistent with the results of Shin et al.~\cite{shin2024dequantizing}, who show that variational QML models, including QKMs as a special case, admit classical decompositions in terms of MPS.

A natural question to ask now is whether $\chi$ can always be kept polynomial in $n$.
In the worst case, the answer is no.
To see this, consider partitioning the system represented by the MPS into two connected subsystems, then the greatest von Neumann entropy $S$ (see Chapter 11.3 of~\cite{Nielsen2000Quantum}) which can be captured with a given $\chi$ is bounded such that
\begin{equation*}
    S\leq\log_2(\chi).
\end{equation*}
In other words, quantum states which are efficiently representable by a MPS satisfy an area law~\cite{Eisert2010AreaLaw}.
In general, an $n$-qubit quantum state can possess a von Neumann entropy across such a partition of at most $S=\lfloor n\slash2\rfloor$.
This means that there exist $n$-qubit quantum states which would require $\chi=2^{\lfloor n\slash2\rfloor}$ to represent exactly, and hence need exponential time to simulate with MPS.

In light of this, one potential route to avoiding classical simulability is to consider quantum feature maps that generate volume-law entangled states, for which $\chi$ must grow exponentially in $n$. 
In this regime, tensor-network simulation becomes exponentially costly. 
However, as discussed in Section~\ref{sec:Entanglement}, such states can present problems for QKMs, and more importantly, the physical accessibility of such states in practice is doubtful. 

Poulin et al.~\cite{Poulin2011Illusion} show that states reachable in polynomial time by local time-dependent Hamiltonians form an exponentially small subset of the total Hilbert space, suggesting that generic volume-law states may require prohibitively long preparation times on quantum devices.
Moreover, this physically accessible subset is expected to predominantly contain area-law states~\cite{Orus2014Practical}, for which tensor-network methods provide efficient classical algorithms.
Accordingly, an important open question is whether one can construct quantum feature maps that produce physically preparable (area-law) quantum states, yet evade efficient classical simulation.
This brings us to a discussion of PEPS.

\paragraph{Projected entangled-pair states.}

PEPS generalise MPS to two-dimensional lattices (see Fig.~\ref{fig:MPSandPEPS}) and are naturally suited to representing quantum states on a square or rectangular lattice with local connectivity.
As in the MPS case, the expressive power of the ansatz is controlled by the bond dimension $\chi$.

In contrast to MPS, the exact contractions of PEPS are \#P-hard~\cite{Schuch2007Computational,Haferkamp2020PEPSContraction}.
In particular, evaluating norms or computing expectation values of local observables with PEPS can encode classical counting problems that are \#P-hard.
This implies that, unlike the 1D case, there is no known polynomial-time classical algorithm for exactly contracting generic PEPS, even when $\chi$ is held fixed.

Nevertheless, a variety of approximate contraction algorithms have been developed~\cite{Verstraete2008MPSandPEPS,Cirac2021MPSandPEPS}.
Here we briefly discuss one such method for finite PEPS, which requires reformulating the 2D contraction as a series of 1D contractions involving both the usual bond dimension $\chi$, and another parameter known as the boundary bond dimension $\tilde{\chi}$.
For a given $\tilde{\chi}$, as discussed in~\cite{Orus2014Practical}, this method exhibits a runtime scaling of 
\begin{equation*}
    \mathcal{O}(n \tilde{\chi}^2 \chi^6),
\end{equation*} 
though the precise exponents can change depending on the chosen contraction algorithm.
While this appears to imply that PEPS contractions can be performed in polynomial time, the setback comes from the fact that after each 1D contraction, $\tilde{\chi}$ must be truncated to maintain a fixed size, otherwise it would grow exponentially in $\mathcal{O}(\sqrt{n})$.

In the context of QKMs, this distinction is important.
To simulate FQKs or PQKs generated by local 2D circuits or local 2D Hamiltonian evolutions using PEPS, one must evaluate overlaps, purities, or expectation values.
Since exact PEPS contraction is \#P-hard and approximate contractions may require large boundary bond dimensions to maintain accuracy, classical simulation of such kernels is either generically intractable for arbitrary 2D states, or involves heavy approximations.

A natural question, analogous to the MPS case, concerns how large $\chi$ must be to represent arbitrary quantum states.
Consider a PEPS defined on an $L\times L$ lattice.
If the system is bipartitioned into two connected regions separated by a boundary of length $\mathcal{O}(L)$, then the von Neumann entropy $S$ across such a partition satisfies
\begin{equation*}
S\leq\mathcal{O}(L\log_2(\chi)),
\end{equation*}
so that quantum states efficiently representable by PEPS obey an area law.
Therefore, since $S$ can be as large as $\lfloor L^2\slash2\rfloor$, there exist states for which $\chi$ must scale exponentially in $\mathcal{O}(L)=\mathcal{O}(\sqrt{n})$ in order to be faithfully represented as a PEPS.

Even when restricting to physically accessible states, the use of PEPS for simulating quantum kernels faces significant challenges.
Although ground states of gapped local 2D Hamiltonians typically satisfy area laws and can be well approximated by PEPS~\cite{Verstraete2008MPSandPEPS,Cirac2021MPSandPEPS,Eisert2010AreaLaw}, contracting these states requires either substantial computational resources or heavy approximations to control boundary bond dimensions.
These necessary approximations and the intrinsic difficulties of 2D tensor-network methods suggest that, unlike in 1D, efficiently simulating even area-law states may be nontrivial on classical hardware.
Consequently, quantum feature maps that prepare physically realisable, area-law states in 2D could still potentially evade efficient classical simulation, offering a route to pursue quantum advantage that leverages the interplay between physically capabilities and the limitations of classical PEPS contraction algorithms.

A final important point to make is the following.
Even though PEPS may not be well equipped to simulate 2D area-law states and hence simulate the associated QKMs, we cannot rule out the existence of other classical algorithms which can.
For example, Jahromi et al.~\cite{Jahromi2019Universal} develop a variant of PEPS which they term graph-based projected entangled-pair states (gPEPS) to estimate ground states of local Hamiltonians on infinite arbitrary lattices.
To further emphasise the possibility of new classical algorithms emerging in the future, Patra et al.~\cite{Patra2024Efficient} utilise a finite-lattice version of gPEPS to simulate IBM's 1121-qubit processor Condor.
In particular, they demonstrate that gPEPS are capable of calculating expectation values with significantly higher accuracy than could be achieved on the physical quantum hardware.

Accordingly, while current PEPS-based methods face significant limitations, and approximate contractions can be computationally demanding, these examples illustrate that advances in classical tensor-network techniques may continue to push the boundary of what is efficiently simulable. 
Consequently, determining whether quantum feature maps based on physically realistic 2D dynamics can evade efficient classical approximation remains an important open problem.
This highlights the ongoing interplay between the expressive power of quantum circuits and the capabilities of emerging classical algorithms.

\subsubsection{Other dequantisation approaches.}

Beyond tensor-network methods, a number of classical techniques have been developed that may be used to dequantise QKMs. 
One prominent family of approaches is based on random Fourier features (RFF) and related sampling techniques~\cite{rahimi2007random}. 
These methods exploit the spectral structure of shift-invariant or approximately band-limited kernels to construct explicit classical feature maps whose inner products approximate the quantum kernel. 
Recent works have shown that many quantum kernels admit efficient RFF-based approximations when their Fourier spectra decay sufficiently fast, leading to small generalisation gaps between the quantum and classical models~\cite{sweke2025potential,sahebi2025dequantization}. 
At the same time, these results also clarify that highly expressive or deep circuits with broad spectral support may resist such approximations.
Though such regimes obviously run the risk of other problems, such as EC.

In addition, several kernel-based dequantisation strategies avoid explicit Fourier representations altogether. 
These include low-rank approximations of the quantum Gram matrix $K$ (e.g., Nystr\"om methods and leverage-score sampling)~\cite{d2025assessing,coelho2025quantum}. 
Such techniques exploit structural properties such as low effective rank, limited entanglement, subsystem separability, or noise-induced concentration, and can significantly reduce classical and quantum resource requirements in practical regimes.

A comprehensive treatment of these dequantisation methods lies beyond the scope of this review. 
Here we simply note that they substantially temper broad claims of exponential quantum advantage for many existing QKMs, while simultaneously clarifying the structural conditions under which classical approximations break down. 
We return to these broader implications in the concluding section.

\subsection{The spectrum of the kernel integral operator}
\label{sec:The spectrum of the kernel integral operator}

In this section, we will discuss challenges for QKMs relating to the associated kernel integral operator.
Formally, we consider a kernel $\mathcal{K}:\mathcal{X}\times\mathcal{X}\to\R$, and a distribution $\mathscr{D}$ over $\mathcal{X}$. 
Let $L_2(\mathcal{X})$ be the space of square-integrable functions $\mathcal{X}\to\R$ with norm $\|\cdot\|_{L_2(\mathcal{X})}=\int_{\mathcal{X}}(\cdot)(x)d\mathscr{D}(x)$ defined with respect to the distribution $\mathscr{D}$.
The integral operator $T_{\mathcal{K}}:L_2(\mathcal{X})\to L_2(\mathcal{X})$ of $\mathcal{K}$ is then defined such that
\begin{equation}
    \label{eq:KernelIntegralOperator}
    (T_{\mathcal{K}}f)(x)=\int_{\mathcal{X}}f(x^\prime)\mathcal{K}(x,x^\prime)d\mathscr{D}(x^\prime).
\end{equation}
The kernel integral operator plays a vital role in Mercer's theorem, which shows that a kernel $\mathcal{K}$ can be expressed in terms of the eigenvalues and eigenfunctions of the associated operator $T_{\mathcal{K}}$, providing important insights from a learning theoretical perspective.

Most relevantly, K\"ubler et al.~\cite{Kubler2021Inductive} relate the spectral properties of the kernel integral operator for an arbitrary kernel to the generalisation error of the trained model $f\in\mathcal{R}_{\mathcal{K}}$ obtained via KRR.
Specifically, they consider a training dataset $\mathcal{D}=\{(\mathbf{x}_i,y_i)\}_{i=1}^{M}$, where the $\mathbf{x}_i$'s are drawn independently from $\mathscr{D}$.
They then assume that the kernel matrix $K_{ij}=\mathcal{K}(\mathbf{x}_i,\mathbf{x}_j)$ has unit trace (i.e. $\textrm{tr}(K)=1$), which can be achieved by scaling the kernel $\mathcal{K}$, and that the target function is $g(x)=y$ (i.e. $g(\mathbf{x}_i)=y_i$). 
Then for any $\epsilon>0$, with probability $1-\epsilon-\lambda_{\textrm{max}}M^4$, it is shown (in Theorem 3 of their Appendices) that
\begin{equation}
    \label{eq:Kubler-GeneralGeneralisationErrorBound}
    \|f-g\|_{L_2(\mathcal{X})}\geq\left(1-\sqrt{\frac{2\lambda_{\textrm{max}}M^2}{\epsilon}}\right)\|g\|_{L_2(\mathcal{X})},
\end{equation}
where $\lambda_{\textrm{max}}$ denotes the largest eigenvalue of $T_{\mathcal{K}}$.
Roughly this result shows that the trained model $f$ obtained via KRR will not be able to learn $g$ (in the sense of achieving small squared error, which is captured by $\|f-g\|_{L_2(\mathcal{X})}$, with high probability) if the largest eigenvalue $\lambda_{\textrm{max}}$ is small (close to 0).
Note that investigations of the spectrum of $T_{\mathcal{K}}$ are often performed by calculating the spectrum of the associated kernel matrix $K$, since the spectrum of $K$ will coincide with the spectrum of $T_{\mathcal{K}}$ in the limit of infinitely many data samples.

We now restrict our focus to the case of FQK $\mathcal{K}=\mathcal{K}_F$.
Given the associated quantum feature map $\rho$, one important characteristic of $\rho$ is the mean density matrix $\rho_{\mathscr{D}}$ defined such that
\begin{equation}
    \label{eq:MeanDensityMatrix}
    \rho_{\mathscr{D}}=\int_{\mathcal{X}}\rho(x)d\mathscr{D}(x).
\end{equation}
In~\cite{Kubler2021Inductive}, it is shown (Lemma 1 of the main text) that the largest eigenvalue $\lambda_{\textrm{max}}$ of the kernel integral operator $T_{\mathcal{K}_F}$ is bounded above by the purity of $\rho_{\mathscr{D}}$ defined in Eq.~\eqref{eq:MeanDensityMatrix}.
That is
\begin{equation}
    \label{eq:Kubler-KernelIntegralOperatorEigenvalueBound}
    \lambda_{\textrm{max}}\leq\textrm{tr}\left(\rho_{\mathscr{D}}^2\right).
\end{equation}

By considering Eqs.~\eqref{eq:Kubler-KernelIntegralOperatorEigenvalueBound} and \eqref{eq:Kubler-GeneralGeneralisationErrorBound} together, we see that $\lambda_{\textrm{max}}$ can only be large (close to 1), and thus allow the trained model $f\in\mathcal{R}_{\mathcal{K}_F}$ to generalise, if the mean density matrix $\rho_{\mathscr{D}}$ is close to a pure state.
Using this idea, under certain conditions on $\mathscr{D}$, the authors of~\cite{Kubler2021Inductive} show that if the purity of the mean density matrix becomes exponentially small with increasing numbers of qubits $n$, then with $M\in\mathcal{O}(\textrm{poly}(n))$ data samples (i.e. polynomially many data samples in the number of qubits) it will only be possible to achieve good generalisation with a finite number of qubits, beyond which good generalisation will not be possible (Theorem 1 of the main text).

This result places restrictions on the kinds of quantum feature maps that could give rise to models capable of generalising well with a polynomial amount of data as the number of qubits grows.
Specifically, we need to design quantum feature maps for which the mean density matrix maintains a good level of purity with increasing $n$.
If this is not the case, then the results of K\"ubler et al.~\cite{Kubler2021Inductive} imply that a number of datapoints scaling exponentially with $n$ will be necessary for learning, which would render kernel methods intractable.

Such outcomes are consistent with results that we have encountered earlier in this section. 
Specifically, the use of highly expressive (e.g. problem-agnostic) embeddings will result in a highly mixed mean density matrix with growing $n$, and hence provide poor generalisation performance in line with Eqs.~\eqref{eq:Kubler-GeneralGeneralisationErrorBound} and \eqref{eq:Kubler-KernelIntegralOperatorEigenvalueBound}.
In particular, K\"ubler et al. state that a quantum advantage might be attainable if the associated RKHS $\mathcal{R}_{\mathcal{K}_F}$ is low dimensional.
This draws parallels with Eq.~\eqref{eq:Huang-QuantumModelComplexityBound}, where the effective dimension $d$—which is related to the purity of the mean density operator $\rho_{\mathscr{D}}$—provides a bound on the generalisation capabilities of the problem-inspired quantum kernel $\mathcal{K}_{\tilde{\rho}}$.
Overall, the results which K\"ubler et al. present support ideas and themes that we have already encountered, specifically those pertaining to restricting the expressivity of the quantum feature maps which one employs in applications of QKMs, and instead opting for problem-inspired embeddings.

\subsection{Bandwidth tuning}
\label{sec:Bandwidth tuning}

In this section we discuss advantages and challenges associated with a hyperparameter for quantum kernels known as the quantum kernel bandwidth.
The quantum kernel bandwidth was originally introduced to help control the expressive power of quantum kernels, mitigate EC, and combat the generalisation challenges associated with fixed quantum feature maps. 
Such an approach was originally suggested by the authors of~\cite{Kubler2021Inductive} and was then formalised in~\cite{Shaydulin2022Importance}, with further analysis and considerations appearing in~\cite{Canatar2023Bandwidth,FlorezAblan2025Similarity}.
Here we discuss the advantages and challenges associated with tuning the quantum kernel bandwidth.

The bandwidth of a classical kernel is a well established concept~\cite{Scholkopf2001Kernels,Silverman1986Density}, describing how ``wide'' the kernel is when we fix the value of one of its inputs and consider how the value of the kernel decays as we vary the second.
To illustrate the idea, consider the radial basis function (RBF) kernel $\mathcal{K}_{\textrm{RBF}}:\mathcal{X}\times\mathcal{X}\to\R$, defined such that
\begin{equation}
    \label{eq:RBFKernel}
    \mathcal{K}_{\textrm{RBF}}(x,x^\prime)=\exp\left(-\gamma\|x-x^\prime\|^2\right),
\end{equation}
where $\|\cdot\|$ denotes the standard Euclidean norm.
In this case the bandwidth is controlled by the hyperparameter $\gamma>0$, with larger values of $\gamma$ resulting in a ``narrower'' (more expressive) kernel, and small values in a ``wider'' (less expressive) kernel.

In~\cite{Shaydulin2022Importance}, Shaydulin et al. formally introduce the \emph{quantum kernel bandwidth}, which we denote here by $\beta>0$. 
Specifically, the quantum kernel bandwidth plays the role of scaling datapoints $x\in\mathcal{X}$ as
\begin{equation}
    \label{eq:Bandwidth}
    x\mapsto\beta x.
\end{equation}
In many cases, tuning $\beta$ results in analogous changes in the value of a quantum kernel as tuning the bandwidth of a classical kernel, controlling the ``width'' (expressivity) of the kernel.

The idea of scaling data to control the expressivity of quantum models was originally considered by Schuld et al.~\cite{Schuld2021Effect}.
In said work, however, the authors didn't formally refer to this idea as bandwidth tuning, nor did they consider the relationships between the bandwidth scaling and increasing qubit numbers.
Nonetheless, extreme values of $\beta$ generally result in overfitting or underfitting, while intermediate values deliver good generalisation performance. 

To demonstrate this, Shaydulin et al.~\cite{Shaydulin2022Importance} provide numerical evidence on a number of standard classical datasets which shows that with no bandwidth tuning (i.e. $\beta=1$), prediction accuracies and kernel values decay exponentially with qubit count, consistent with the predictions of~\cite{Huang2021Power,Kubler2021Inductive,Thanasilp2024Exponential}.
However, by tuning the bandwidth, the decay of kernel values can be mitigated and prediction accuracies become competitive with the best accuracies achieved with classical models, in some cases even improving with qubit count.
This does not contradict the results of~\cite{Huang2021Power} however, since the authors of said work did not consider tuning the bandwidth in their analysis.

Canatar et al.~\cite{Canatar2023Bandwidth} investigate the idea that quantum kernel bandwidth might help to bolster generalisation by establishing a connection between bandwidth tuning and the spectrum of the kernel integral operator.
In particular, the authors demonstrate that varying the bandwidth parameter can shift the model from a regime in which generalisation to any target function is provably impossible, to one in which strong generalisation is achieved for well-aligned target functions (in the sense of~\cite{Canatar2021Alignment}).
Note that we will not discuss kernel alignment here though, since such a technique is often employed in a variational context and hence lies outside the scope of this review.

To support their claims, Canatar et al. consider the following learning problem which was used in~\cite{Huang2021Power} as an example of a simple problem which classical methods can easily learn, but for which QKMs would require exponentially many data samples to achieve the same.
Specifically, they consider a training dataset $\mathcal{D}=\{(\mathbf{x}_i,y_i)\}_{i=1}^{M}$ where the input data samples come from $\mathcal{X}=\{0,\pi\}^n\subset\R^n$ with associated labels given by $y_i=\cos((\mathbf{x}_i)_n)$, where $(\mathbf{x}_i)_n$ denotes the last component of $\mathbf{x}_i\in\mathcal{X}$.
Since this problem is equivalent to predicting the final component of the input, a linear regression model could easily be used to predict the labels using just $M=n$ training data samples with high probability.

To approach the problem with QKMs, consider using a quantum feature map $\rho:\mathcal{X}\to\mathcal{H}_n$ defined such that $\rho(x)=|\psi(x)\rangle\langle\psi(x)|$ where $\ket{\psi(x)}=\otimes_{j=1}^{n}\left(e^{-iXx_j}\ket{0}\right)$.
In other words, the encoding is just a product of 1-qubit unitaries which each rotate their qubit initially in the state $\ket{0}$ about the $x$-axis of the Bloch sphere with an angle proportional to the associated component of the input data sample.
A simple computation shows that for any $x,x^\prime\in\{0,\pi\}^n$ we have that the associated FQK is given by $\mathcal{K}_F(x,x^\prime)=\Pi_{j=1}^{n}\cos^2\left((x_j-x^\prime_j)\slash2\right)=\delta_{x,x^\prime}$, and so the kernel matrix is just the $M\times M$ identity matrix.
Accordingly, the trained model will perfectly fit the training data but will not be capable of generalising unless $M=2^n$ (i.e. all samples from $\{0,\pi\}^n$ are provided as training data).

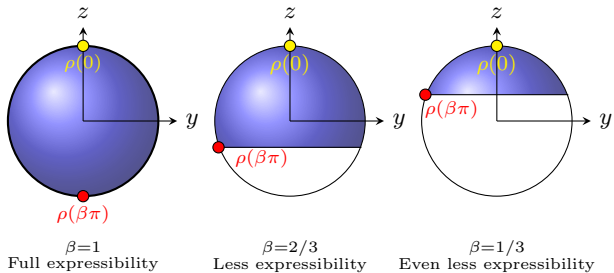
\begin{figure}
    \centering
    \begin{tikzpicture}
      \begin{scope}
        \shade[ball color=blue!75, opacity=0.75] (0,0) circle (1);
      \end{scope}
      \draw[-stealth] (0,0) --++ (0,1.25) node[above] {$z$};
      \draw[-stealth] (0,0) --++ (1.25,0) node[right] {$y$};
      \draw[thick] (1,0) arc (0:360:1);
      \node at (0,-1.8) {$\substack{\beta=1\\\textrm{Full expressibility}}$};
      \draw[fill=yellow] (0,1) circle (0.075cm) node[below] {\scriptsize\textcolor{yellow}{$\rho(0)$}};
      \draw[fill=red] (0,-1) circle (0.075cm) node[below] {\scriptsize\textcolor{red}{$\rho(\beta\pi)$}};
      \begin{scope}
        \clip (-1+2.75,-0.35) rectangle (1+2.75,1);
        \shade[ball color=blue!75, opacity=0.75] (0+2.75,0) circle (1);
      \end{scope}
      \draw (1.8,-0.35) --++ (1.9,0);
      \draw[-stealth] (0+2.75,0) --++ (0,1.25) node[above] {$z$};
      \draw[-stealth] (0+2.75,0) --++ (1.25,0) node[right] {$y$};
      \draw (1+2.75,0) arc (0:360:1);
      \node at (0+2.75,-1.8) {$\substack{\beta=2\slash3\\\textrm{Less expressibility}}$};
      \draw[fill=yellow] (0+2.75,1) circle (0.075cm) node[below] {\footnotesize\textcolor{yellow}{$\rho(0)$}};
      \draw[fill=red] (1.8,-0.35) circle (0.075cm); 
      \node at (1.8+0.575,-0.35-0.15) {\scriptsize\textcolor{red}{$\rho(\beta\pi)$}};
      \begin{scope}
        \clip (-1+2.75+2.75,0.35) rectangle (1+2.75+2.75,1);
        \shade[ball color=blue!75, opacity=0.75] (0+2.75+2.75,0) circle (1);
      \end{scope}
      \draw (1.8+2.75,0.35) --++ (1.9,0);
      \draw[-stealth] (0+2.75+2.75,0) --++ (0,1.25) node[above] {$z$};
      \draw[-stealth] (0+2.75+2.75,0) --++ (1.25,0) node[right] {$y$};
      \draw (1+2.75+2.75,0) arc (0:360:1);
      \node at (0+2.75+2.75,-1.8) {$\substack{\beta=1\slash3\\\textrm{Even less expressibility}}$};
      \draw[fill=yellow] (0+2.75+2.75,1) circle (0.075cm) node[below] {\footnotesize\textcolor{yellow}{$\rho(0)$}};
      \draw[fill=red] (1.8+2.75,0.35) circle (0.075cm);
      \node at (1.8+2.75+0.35,0.35-0.2) {\scriptsize\textcolor{red}{$\rho(\beta\pi)$}};
    \end{tikzpicture}
    \caption{Reducing the quantum bandwidth $\beta$ reduces the expressibility of the associated quantum feature map.
    Above we see the $yz$-plane of the Bloch sphere, with the $x$-axis pointing out of the page. 
    The figure shows the states prepared by the 1-qubit feature map $\rho(\beta x)=e^{-iX\beta x}|0\rangle\langle0|e^{iX\beta x}$ for $x\in\{0,\pi\}$ with different values of $\beta$. 
    As $\beta$ gets smaller, the states $\rho(0)$ and $\rho(\beta\pi)$ get closer together, hence reducing the expressibility of the feature map.}
    \label{fig:EffectOfBandwidthOnExpressibility}
\end{figure}

If we instead introduce the quantum kernel bandwidth by scaling the inputs $x\mapsto\beta x$ however, then the FQK becomes $\mathcal{K}(x,x^\prime)=\Pi_{j=1}^{n}\cos^2\left(\beta(x_j-x^\prime_j)\slash2\right)$, resulting in a kernel matrix which has non-zero off-diagonal entries.
In this case with some $\beta\in(0,1)$, Canatar et al.~\cite{Canatar2023Bandwidth} provide numerical evidence showing that the trained model can generalise and accurately predict the labels of unseen inputs.
Informally, this is a result of the fact that smaller values of $\beta$ reduce the expressibility of the feature map, so the states $\rho(\beta x)$ and $\rho(\beta x^\prime)$ get closer together and are no longer orthogonal when $\beta<1$ (see Fig.~\ref{fig:EffectOfBandwidthOnExpressibility}).

The authors extend this analysis by considering the same bandwidth-equipped quantum kernel together with the uniform distribution $\mathscr{D}=\textrm{Unif}([-\pi,\pi]^n)$ over $\mathcal{X}=[-\pi,\pi]^n$. 
They derive closed-form expressions for the associated kernel integral operator $T_{\mathcal{K}_F}$ in terms of the bandwidth $\beta$. 
In general, with reference to Eq.~\eqref{eq:Kubler-KernelIntegralOperatorEigenvalueBound}, the eigenvalues of the FQK decay exponentially with $n$ due to the mean density matrix becoming more mixed. 
However, they show that this behaviour can be mitigated: if the bandwidth is scaled as $\beta(n)=a n^{-\xi}$ with $a\in\mathcal{O}(1)$ and $\xi \geq \tfrac{1}{2}$, the purity of the mean density matrix remains constant (and may even increase) with $n$. Consequently, the largest eigenvalue of the kernel integral operator remains constant, while the remaining eigenvalues decay only polynomially. This implies that generalisation can persist in the large-$n$ limit, in contrast to the result of K\"ubler et al.~\cite{Kubler2021Inductive}, where generalisation is predicted to break down beyond a finite number of qubits.

Clearly, these results are promising.
Tuning the bandwidth of a quantum kernel can mitigate issues with the spectrum of the kernel integral operator, and enable generalisation at large qubit numbers.
However, subsequent work has highlighted potential limitations of this approach. 
In particular, Slattery et al.~\cite{Slattery2023Numerical} provide numerical evidence that, when FQKs are tuned to improve generalisation on classical datasets, the resulting bandwidth-tuned kernels become well approximated by classical kernels.
To demonstrate this, the authors compute the classical-quantum geometric difference $g_{CQ}$ (see Section~\ref{subsec:Necessary conditions for quantum advantage}) between the optimally bandwidth-tuned quantum kernels and a range of classical kernels.
From these results, the authors observe that this geometric difference is small, indicating that the tuned quantum kernels are structurally similar to classically simulable models.

Fl\'orez-Alban et al.~\cite{FlorezAblan2025Similarity} provide further evidence in this direction, showing how bandwidth tuned FQKs and PQKs have kernel integral operators with similar spectrums, low geometric differences, and kernel matrices with low Frobenius distances to classical RBF and polynomial kernels.
Of course these studies are empirical in nature, with insights deriving from specific problem instances.
However, if this behaviour persists more generally, it raises concerns that bandwidth tuning, while alleviating kernel concentration and improving generalisation, may simultaneously erode the possibility of obtaining a genuine quantum advantage on classical learning tasks.

Worth noting is that the problems considered in~\cite{Slattery2023Numerical,FlorezAblan2025Similarity} are widely classical in nature, raising a question about whether the same outcome would be observed when using a problem-inspired quantum feature map.
For example, Slattery et al.~\cite{Slattery2023Numerical} state that they do not expect to observe a decaying $g_{CQ}$ for the problem-inspired feature map used by Liu et al.~\cite{Liu2021Rigorous} for the discrete logarithm problem, which we briefly discuss in the next section.

\section{Structured problems enabling potential advantage in quantum kernel methods}
\label{sec:Structured problems enabling potential advantage in quantum kernel methods}

In this section, we review prior work on structured problem settings for QKMs.
These works consider learning tasks whose structure is intentionally inspired by, or compatible with, a specific quantum kernel.
These studies therefore provide routes to quantum advantage that are either established through fully rigorous analyses or supported by arguments relying on additional assumptions or partial theoretical justifications. 
Despite the latter’s less comprehensive formal grounding, such results are widely regarded within the community as promising directions toward advantage.
Note that here we do not aim to provide precise mathematical details, instead we provide high-level descriptions of the problems under consideration and the sources of the proposed advantages achieved on these problems.
For further details we refer the reader to the cited papers.

\subsection{IQP-based dataset}

Havl\'i\v cek et al.~\cite{Havlicek2019Supervised} introduce a dataset designed to align with quantum feature maps generated by commuting quantum circuits, serving as one of the first indications that advantages may be attainable with QKMs. 
In this construction, classical data is embedded into the parameters of an instantaneous quantum polynomial (IQP) circuit~\cite{Bremner2010IQP}, and the resulting kernel values are determined using the FQK. 
The dataset exemplifies a potential route to quantum advantage: under standard complexity-theoretic assumptions, the corresponding IQP circuits are believed to be hard to simulate classically. 
However, assessing the robustness of this advantage is challenging, as it is tied to the classical hardness of simulating the feature map. 
In particular, the authors do not rule out the possibility that classical learners could achieve good performance on the IQP-based dataset, but instead argue that the encoding and conjectured hardness of kernel evaluation may enable an advantage, without formally establishing classical intractability of the learning task.

\subsection{Shor's algorithm and the discrete logarithm}

One of the first rigorously provable quantum advantages obtained with QKMs was provided by Liu et al.~\cite{Liu2021Rigorous}. 
In this work, the authors consider a learning problem closely tied to the discrete logarithm problem (DLP), a cornerstone problem in computational number theory and cryptography.

Under standard and widely accepted complexity-theoretic assumptions, solving the DLP is believed to be intractable for all classical polynomial-time algorithms. 
Leveraging this assumption, Liu et al. show that any classical learner cannot predict the labels with accuracy inverse-polynomially better than random guessing. 
In contrast, they construct a quantum kernel whose feature map efficiently encodes information about the discrete logarithm into quantum states using Shor's algorithm~\cite{Shor1997Factoring} as a subroutine, allowing a quantum kernel machine to classify the data with high accuracy using polynomial resources. 
This establishes a provable quantum advantage in a learning-theoretic sense, conditioned on the hardness of the discrete logarithm problem.

Relatedly, Huang et al.~\cite{Huang2021Power} extend the construction of~\cite{Liu2021Rigorous} to PQKs, demonstrating that the discrete logarithm-based advantage persists even when the kernel is defined through local reduced density matrices rather than full-state fidelities. 
This result shows that provable quantum advantages are not restricted to global fidelity kernels, but can also arise in more experimentally accessible kernel constructions.

\subsection{Grover-based preprocessing and pattern matching}

A different type of provable advantage was later established by Muser et al.~\cite{Muser2024Provable}, who consider kernel-based quantum learners augmented with quantum preprocessing inspired by Grover’s search algorithm~\cite{Grover1996Fast}. 
Using their framework, this work studies learning tasks related to pattern matching and structured search problems.

In their construction, classical data are first embedded into quantum states using a preprocessing routine that amplifies features associated with marked or relevant patterns using amplitude amplification. 
The resulting quantum states are then compared using kernel estimations, effectively defining a quantum kernel that encodes information about the presence or absence of specific patterns within the data. 
The authors prove that, for certain families of pattern-matching problems, this quantum preprocessing leads to a kernel that can be robustly estimated on a quantum computer with polynomially many shots, while any classical learner would require super-polynomial resources to achieve comparable performance.

Importantly, the advantage in this setting arises not solely from the kernel estimation itself, but from the combination of quantum preprocessing and kernel-based learning. 
This highlights a broader paradigm in which quantum kernels can inherit provable advantages from well-understood quantum algorithmic primitives, such as Grover's search algorithm, when applied to suitably structured learning problems.

\subsection{Quantum phase recognition}

Wu et al.~\cite{Wu2023Phase} consider quantum phase recognition (QPR) in many-body systems, proposing a quantum kernel based on ground-state properties of parameterised Hamiltonians. 
They prove, under widely accepted complexity assumptions, that a variety of QPR problems exist which cannot be solved by classing learning algorithms with access to classical resources.
In contrast, they also prove that QKMs can be used to efficiently solve such problems, suggesting the possibility of quantum advantages in this context.

The data embedding they suggest for solving QPR problems map classical Hamiltonian parameters to quantum states via variational ground-state preparation, after which kernel values are computed using FQKs. 
While this work is certainly impactful and addresses highly relevant QPR problems, one limitation of their framework is that their numerical evidence was obtained using the exact ground-states of the Hamiltonians under consideration.
To justify the use of this feature map, they suggest that such a mapping can be achieved variationally, however this claim was not rigorously justified.
Given the known challenges of variational quantum approaches for complex many-body systems, this raises concerns about the practical feasibility of their proposed solution.

\subsection{Data with group structure}

Glick et al.~\cite{Glick2024Covariant} consider a class of problems involving datasets with inherent group structure.
Specifically, they consider problems which they call labelling cosets with error, in which a group $G$ and a subgroup $S$ of $G$ are given.
Using $S$ and two distinct group elements, two cosets $S_{\pm}$ are generated.
The training dataset then contains elements of the cosets which have been slightly perturbed, together with a binary label indicating which coset these elements belong to prior to being perturbed.
The learner is then tasked to classify which coset unseen perturbed elements belonged to.

To tackle such problems, the authors exploited covariant quantum kernels with feature maps constructed from unitary representations of group elements applied to a variational initial state. 
The resulting kernels respect symmetry-induced equivalence classes and are explicitly tailored to problems where invariance or equivariance is essential. 
Since the DLP considered in~\cite{Liu2021Rigorous} is an example of such a problem, the authors infer that there exists instances of such problems where the use of covariant quantum kernels could provide a rigorous speed-up.

While this direction of research certainly seems promising, we note that the variational nature of the initial state in the definition of their kernels may pose trainability issues in practice.
Nonetheless we include a discussion of the work performed by Glick et al., since the covariant kernels they consider are great examples of the kind of problem-informed quantum kernels that we expect to provide practical advantages in the future.

Henderson et al.~\cite{Henderson2025Advantage} further analyse these constructions and prove that such symmetry-structured kernels do not suffer from EC, showing that the variance and mean of kernel values do not decay exponentially in $n$. 
While this represents a significant structural advantage over generic quantum kernels, as a result of the variational nature of the initial states, it may not constitute a full quantum-classical separation, as comparable symmetry-aware classical kernels may exist for certain groups.

\subsection{Maximal geometric difference}

Huang et al.~\cite{Huang2021Power} propose a method for artificially constructing datasets which exhibit the largest possible separation between classical and quantum model complexities.
The dataset is constructed to ensure that $s_C=g_{CQ}^2s_Q$ and hence saturates the bound in Eq.~\eqref{eq:GeometricDifference-ModelComplexitySeparation}.
In Supplementary Section 7 of~\cite{Huang2021Power}, the authors discuss how they expect that their artificially constructed datasets could provide the first genuine quantum advantages in a classification problem using QML, since the dataset provably provides the largest performance separation between quantum and classical models.
The authors take this one step further and claim that if such a dataset cannot provide a quantum advantage, then they expect that it is likely no quantum advantage for classification in QML will exist.

\section{Benchmarking, comparative, and hardware implementation studies}
\label{sec:Benchmarking, comparative, and hardware implementation studies}

In this section, we review applications of QKMs to classical datasets, considering both numerical simulations and experimental implementations. 
We focus on studies that tackle real-world tasks, compare quantum and classical kernel performance, or report hardware-based implementations.
While a broad range of the papers cited previously in this review, including~\cite{Schuld2019Feature,Huang2021Power,Wang2021Towards,Banchi2021Generalization,Shaydulin2022Importance,Heyraud2022NoisyQKMs,Gan2023Unified,Canatar2023Bandwidth,Peters2023Generalization,Jerbi2023Beyond,Wu2023Phase,Slattery2023Numerical,Glick2024Covariant,Gentinetta2024Complexity,Miroszewski2024QKMShots,Thanasilp2024Exponential,Suzuki2024QuantumFisher,Zhou2024QKE-QSVR,FlorezAblan2025Similarity}, contain numerical analyses, we do not attempt an exhaustive survey here.
Instead, we highlight representative studies with the aim of summarising the principal research directions and distilling the common trends and conclusions that emerge.

\subsection{Applications of quantum kernels to real-world datasets}
\label{sec:Applications of quantum kernels to real-world datasets}

A growing body of work has explored the application of QKMs to real-world datasets spanning regression, image classification, biomedical analysis, remote sensing, and finance. 
These studies typically focus on assessing the empirical performance of QKMs with KRR or SVMs applied to classical data, often under severe constraints on system size and feature dimensionality due to the inherent restrictions imposed by classical simulations.

Early investigations demonstrated the feasibility of applying quantum kernels beyond synthetic benchmarks. 
For instance, Paine et al.~\cite{Paine2023QKMDifferential} studied regression and differential equation solving using quantum kernels, illustrating that kernel-based quantum models can be adapted to continuous-output tasks. 
Similarly, Beaulieu et al.~\cite{Beaulieu2022ManufacturingDefects} applied QSVMs to image classification of real-world manufacturing defects, while Ragab et al.~\cite{Ragab2022Biomedical} and Krunic et al.~\cite{Krunic2022Health} consider biomedical and healthcare datasets, including electronic health records.

More recent works have expanded this empirical program to larger and more structured datasets. 
Zhuang et al.~\cite{Zhuang2024NonHemolytic} investigate peptide classification, while Miroszewski et al.~\cite{Miroszewski2023Clouds} and Wijata et al.~\cite{Wijata2024Soil} applied QKMs to multispectral and hyperspectral satellite imagery. 
In the financial domain, Miyabe et al.~\cite{Miyabe2023Financial} explore quantum multiple kernel learning for classification tasks, combining several quantum kernels within a single learning framework.

These application-driven studies demonstrate that QKMs can be deployed on real-world datasets and integrated into standard machine learning workflows. 
Across these works, quantum kernels generally achieve competitive performance with classical methods on small to moderately sized datasets, particularly when the data exhibit clear underlying structure. 
However, consistent advantages over well-tuned classical baselines are rare, and comparable accuracy is often attainable with classical kernels, consistent with the findings of Slattery et al.~\cite{Slattery2023Numerical}. 
These results reinforce recurring challenges such as limited scalability and the difficulty of isolating genuine quantum advantages in practical learning scenarios.

\subsection{Comparisons between classical and quantum kernel models}
\label{sec:Comparisons between classical and quantum kernel models}

A number of recent studies have undertaken systematic comparisons between quantum and classical kernel methods, with the explicit goal of assessing whether quantum kernels offer measurable advantages when applied to controlled and reproducible benchmarking protocols. 
These works typically emphasise fair baselines, careful hyperparameter optimisation, and transparent reporting practices, addressing concerns raised in earlier empirical studies.

Comprehensive benchmarking efforts, such as those by Schnabel et al.~\cite{Schnabel2025Scrutiny}, Alvarez-Estevez et al.~\cite{AlvarezEstevez2025Benchmarking}, and Abdulsalam et al.~\cite{Abdulsalam2025Comparative}, compare quantum SVMs against a range of classical kernels across multiple datasets and classification tasks. 
A recurring conclusion across these studies is that, when classical baselines are carefully tuned, quantum kernels rarely demonstrate consistent or statistically significant performance improvements. 
In many cases, classical kernels with comparable effective complexity achieve similar or superior accuracy.

Hyperparameter selection has emerged as a particularly important confounding factor in such comparisons. 
Egginger et al.~\cite{Egginger2024Hyperparameter} systematically investigate the impact of hyperparameter tuning on quantum kernel performance, demonstrating that kernel bandwidth, feature scaling, and regularization choices can dominate the observed performance. 
Similar conclusions were made by Schnabel et al.~\cite{Schnabel2025Scrutiny}.
These findings suggest that reported advantages in earlier studies may sometimes be attributable to unequal hyperparameter optimisation, rather than intrinsic differences between quantum and classical kernels.

Methodological considerations are further emphasised in the critical analysis by Bowles et al.~\cite{Bowles2024Benchmarking}, which highlights common pitfalls in benchmarking quantum machine learning models, including dataset leakage, mismatched baselines, and selective reporting. 
Their analysis underscores the importance of rigorous experimental design and motivates standardised benchmarking practices, a theme echoed in subsequent comparative studies.

Taken together, these works paint a consistent picture: while quantum kernels can be competitive with classical kernels on certain tasks and datasets, robust empirical evidence for a generic performance advantage remains limited. 
Instead, performance appears to depend sensitively on dataset structure, kernel design, and hyperparameter choices. 
These benchmarking studies therefore reinforce the broader conclusion that identifying regimes of genuine quantum advantage requires not only novel kernel constructions, but also careful experimental methodology and fair classical comparisons.

\subsection{Hardware implementations of quantum kernel methods}
\label{sec:Hardware implementations of quantum kernel methods}

While much of the research on QKMs has focused on theory and synthetic tasks, several recent studies demonstrate proof-of-principle implementations on real quantum devices. 
These experiments, conducted on superconducting, ion-trap, or photonic devices, estimate kernel entries from quantum circuits and apply them to small-scale classification tasks. 
Once again, here we do not aim to provide a comprehensive survey; rather, we highlight these works to give an overview of recent hardware results, illustrating how hardware and shot noise, together with shallow circuit depth, affect kernel quality and classifier performance.

\paragraph{Superconducting hardware.}
Early experimental evidence was provided by Havl\'i\v cek et al.~\cite{Havlicek2019Supervised}, who demonstrated both a variational classifier and a quantum SVC using a superconducting quantum processor. 
In their experiments, kernel matrix elements were estimated using shallow circuits on a five-qubit device, with only two qubits actively used in the kernel construction and 50,000 measurement shots per entry to suppress sampling noise. 
While limited in scale, this work establishes the basic feasibility of estimating quantum kernels on near-term devices.

Subsequent studies explore more demanding settings. 
Peters et al.~\cite{Peters2021Noisy} implemented a non-variational QKM on a 17-qubit superconducting processor, applying it to classify uncompressed 67-dimensional classical data. 
By employing error mitigation techniques tailored to quantum kernel estimation, they reported classification performance on hardware that was comparable to noiseless simulation, highlighting a degree of robustness of kernel-based approaches to hardware noise. 
Similarly, Wu et al.~\cite{Wu2021Application} applied a quantum kernel algorithm to a high-energy physics classification task using data from the Large Hadron Collider.
They reported results from both simulation and superconducting hardware runs that demonstrate the practical viability of quantum kernel estimation for domain-relevant datasets, albeit at small scale.

More recently, Agnihotri et al.~\cite{Agnihotri2026Practical} performed a systematic evaluation of quantum SVMs for radar micro-Doppler classification on IBM superconducting processors. 
Their study compares noiseless simulation with hardware execution, analyses the impact of shot noise and decoherence, and shows that while noise degrades kernel quality, meaningful classification performance can still be obtained on current devices.

\paragraph{Trapped-ion hardware.}
Suzuki et al.~\cite{Suzuki2024TrappedIon} investigated QKMs for both classification and regression using a trapped-ion quantum computer. 
Employing shallow circuits and a small number of qubits, they found that hardware performance closely tracked noiseless simulations across several benchmark datasets.
Their results hence suggest that trapped-ion platforms may offer favourable noise characteristics for quantum kernel estimation in low-depth regimes.

\paragraph{Photonic hardware.}
Beyond qubit-based architectures, several works demonstrated QKMs on photonic platforms. 
Anai et al.~\cite{Anai2024Continuous} implemented a continuous-variable quantum kernel on a programmable photonic quantum processor, showing that kernel-based classification can be realised experimentally despite imperfections inherent to photonic systems. 
Yin et al.~\cite{Yin2025Experimental} further demonstrated an integrated photonic implementation of a QKM, reporting classification performance that in some cases exceeded that of standard classical kernels.

Hardware demonstrations across platforms and domains show that quantum kernels can be reliably estimated on real devices and integrated into QML workflows. Although noise and finite sampling degrade performance compared to ideal simulations, shallow circuits with error mitigation often yield results close to noiseless cases. These experiments thus support QKMs for promising near-term quantum applications, while highlighting the need to overcome scalability and robustness challenges.

\section{Conclusions and perspectives}
\label{sec:Conclusions and perspectives}

This review has outlined the theoretical foundations, practical implementations, and current limitations of non-variational QKMs. 
By tracing the path from classical kernel theory to quantum feature maps, we have clarified both the promise and distinctive challenges introduced by quantum embeddings. 
The necessary conditions for quantum advantage, together with the challenges posed by EC, dequantisation pathways, and the spectra of the associated kernel integral operators highlights how demanding the realisation of practical advantage is in this setting.
At the same time, structured problem classes and carefully designed feature maps, in addition to the hardware implementations that closely resemble noiseless simulations, illustrate that advantage is not excluded in principle. 
Together, these perspectives provide a consolidated framework for assessing quantum kernel models and identify the conceptual, methodological, and technical advances required to move from theoretical promise to demonstrable quantum advantage in machine learning.

Looking ahead, we highlight several directions for future research aimed at either uncovering viable routes to practical quantum advantage with non-variational QKMs, or clarifying their limitations. 
A central theme in the literature is the pivotal role of a standardised framework for fair comparisons between quantum and classical kernel models. 
Progress in this direction requires us to move beyond purely empirical comparisons, and explicitly verify the necessary conditions for quantum advantage discussed in Section~\ref{sec:Assessing quantum advantage in quantum kernel methods}. 
Similarly, it is important to consider whether a given quantum kernel exhibits concentration effects, and whether the underlying entanglement structure may admit classical simulations.
Ultimately, the challenge is not merely to demonstrate classical intractability and practical feasibility, but to ensure that the structural properties of proposed constructions yield genuine improvements in learning performance. 

Dequantisation results indicate that 1D architectures and circuits are unlikely to yield quantum advantage, given the efficiency of classical MPS methods in such settings. 
In contrast, 2D architectures and circuits with non-trivial connectivity remain comparatively less well understood, and their classical simulability is still an open area of investigation. 
While PEPS provide powerful approximations, their contraction and general use remain computationally demanding. 
This suggests, though does not rigorously establish, that quantum feature maps with inherent 2D entanglement structures may offer more viable routes to advantage. 
The possibility appears particularly relevant for FQKs, where entanglement has not yet been shown to induce EC effects, although further study is required. 
These considerations motivate systematic investigations into how lattice geometry, connectivity, and circuit depth influence both entanglement scaling and classical simulability. 

In terms of problem instances which permit problem-inspired quantum feature maps, we suggest that datasets capturing properties of 2D local Hamiltonians with area-law entanglement scaling could be fruitful.
For example, such datasets may contain inputs describing specific Hamiltonians or Hamiltonian evolutions in terms of some parameters, with labels capturing physical properties.
Examples of such properties include quantum phases, non-local expectation values, entanglement entropies, mutual information, ground-state energies, or dynamical quantities such as out-of-time-ordered correlators~\cite{Xu2023Tutorial} and Loschmidt echoes~\cite{Wisniacki2012Loschmidt}. 
In these settings, a natural class of data-encoding unitaries to explore would be the associated time-evolution operators, which encode information about the underlying Hamiltonians in an experimentally accessible and physically meaningful manner. 

The choice of initial states for such quantum feature maps is less obvious however, and remains an important open design question.
A similar line of work aimed at providing non-variational approaches to choosing initial states would be beneficial, particularly in the context of the covariant quantum kernels~\cite{Glick2024Covariant} given that such kernels have been shown not to exhibit EC~\cite{Henderson2025Advantage}.

Existing limitations in QKMs, including many of those discussed in Section~\ref{sec:Challenges for quantum kernel methods}, are often derived under assumptions about the data distribution. 
Such analyses typically rely on either factorised or independently sampled distributions: assumptions which may not hold for structured datasets such as images, time series, or biological measurements. 
It remains unclear to what extent concentration arguments capture behaviour on such structured data. 
These considerations highlight the need for caution in generalising negative results too broadly, especially in domains such as medicine or finance where data distributions often deviate from theoretical assumptions. 
As such, clarifying how data distributions shape kernel concentration and spectral properties remains an important open challenge.

Relatedly, while growing evidence suggests that problem‑agnostic quantum feature maps rarely outperform strong classical baselines on standard classical datasets, active work continues on tailoring maps to problem-specific structures. 
Automatic design strategies, such as the genetic algorithm inspired approach proposed by Altares-L\'opez et al.~\cite{AltaresLopez2021Automatic}, illustrate that non‑variational constructions can adapt embeddings to dataset geometry. 
More broadly, datasets with intrinsic quantum or physical structure may provide more compelling settings. 
In these cases, feature maps may be engineered to encode locality, symmetry, conserved quantities, or dynamical evolution, embedding task‑relevant structure directly into the quantum state in ways that may be less accessible to classical replication. 
In this regard, embedding intrinsic structure into quantum states through problem-inspired feature maps may provide one of the most compelling routes to quantum advantage.

An additional consideration concerns bandwidth and spectral scaling. 
When bandwidth decreases too rapidly with qubit count, geometric distinctions between quantum and classical kernels may disappear. 
In contrast, intermediate regimes involving slower rates of decay of leading eigenvalues may provide scenarios in which quantum kernels can preserve meaningful structure while keeping resource demands manageable. 
Systematically clarifying whether such intermediate regimes exist will be crucial for determining whether quantum kernels can achieve predictive advantages without incurring exponential costs.

Finally, a central open question concerns hardware-specific resource estimates for large-scale implementations of QKMs. 
While prior studies have incorporated simplified or theoretically motivated noise models into analyses of quantum kernel performance, systematic investigations that integrate architecture-dependent constraints—such as connectivity, crosstalk, calibration drift, state preparation, measurement asymmetries, and error-mitigation overhead—remain limited. 
This gap is particularly significant for QKMs, where the quadratic scaling in the number of required kernel values, combined with shot noise and potential exponential concentration effects, can substantially amplify hardware-induced costs. 
Rigorous, hardware-aware scaling studies would therefore help clarify whether expected advantages survive under realistic experimental conditions, and could identify device-specific strategies for circuit design and error mitigation that improve practical feasibility.

Taken together, these directions suggest that the future of non‑variational QKMs lies not in generic, problem‑agnostic embeddings, but in carefully structured feature maps that align entanglement geometry, spectral properties, symmetry, and hardware considerations with the intrinsic structure of the learning task. 
Ultimately, realising meaningful quantum advantage with non-variational QKMs will hinge on whether task-aligned feature maps can simultaneously evade EC and dequantisation, withstand realistic noise, and deliver verifiable predictive gains over classical baselines.

\bibliographystyle{unsrt}

\bibliography{Bibliography}

\clearpage

\appendix

\end{document}